\documentclass[a4paper,11pt]{article}
\pdfoutput=1 

\usepackage{setspace}
\usepackage{jheppub}
\usepackage{xcolor}

\usepackage[T1]{fontenc} 
\usepackage{bbold}
\usepackage{amsmath}
\usepackage{amssymb}
\usepackage{mathtools}
\usepackage{stmaryrd}
\let\a=\alpha 
\let\b=\beta 
\let\g=\gamma 
\let\d=\delta 
\let\e=\epsilon

\let\l=\lambda 
\let\m=\mu

\let\s=\sigma 
\let\t=\tau

\let\y=\psi
\let\w=\omega  

\let\D=\Delta 
 
\let\L=\Lambda

\let\Y=\Psi
 
\let\W=\Omega

\let\del=\partial
\let\<=\langle
\let\>=\rangle

\newcommand{\DD}{\mathcal{D}}

\newcommand{\NN}{\mathcal{N}}
\newcommand{\OO}{\mathcal{O}}

\def\be#1\ee{\begin{align}#1\end{align}}

\newcommand{\op}[1]{\operatorname{#1}}
\newcommand{\al}[1]{\begin{align}#1\end{align}}
\newcommand{\spl}[1]{\begin{split}#1\end{split}}
\newcommand{\ga}[1]{\begin{gathered}#1\end{gathered}}
\def \bmat {\begin{pmatrix}}
\def \emat {\end{pmatrix}}

\usepackage{graphicx}
\usepackage{multirow}
\usepackage{subcaption}
\usepackage{diagbox}

\title{Gravity from quantum mechanics of finite matrices}

\author[a]{Shota Komatsu}
\author[b]{Adrien Martina}
\author[b]{Joao Penedones}
\author[b,c]{Noé Suchel}
\author[b]{Antoine Vuignier}
\author[b]{Xiang Zhao}
\affiliation[a]{CERN, Theoretical Physics Department,
CH-1211 Geneva 23, Switzerland}
\affiliation[b]{Fields and Strings Laboratory, Institute of Physics, Ecole Polytechnique Federale de Lausanne (EPFL),
CH-1015 Lausanne, Switzerland}
\affiliation[c]{Laboratoire de Physique Théorique,
de l’Ecole Normale Supérieure, PSL University, CNRS, Sorbonne Universités, UPMC Univ. Paris 06
24 rue Lhomond, 75231 Paris Cedex 05, France
}

\abstract{
We revisit the Berenstein-Maldacena-Nastase (BMN) conjecture relating M-theory on a PP-wave background and Matrix Quantum Mechanics (MQM) of $N\times N$ matrices. 
In particular, we study the BMN MQM at strong coupling and finite $N$ and derive an effective Hamiltonian that  describes non-relativistic free particles in a harmonic trap.
The energy spectrum predicted by this Hamiltonian matches the supergravity excitation spectrum around the PP-wave background, if we further assume the existence of  bound states.
Our derivation is based on the strong coupling expansion of the wavefunction and supersedes the naive path integral approach that can lead to incorrect results, as we demonstrate in a simple toy model.
We conclude with  open questions about various regimes of the theory when we vary the size of the matrices, the coupling and the temperature.

}

\begin{document} 
\maketitle
\flushbottom

\section{Introduction and summary}

The geometric nature of spacetime dynamics may just be an emergent approximate  description of some systems with many strongly coupled  degrees of freedom.
This is suggested by the holographic principle \cite{tHooft:1993dmi,Susskind:1994vu} and realized in the  examples provided by the AdS/CFT correspondence \cite{Maldacena:1997re,Gubser:1998bc,Witten:1998qj}.
We would like to study these examples to distill the basic mechanism that leads to the emergence of a gravitational description of some quantum systems.
In this paper, we focus on Matrix Quantum Mechanics (MQM) to avoid difficulties associated with continuum Quantum Field Theory. Our goal is to develop an intuitive picture of a quantum system with finitely many degrees of freedom (d.o.f.) in the regime where it makes contact with a gravitational description.

\paragraph{Matrix Quantum Mechanics:}

The system we choose to study is the model introduced by Berenstein, Maldacena and Nastase (BMN) \cite{Berenstein:2002jq}. 
The Hamiltonian is (following the conventions of \cite{Maldacena:2002rb}):
\begin{align} 
H= 
&R \,{\rm Tr} \left[\frac{1}{2} \sum_{I=1}^9 (P^I)^2 
-\frac{1}{4\ell_P^6}[X^I,X^J]^2 - \frac{1}{2\ell_P^3}\hat\Theta^\top\gamma^I [X^I,\hat\Theta] \right] +\label{eq: BMN Hamiltonian with l_P} \\
&+\frac{R}{2} \,{\rm Tr} \left[ \left(\frac{\mu}{3R}\right)^2 \sum_{i=1}^3 (X^i)^2 
+\left(\frac{\mu}{6R}\right)^2 \sum_{p=4}^9 (X^p)^2 
+i\frac{\mu}{4R}\hat\Theta^\top\gamma^{123} \hat\Theta
+i\frac{2\mu}{3R\ell_P^3}\epsilon_{ijk} X^i X^j X^k
\right]
\nonumber
\end{align}
where $R, \ell_P$ and $1/\mu$ are parameters with dimensions of length. The matrices $X^I$ are $N\times N$ hermitian matrices. The indices $I,J\in \{1,2,\dots,9\}$ while $i,j,k\in \{1,2,3\}$ and $p\in \{4,5,\dots 9\}$.
The Majorana fermions  $\Theta_\a$ (with $\a\in \{1,\ldots,16\}$) are also hermitian $N\times N$ matrices.
The gamma matrices are $16\times 16$, real and symmetric
and $\gamma^{123} = \g^1\g^2\g^3$.  
Setting $\mu=0$ one obtains the MQM of Banks, Fischler, Shenker and Susskind (BFSS)  \cite{Banks:1996vh}.

The BMN model is supersymmetric and has global symmetry $SO(3) \times SO(6) \times SU(N)$, as we review in section \ref{Sec: free gravitons from strong coupling}.
We  consider the ungauged model and treat $SU(N)$ as a global symmetry. Of course the physical states of the gauged model are just the $SU(N)$ singlets of the ungauged model. 
The model only has two dimensionless parameters, namely $N$ and the coupling 
\begin{align}
g^2\equiv \frac{R^3}{\ell_P^6 \mu^3}\,.
\label{eq: BMN coupling g}
\end{align}
To see that we rescale $X^I\to  \sqrt{\frac{R}{\mu}} X^I = g^{1/3}X^I \ell_P$. This leads to
\begin{align}\spl{
H/\mu &= {\rm Tr} \left[\frac{1}{2} \sum_{I=1}^9 (P^I)^2 -\frac{g^2}{4}[X^I,X^J]^2 - \frac{g}{2}\hat{\Theta}^\top\gamma^I [X^I,\hat{\Theta}] \right] 
 \\
&+\frac{1}{2} \,{\rm Tr} \left[  \frac{1}{3^2} \sum_{i=1}^3 (X^i)^2 
+ \frac{1}{6^2}  \sum_{p=4}^9 (X^p)^2 
+i\frac{1}{4}\hat{\Theta}^\top\gamma^{123} \hat{\Theta}
+i\frac{2g}{3}\epsilon_{i j k} X^i X^j X^k
\right]
}
\label{eq: BMN Hamiltonian}
\end{align}
where now $\mu$ is the only dimensionful parameter.
For any $g$, there are degenerate ground states (with zero energy) labeled by integer partitions of $N$. These are easy to describe at weak coupling as we review  in section \ref{Sec: free gravitons from strong coupling}.

In this paper, we study the theory in the strong coupling limit $g\to \infty$ at fixed $N$. In this limit, the off-diagonal elements $X_{ab}$ and $\hat{\Theta}_{ab}$ for $a\neq b$ become fast oscillators with frequency of  order $g(X_{aa}-X_{bb})$. On the other hand, the diagonal elements remain slow with frequency of order $1$. In section \ref{Sec: free gravitons from strong coupling}, we use this fact to systematically derive the following effective Hamiltonian for the slow d.o.f. after integrating out the fast d.o.f., 
\begin{align}\spl{
H_\text{eff}   &= \frac{\mu}{2} \sum_{a=1}^N \left[  \sum_{I=1}^9 (p_a^I)^2  +      \frac{1}{3^2} \sum_{i=1}^3 (r_a^i)^2 
+ \frac{1}{6^2}  \sum_{p=4}^9 (r_a^p)^2 
+\frac{i}{4} \theta_a^\top   \g^{123} \theta_a
\right] +o(1/g)
}
\label{eq: eff BMN Hamiltonian}
\end{align}
where $r_a^I \sim X^I_{aa}$ and $\theta_a \sim \hat{\Theta}_{aa}$. 
Therefore, we find $9N$ bosonic and $8N$ fermionic decoupled harmonic oscillators at strong coupling. It is convenient to think about this as $N$ non-interacting particles in a 9 dimensional harmonic trap. Each particle has $2^8=256$ internal states described by the fermionic oscillators. This result itself is not new, see \cite{Lee:2003kf} for results of the BMN model and \cite{Danielsson:1996uw,Sethi:1997pa,Becker:1997wh,Becker:1997xw,Lin:2014wka} for similar results of the BFSS model.\footnote{Aslo see \cite{Paban:1998ea,Paban:1998qy,Sethi:1999qv} for results from using supersymmetry as constraints to determine the leading-order effective BFSS Hamiltonian through the operator approach and subleading corrections through the effective action approach (i.e. without fixing operator ordering). We thank David Berenstein for point out these references to us.} In fact, it can be easily ``derived'' using a naive path integral approach where one integrates out the off-diagonal variables in a quadratic approximation. However, we discovered that this naive approach gives wrong effective Hamiltonians in other models that we discuss in section \ref{Sec: toy models}. Therefore, to put $H_\text{eff}$ on a firm footing and make comparison with a gravity description, it was important for us to properly derive \eqref{eq: eff BMN Hamiltonian}. In addition, having such a systematic strong coupling expansion is crucial for the analysis at the next order in $1/g$, which will lead to interactions between the slow d.o.f.\footnote{See \cite{Lee:2003kf} for the result in the naive path integral approach.} and encode the gravitational interaction in the bulk. We plan to report the outcome of the analysis in the near future.

The effective Hamiltonian \eqref{eq: eff BMN Hamiltonian} only describes the states where all particles are well separated in the trap, i.e. $\sum_{I=1}^9(r_a^I-r_b^I)^2 \sim 1$. In this regime there is only one ground state. It is well-known that the BMN model has other degenerate ground states, which  are obtained by forming $q$ bound states of $k_j$ particles so that $N=\sum_{j=1}^q k_j$.
These bound states and the excitations around them are difficult to study directly in the strong coupling regime but they must have a state of zero energy and their size should scale like $g^{-\frac{1}{3}}$.\footnote{One may think of a bound state of $k$ particles as a small deformation (due to the harmonic trap) of the normalizable ground state of the BFSS model with $k\times k$ matrices.} 
The bound states act as new particles (also with 256 internal states) moving in the same harmonic trap. Their mass is $k$ times larger but they oscillate at the same frequency. Therefore they give rise to the same energy spectrum as the original particles in the limit $g\to\infty$. Nevertheless, they are important to account for the correct degeneracy of each energy level. 
Exciting other internal degrees of freedom of the bound states costs an energy  of order $\mu g^\frac{2}{3}$. It would be interesting to confirm or disprove this scenario with a first-principle computation in the BMN MQM.

\paragraph{Gravity Dual:} Various conjectures propose different connections between the BMN model and gravitational descriptions. Roughly speaking, there exist two distinct categories of the conjectures. The first category focuses on the 't Hooft limit ($g^2N$ fixed and $N\to\infty$) and can be viewed as variants of the gauge/gravity duality. The strong coupling regime in this limit is conjectured to provide holographic descriptions of backreacted geometries (often called Lin-Maldacena vacuum geometries) \cite{Lin:2005nh} and  black holes \cite{Costa:2014wya} (at finite temperature). 

The second category of the conjectures, which is the main focus of this paper, is similar to the Matrix Theory conjecture \cite{Banks:1996vh} or its stronger version proposed by Susskind \cite{Susskind:1997cw}. The conjectures state that the BMN model at finite $g^2$ provides a UV completion of M-theory on the PP-wave background \cite{Kowalski-Glikman:1984qtj},
\begin{align}
    &ds^2 = -2dt dx^- +dx^i dx^i + dx^pdx^p -\left(
    \frac{\mu^2 }{3^2}x^ix^i + \frac{\mu^2 }{6^2}x^px^p
    \right)dt^2
    \label{PW}
    \\
    &F_4 = \mu \,dt\wedge dx^1 \wedge dx^2 \wedge dx^3\, \nonumber
\end{align}
where $i\in \{1,2,3\}$ and $p\in \{4,5,\dots,9\}$ as before.
In its weaker form, it relates the $N\to\infty$ limit of the BMN model\footnote{Note that this is different from the 't Hooft limit, where $g^2N$ is fixed and $N\to\infty$. Instead, here we take $N\to\infty$ with $g^2$ fixed.} to the uncompactified PP-wave background while, in its stronger form, it postulates a relation between the BMN model at finite $N$ and the Discrete Lightcone Quantization (DLCQ) of M-theory on the PP-wave background.
DLCQ is implemented by 
compactifying the null direction $x^- \sim x^- + 2\pi R$. 
This quantizes the momentum  $-p_-=N/R$.
The stronger form of the BMN conjecture is that the sector of M-theory with $-p_- =N/R$, is described by  the Hamiltonian \eqref{eq: BMN Hamiltonian with l_P}, in the $SU(N)$ singlet sector.\footnote{
It is not clear if there is an M-theory interpretation of the $SU(N)$ non-singlet states of the MQM. Perhaps they are heavy at strong coupling as suggested in \cite{Maldacena:2018vsr} but we see no evidence of that in the regime $g\to \infty$ at finite $N$. See section \ref{Sec: free gravitons from strong coupling} for more details.} The parameter $\ell_P$ is identified with the 11D Planck length.

All these conjectures together make the BMN duality extremely rich. As we vary the parameters $N$, $g^2$ and dimensionless temperature $T/\mu$, the BMN MQM is supposed to describe gravitons, M2-branes, M5-branes, Lin-Maldacena geometries \cite{Lin:2005nh} and  black holes \cite{Costa:2014wya}. 
The main focus of this paper is the strong coupling limit $g\to \infty $ of the BMN model at finite $N$, which corresponds to $\ell_P \to 0$ in the gravitational description. In this limit, it is natural to expect\footnote{Attentive readers may question the validity of 11D supergravity in the presence of a compact direction with zero proper length. We will discuss this important issue in section \ref{sec:discussion}.} that the DLCQ of M-theory on the PP-wave background can be approximated well by the DLCQ of the 11D supergravity.
The  wave operator on the PP-wave background leads to the non-relativistic Schrodinger equation with an harmonic potential,
\begin{align}
    \nabla^2 \phi = 0 \qquad
    \Rightarrow \qquad 
    i\partial_t \psi = -\frac{R}{2k} \frac{\partial^2}{\partial x^I \partial x^I} \psi
    + \frac{k}{2R}\left(
    \frac{\mu^2 }{3^2}x^ix^i + \frac{\mu^2 }{6^2}x^px^p
    \right)\psi\,,
\end{align}
where $\phi(x^-,t,x^I)= \psi(t,x^I)e^{-i \frac{k}{R}x^-} $ carries $k$ units of momentum along the compact direction $x^-$.
Notice that for $k=1$ this is precisely the  quadratic part of the Hamiltonian \eqref{eq: BMN Hamiltonian with l_P}. Different values of $k$ change the width of the wave-functions but do not change the energy spectrum.

The paper \cite{Kimura:2003um} computed the full spectrum of (linearized) 11D supergravity on the PP-wave background \eqref{PW}. 
The energy of a single (super)graviton is given by the Hamiltonian \eqref{eq: eff BMN Hamiltonian} for one particle ($N=1$). The 256 states of the fermionic oscillator correspond to the 256 states of the graviton multiplet in 11 dimensions. 
To obtain the generic state with total momentum $-p_-=\frac{N}{R}$, we can create $q$ gravitons with momentum $-p_-=k_j/R$ such that $N=\sum_{j=1}^q k_j$.
This exactly reproduces the energy spectrum of the BMN model at strong coupling, under some reasonable assumptions that we spell out in section \ref{sec:comparison SUGRA}.
Notice that to match the degeneracy, we must consider only singlets of $SU(N)$,
which acts as the permutation group on the $N$ particles in the harmonic trap.
In other words, a graviton with $k$ units of momentum along $x^-$ corresponds to a bound state of $k$ particles in the BMN model at strong coupling.

This precise matching of the two descriptions at finite $N$ and $g\to \infty$ is encouraging, and at the same time raises numerous interesting questions for the future:
\begin{itemize}
\item What happens when we decrease $g$ and start to include interactions between gravitons? 
Will the energy spectrum match in the two descriptions to all orders in the $1/g$ expansion?\footnote{The (naive) computations of \cite{Lee:2003kf} suggest this works at the next order.}
\item What happens when we consider large $N$ so that gravitational backreaction cannot be neglected? Maldacena and Lin constructed spacetime geometries that are dual to the vacua of the BMN model \cite{Lin:2005nh}. Can we use the intuitive picture of particles in a harmonic trap to understand these geometries as some form of collective state of $N$ interacting particles?
\item What happens at finite temperature? There is a regime described by a black hole geometry \cite{Costa:2014wya}. 
Can we understand this regime from the point of view of particles in the trap?
\end{itemize}

\paragraph{Structure of the paper:} 
In section \ref{Sec: toy models}, we discuss the derivation of an effective Hamiltonian for slow variables in several toy models.  We explain in detail why the naive path integral approach fails. In section \ref{Sec: free gravitons from strong coupling}, after reviewing some basic properties of the BMN model, we present our computation of its effective Hamiltonian at strong coupling.
In section \ref{sec:minimalBMN}, we perform a similar study of a related model known in the literature as the minimal BMN model. This model is sufficiently simple that we can perform numerical tests of our results using Hamiltonian truncation methods. We conclude in section \ref{sec:discussion} with a critical assessment of the strong form of the BMN duality described above and the prospects for future tests. In particular, we discuss the validity of  DLCQ of supergravity.




\section{Methodology and pedagogical toy models}
\label{Sec: toy models}

In this section we illustrate the methodology to obtain the effective description of the BMN model in the strong coupling limit. The main idea is that in the strong coupling limit the system's degrees of freedom separate into slow variables and fast variables. Since we are interested in the dynamics in the low energy regime we can integrate out the fast variables and obtain an effective description. We will discuss two commonly used methods for achieving this goal, namely the effective action approach using the Lagrangian and the Born-Oppenheimer approach using the Hamiltonian. We will apply the two approaches to a toy model and compare the results with that from numerical study. This comparison is useful for understanding the subtleties and validity of the two methods.  We will see that while seemingly much simpler than the Hamiltonian approach, the Lagrangian approach is in fact quite subtle. The key observation is that there are \emph{fast} modes in the \emph{naively slow} variables, and in the path integral  approach we also need to integrate them out. The toy model considered in this section will demonstrate this clearly. The results for the BMN model will be given in Section \ref{Sec: free gravitons from strong coupling}.


Let us now consider the following toy model describing a single particle in two dimensions
\begin{align}
   H = \frac12 p_x^2+ V(x) +
   \frac{1}{2} \left[p_y^2+ g^2 \omega^2(x) y^2 -  g \omega(x)\right] \,,
   \label{eq: toy model with general potential}
\end{align}
where $V(x)$ is a generic $g$-independent potential for $x$ and we shall assume that $\omega(x) >0$ for all $x$. We are interested in the limit $g\to\infty$, where the potential in the $y$ direction is much steeper than that in the $x$ direction.  Therefore, we would (naively) treat $x$ as the slow variable and $y$ as the fast variable which sees $x$ as a constant. In the low energy regime, we can derive an effective description of the system in which the $y$ variable is put on its ground state.

This toy model is constructed to mimic certain features of the BMN model: $x$ is analogous to the diagonal matrix elements whereas $y$ is analogous to off-diagonal ones; the potential term $-g\w(x)$ which cancels the ground state energy of the $y$ oscillator is to simulate the effect of supersymmetry.\footnote{One can easily come up with a supersymmetric toy model. One example is the supermembrane toy model (having flat directions) discussed in \cite{deWit:1988xki} and studied numerically in \cite{Balthazar:2016utu}. The mass deformation version (without flat directions) was studied in \cite{Boulton:2010nd,Motycka:2014vra}. However, for our purpose it is enough to study the even simpler bosonic toy model \eqref{eq: toy model with general potential}.}

\subsection{The naive path integral approach}

\label{subsec: L approach done wrong}
Let us first use the Lagrangian approach to find the effective action in the large $g$ limit. The idea is to use the path integral formulation of quantum mechanics and naïvely integrate out the fast variables $y$. We then define the effective action as
\begin{equation}
    \int \mathcal{D}x \mathcal{D}y e^{-S_E[x,y]} \equiv \int \mathcal{D}x  e^{-S_{E,\text{eff}}[x]}\,,
    \label{eq: Path int wrong starting point}
\end{equation}
where the Wick rotated Euclidean action reads
\al{
S_E[x,y]=
\frac{1}{2} \int_{-\beta/2}^{\beta/2} d\tau\,  y(\tau) \left( -\partial_\tau^2 + g^2 \w^2(x)  \right) y(\tau) + \frac{1}{2} \int_{-\beta/2}^{\beta/2} d\tau \left( \dot{x}^2 + 2V(x)-g\w(x) \right)
\,.
}
Since the Lagrangian is quadratic in $y$, its path integral is Gaussian and given by a functional determinant. There are many ways to compute the determinant and we show a solution using the Gelfand-Yaglom theorem in appendix \ref{App: Calculation details}, assuming  $x(\tau)$ is a slow varying function.
The main result is that this integration procedure yields 
\be 
S_{E,\text{eff}}[x] =  \int_{-\beta/2}^{\beta/2} d\tau \left( \frac{1}{2}\dot{x}^2 +V(x)  + O(1/g) \right)\,, 
\ee 
corresponding to the effective Hamiltonian 
\be 
H_\text{eff} = \frac{1}{2} p_x^2+ V(x) + O(1/g)\,. 
\label{eq: result from Lagraingian approach done wrong}
\ee 
While the result looks like what one might have expected, it is in fact incorrect as predicted by the title of this subsection. 
What goes wrong is that we cannot treat $x(\tau)$ as a slow varying function, in other words, $x$ is not really a slow variable because it contains fast modes.
\footnote{We thank Gabriel Cuomo for raising this point.} 
We will give the correct solution in Section \ref{subsec: L approach done right}.

\subsection{The Hamiltonian approach}
\label{subsec: H approach toy model}
Next we switch to the Born-Oppenheimer approach explicitly using the Hamiltonian. Similar methods were developed in the literature in order to study the asymptotic expansion of the BFSS model \cite{Banks:1996vh} in large distance and low velocity, see e.g. \cite{Danielsson:1996uw,Halpern:1997fv,Graf:1998bm,Frohlich:1999zf,Lin:2014wka} and also \cite{Smilga:1986rb,Smilga:2002mra}. 


The starting point is the Schrodinger equation and our goal is to find the effective Hamiltonian acting on the reduced wavefunction which only depends on the slow variables
\al{
H \Y(x,y) = E \Y(x,y) 
\overset{g\to\infty}{\Longrightarrow}
H_\text{eff} \y(x) = E \y(x) \,.
}
Intuitively we expect the $y$ variable to be on its ground state plus corrections suppressed by $g$. By expanding the Schrodinger equation in large $g$ and using the following ansatz for the wavefunction
\begin{align*}
    \Psi(x,y)  =  \psi(x) \   \W_x(y) \Big(1+ \sum_{m,n} g^{-m} f_{mn}(x) y^n\Big)
    = \Y^{(0)}(x,y)  + g^{-1}  \Y^{(-1)}(x,y)  + \ldots
    \,,
\end{align*}
where $\W_x(y)$ denotes the fast-mode ground state, we can integrate out $y$ by imposing the coefficient in front of each $y^n \W_x(y)$ term to vanish. The remaining $y$-independent part of the Schrodinger equation becomes a differential equation solely for $\y(x)$, from which we can extract the effective Hamiltonian. See Section \ref{subsubsec: general Born-Oppenheimer strategy} for a systematic explanation of the strategy.


Let us now apply this idea to the toy model \eqref{eq: toy model with general potential}. Since the fast mode $y$ has a harmonic potential with frequency of $O(g)$, it is convenient to rescale $y\to y/\sqrt{g}$ such that after rescaling $p_y, y=O(1)$, and the power counting of $g$ in the Hamiltonian becomes manifest
\al{
H = g\frac12\left(p_y^2 + \w(x)^2 y^2 -\w(x) \right) + \frac12 p_x^2 + V(x)
= g H^{(1)} + H^{(0)}
\,.
}
Notice that $H^{(1)}$ is simply a harmonic oscillator in $y$, thus its ground state is a (normalised) Gaussian wavefunction
\al{
\W_x(y) = \left(\frac{\w(x)}{\pi}\right)^{\frac{1}{4}} e^{-\frac12 \w(x) y^2}\,,
\qquad
\int dy |\W_x(y)|^2=1\,.
}

Next we solve the Schrodinger equation order by order in $g$. Since we are interested in the low energy regime with $E=O(1)$, the Schrodinger equation at $O(g)$ reads
\al{
g H^{(1)} \Y^{(0)}(x,y)
\equiv
g H^{(1)} \y(x) \W_x(y)
= 0\,,
}
and is automatically solved since the ground state energy of $H^{(1)}$ is zero. At $O(1)$, it is convenient to project the Schrodinger equation into the fast-mode ground state $\W_x(y)$ \cite{Lin:2014wka}
\al{\spl{
\underbrace{\int dy\, 
\W_x(y) H^{(1)} \Y^{(-1)}}_{=0}
+
\int dy\, 
\W_x(y)H^{(0)}\Y^{(0)}
=
E \int dy\, \W_x(y) \Y^{(0)}
=
E\y(x)
\,,
}}
where the first term vanishes by construction. Therefore, the leading-order effective Hamiltonian is
\al{
H_\text{eff}^{(0)} \y(x) \coloneqq \int dy\, \W_x(y) H^{(0)} \Y^{(0)} = 
\left(\frac12 p_x^2 + V(x) + \frac{1}{16}\left(\frac{\w'(x)}{\w(x)}\right)^2\right) \y(x)\,.
}
Clearly this is different from the Lagrangian approach's result \eqref{eq: result from Lagraingian approach done wrong}. The extra term comes from acting $p_x^2$ on $\W_x(y)$,\footnote{This extra term is related to adiabaticity of QM systems, see e.g. \cite{Asplund:2010xh} for discussions.} which indicates that $\w(x)$, or more generally the $x$ variables, cannot be treated as completely frozen. In other words, \emph{there are also fast modes in the $x$ variable}. We will show how this changes the effective action computation in Section \ref{subsec: L approach done right}.

It is also instructive to compute subleading order corrections. For simplicity let us be more specific and set\footnote{
It was shown in \cite{Hoppe:1980,Simon:1983jy} that quantum mechanical systems whose potential has classically flat directions do not necessarily have a continuous spectrum. The simplest example considered was $H=\frac{1}{2}(p_x^2 +p_y^2 + x^2 y^2)$. It can be viewed as two harmonic oscillators in $x$ and $y$ whose frequencies and zero-point energies depend on each other, thus for finite energy neither $x$ or $y$ can be large and the classical flat direction is lifted quantum mechanically. Our model (before rescaling of $y$) reads 
\begin{align}
H = \frac{1}{2} \left(p_x^2+p_y^2+x^2+y^2 +   g^2 x^2 y^2  -1
   -  \sqrt{1+g^2x^2}\right)\,,
\label{eq: un-rescaled toy model}
\end{align}
and essentially it restores the flat direction in $x$ by cancelling the ground state energy, thus allowing for the separation of fast and slow variables when $g|x|$ is large. Then we put everything in a harmonic potential much weaker than the quartic potential for large $g$ in order to retain a discrete spectrum.
}
\al{
\label{eq:omega and V}
V(x) = \frac12 (x^2-1)\,,
\qquad
\w(x) = \frac{\sqrt{1+g^2 x^2}}{g}\,,
}
which after rescaling $y\to y/\sqrt{g}$ leads to the following Hamiltonian
\begin{align}
\ga{
   H = g H^{(1)}+H^{(0)}+ g^{-1} H^{(-1)} + O(1/g^{2})\,, \\
   \quad
   H^{(1)}=\frac12\left(p_y^2+x^2y^2-|x|\right)\,,
   \quad
   H^{(0)}=\frac12\left(p_x^2+x^2-1\right)\,,
   \quad
   H^{(-1)}= \frac{y^2}{2} - \frac{1}{4 |x|}\,.
}
\end{align}
The $O(1)$ effective Hamiltonian can be obtained in exactly the same way as above. For the next order we use the projected Schrodinger equation at $O(g^{-1})$ to compute $H_\text{eff}^{(-1)}$ (recall $\int dy\, \W_x(y) \Y^{(0)} = \y(x)$)
\al{\ga{
\int dy\,
\W_x(y)\left(H^{(0)}\Y^{(-1)} + H^{(-1)} \Psi^{(0)} = E^{(0)} \Y^{(-1)} + E^{(-1)}\Y^{(0)}\right)
\ \Rightarrow
\\
H^{(-1)}_\text{eff} \y(x) \coloneqq
\int dy\, \left(\W_x(y)H^{(0)}\Y^{(-1)} + \W_x(y)H^{(-1)}\Y^{(0)}-
E^{(0)} \W_x(y) \Y^{(-1)}\right)
= E^{(-1)} \y(x)\,.
\label{eq: H_eff^-1 for toy model}
}}
Here $\Y^{(-1)}$ can be determined through solving the Schrodinger equation at the previous order
\al{
H^{(1)}\Y^{(-1)} = - H^{(0)}\Y^{(0)} + E^{(0)}\Y^{(0)}\,.
\label{eq: Psi -1 eqn for toy model}
}
To this end we make the following ansatz
\al{
\Y^{(-1)}(x,y) =
\y(x) \W_x(y) \left(\sum_{m=1}^{\bar{m}}c^{(-1)}_m(x) y^m \right) + \chi^{(-1)}(x) \W_x(y)\,,
}
where $\bar{m}$ is determined by the RHS of \eqref{eq: Psi -1 eqn for toy model}, the last term is in the kernel of $H^{(1)}$ and $\chi^{(-1)}(x)$ will be determined later. Plugging the ansatz into \eqref{eq: Psi -1 eqn for toy model}, and expanding near $y=0$, we turn the differential equation into a polynomial equation of the form $\sum_k d_k(x) y^k = 0$, where $d_k(x)$ are functions of the coefficients $c_m^{(-1)}(x)$. Since this holds for any $y$, we simply need to solve $d_k(x)=0$ for each $k$ separately. This fixes all $c_m^{(-1)}(x)$ and hence $\Y^{(-1)}(x,y)$ up to $\chi^{(-1)}(x)$. To determine $\chi^{(-1)}(x)$, we impose the normalisation condition
\al{
    \<\Y^{(0)}+g^{-1}\Y^{(-1)},\Y^{(0)}+g^{-1}\Y^{(-1)}\>_{x,y} \overset{!}{=} \<\y,\y\>_x + O(1/g^2)\,,
    \label{eq: normalisation condition for toy model}
}
where the inner product on the LHS involves integral over both $x$ and $y$ while the RHS only involves $x$. This gives 
\al{
\chi^{(-1)}(x) = \frac{16x \y'(x)-5\y(x)}{128 |x|^3}\,,
}
and hence completely fixes $\Y^{(-1)}(x,y)$. Plugging the solution of $\Y^{(-1)}(x,y)$ back into \eqref{eq: H_eff^-1 for toy model}, we obtain the following result
\begin{align}
   H_{\rm eff}
   &= \frac{1}{2} \left(p_x^2 +x^2  + \frac{1}{8x^2}-1 \right)
   +\frac{1}{16g}
   \left(
   p_x\frac{-1}{|x|^3}p_x + \frac{53}{32 |x|^5}
   \right)+ O(1/g^{2})\,.
   \label{eq: final H_eff of toy model}
\end{align}  
Notice that at $O(1/g)$, the ordering of the operators is unambiguously determined.

When fixing the normalisation using \eqref{eq: normalisation condition for toy model}, there is a freedom from using integration by part
\begin{align*}
\int dx\, f(x) \sim \int dx\, \left(f(x) + \del_x g(x)\right)\,.
\end{align*}
Using it to remove derivatives on $\y(x)$ from $\chi^{(-1)}(x)$ gives
\al{
\tilde{\chi}^{(-1)}(x) = \frac{11 \y(x)}{128|x|^3}\,,
}
and the corresponding effective Hamiltonian reads
\begin{align}
   \tilde{H}_{\rm eff}
   &= \frac{1}{2} \left(p_x^2 +x^2  + \frac{1}{8x^2}-1 \right)
   +\frac{1}{16g}
   \left(
   p_x\frac{3}{|x|^3}p_x + \frac{2}{|x|}-\frac{339}{32 |x|^5}
   \right)+ O(1/g^{2})\,.
   \label{eq: final H_eff of toy model option 2}
\end{align}
One may worry that there is a physical ambiguity for the subleading terms in the effective Hamiltonian, but in fact the two effective Hamiltonian's give the same spectrum up to $O(1/g^2)$. To see this, consider the $O(1/g)$ terms in the effective Hamiltonian as perturbation to $H_{\rm eff}^{(0)}$ and their contributions to the energy spectrum is
\al{
\d E = \frac{1}{g} \sum_n \<n| H_{\rm eff}^{(-1)}|n\>\,,
}
where $|n\>$ are energy eigenstates of $H_{\rm eff}^{(0)}$. If the difference $\d H_{\rm eff}\equiv H_{\rm eff} - \tilde{H}_{\rm eff}$ can be written as a commutator between $H_{\rm eff}^{(0)}$ and another operator $\OO$ then up to $O(1/g^2)$ the spectra are the same
\al{
\d E -\d \tilde{E} = \<n| \d H_{\rm eff}^{(-1)}|n\> =
\<n| [H_{\rm eff}^{(0)}, \OO] |n\>
=
E^{(0)}_n\<n| \OO |n\> - \<n| \OO |n\> E^{(0)}_n
= 0\,,
}
which is in fact one of the common constraints used in matrix quantum mechanics bootstrap \cite{Han:2020bkb}. One can verify that this is indeed the case:
\al{
\d H_{\rm eff} = \frac{1}{g} \left[H_{\rm eff}^{(0)}, \left[p_x^2, \frac{1}{16 |x|} \right] \right] + O(1/g^2)\,,
}
or replace $(16|x|)^{-1}$ by $(16 \w(x))^{-1}$ for the general potential. The fact that the two effective Hamiltonian's differ by a commutator  can be understood as that they are related by an infinitesimal unitary transformation:
\al{
\tilde{H}_{\rm eff} = U^\dagger H_{\rm eff} U\,,
\quad
U = \mathbb{1}+\frac{1}{g} \e\,,
}
where the unitarity condition gives $\e^\dagger = -\e$. Up to $O(1/g^2)$ we get
\al{
\tilde{H}_{\rm eff} = H^{(0)} _{\rm eff} 
+ \frac{1}{g}
\left(
H^{(-1)} _{\rm eff} 
+
[H^{(0)} _{\rm eff},\e]
\right) + O(1/g^2)\,.
}
Therefore, the superficial ambiguity introduced by the freedom of using integration by part is not physical.

Finally let us comment on the validity regime of \eqref{eq: final H_eff of toy model}. From the original toy model Hamiltonian \eqref{eq: un-rescaled toy model} we see that the oscillator frequencies for $x$ and $y$ are
\al{
\w_x = \sqrt{1+ g^2 y^2}\,,\qquad
\w_y = \sqrt{1+ g^2 x^2}\,,
}
and the typical scale of $x$ and $y$ are $x\sim \w_x^{-\frac12},\, y\sim \w_y^{-\frac12}$. In the large $g$ limit, when $x=O(1)$ we have $1\sim\w_x \ll \w_y \sim g$ and $y\sim g^{-1/2}$, so indeed there is a separation of scale. The effective description breaks down when $\w_x \sim \w_y$ which leads to 
\al{
x\sim y \sim (1+g^2x^2)^{-\frac14}\,,
}
from which we deduce
\al{
x \sim y \sim g^{-1/3}\,.
}
Therefore, the effective description no longer applies in the above regime.

\subsubsection{The spectrum and the gluing of wavefunctions}
In order to test the result \eqref{eq: final H_eff of toy model} we will solve for its spectrum and compare with the numerical results from solving the original Hamiltonian (see Section \ref{subsec: numerical check of toy model}). For the leading order effective Hamiltonian, the wavefunctions that decay at infinity ($x\to \infty$) are given by
\begin{align}\spl{
    \psi_E(x)  &=x^\frac{2+\sqrt{6}}{4} 
    e^{-x^2/2}\,
    U\left(\frac{2+\sqrt{6}-4E}{8},\frac{\sqrt{6}+4}{4},x^2\right) \\
    &\sim \frac{\Gamma\left(-\frac{\sqrt{6}}{4}\right)}{\Gamma\left(\frac{2-\sqrt{6}-4E }{8}\right)} 
    x^\frac{2+\sqrt{6}}{4} +
    \frac{\Gamma\left(\frac{\sqrt{6}}{4}\right)}{\Gamma\left(\frac{2+\sqrt{6}-4E }{8}\right)} 
    x^\frac{2-\sqrt{6}}{4} ,\qquad {\rm as}\quad x\to 0\,,
    \label{eq:wavefunction_bosons_large_g}
}\end{align}
where $U(a,b,z)$ is the Tricomi confluent hypergeometric function. To obtain the energy spectrum we need another boundary condition at $x=0$. However, for $x\sim g^{-1/3}$ the derivation of the effective Hamiltonian breaks down and we need to return to the original Hamiltonian.
Fortunately, we only need very limited information about the original Hamiltonian. More precisely, we can introduce rescaled coordinates $x=g^{-\frac{1}{3}} \tilde{x}$, $y=g^{-\frac{1}{3}} \tilde{y}$. This leads to 
\begin{align}
   H = g^\frac{2}{3}\frac{1}{2} \left(\tilde{p}_x^2+\tilde{p}_y^2+  \tilde{x}^2 \tilde{y}^2  -|\tilde{x}|
   \right) -\frac12 +O(g^{-\frac{2}{3}})\,.
\label{eq: toy model in BFSS-like form}
\end{align}
Thus, we should look for solutions of $\left(\tilde{p}_x^2+\tilde{p}_y^2+  \tilde{x}^2 \tilde{y}^2  -|\tilde{x}|
   \right)\Psi=0$.
This is a hard problem but we only need to know the solution in the asymptotic region $\tilde{x}\to\infty$, which can be found in a way similar to that for determining \eqref{eq: final H_eff of toy model} and \eqref{eq:wavefunction_bosons_large_g}
\al{
\Psi_\pm(\tilde{x},\tilde{y})\to  \left(\tilde{x}\right)^{\frac14}e^{-\frac12 \tilde{x} \tilde{y}^2}
\left(\tilde{x}^\frac{2+\sqrt{6}}{4}+ w_\pm \tilde{x}^\frac{2-\sqrt{6}}{4}\right)\,,
}
where the coefficient $w_\pm$ corresponds to even and odd wavefunctions under $\tilde{x}\to -\tilde{x}$.

Matching the two wavefunctions, we obtain
\begin{align}
    g^\frac{1}{\sqrt{6}} \Gamma\left(\frac{\sqrt{6}}{4}\right)\Gamma\left(\frac{2-\sqrt{6}-4E }{8}\right) =
    w_\pm \Gamma\left(-\frac{\sqrt{6}}{4}\right)\Gamma\left(\frac{2+\sqrt{6}-4E }{8}\right)
\end{align}
which gives
\begin{align}
    E_n^\pm = \frac{2+\sqrt{6}}{4}+2n -  g^{-\frac{1}{\sqrt{6}}} 
     \frac{2(-1)^n \Gamma\left(-\frac{\sqrt{6}}{4}\right)}{n! \Gamma\left(\frac{\sqrt{6}}{4}\right)\Gamma\left(-\frac{\sqrt{6}}{4}-n\right)}w_\pm + (\text{subleading})
    \,,\quad n=0,1,2,\dots
\label{eq:spectrum_bosons_large_g}
\end{align}
This is the prediction for the energy spectrum of the theory as $g\to \infty$. 

\subsection{Numerical check}
\label{subsec: numerical check of toy model}
We will now solve the the Schrodinger equation of the toy model \eqref{eq: toy model with general potential} with \eqref{eq:omega and V} numerically to check the analytic predictions. 
Our strategy is simple. We discretize space on a grid of $M$ points to get a finite dimensional Hamiltonian written in position basis. Then we diagonalize it to obtain the spectrum and  wavefunctions. Because of the symmetry $x \to -x$ the spectrum splits into even (+) and odd (-) sectors, with eigenstates $\Psi_n^{(\pm)}(x,y)$ having energy $E_n^{(\pm)}$. We define
\begin{equation}
    \psi_n^{(\pm)}(x) =  \operatorname{sgn}\Psi_n^{(\pm)}(x,0)
    \times \left(\int dy |\Psi_n^{(\pm)}(x,y)|^2 \right)^{1/2}.
\end{equation}
We then increase $M$ such that the discretization becomes more and more fine grained and compare the result with the predictions \eqref{eq:wavefunction_bosons_large_g} and \eqref{eq:spectrum_bosons_large_g}. This is shown in Figure \ref{fig:spectrum} and \ref{fig:fit_Eg}. Details on our implementation are given in appendix \ref{App: numerics}. 

\begin{figure}[h!]
    \centering
    \includegraphics[scale=0.5]{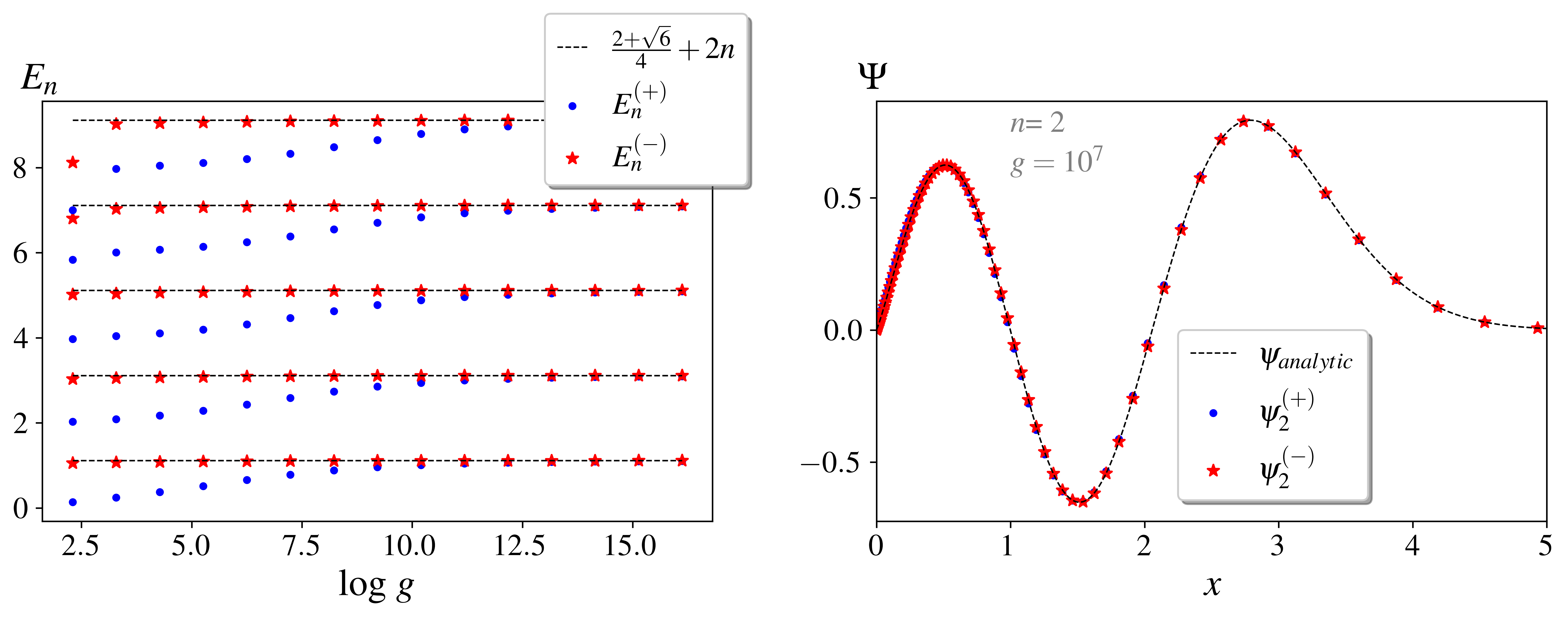}
    \caption{(Left) First 10 eigenvalues with $M=100$ grid points. The ground state energy at $g=10^7$ is  $E_0(g=10^7)= 1.10317...$, which is to compare with the prediction $E_0(\infty)=1.11237...$ (dashed lines). (Right) Numerical wavefunctions at $g=10^7$ for the $n=2$ states compared to the analytical prediction at $g= \infty$.}
    \label{fig:spectrum}
\end{figure}

\begin{figure}[h!]
    \centering
    \includegraphics[scale=0.5]{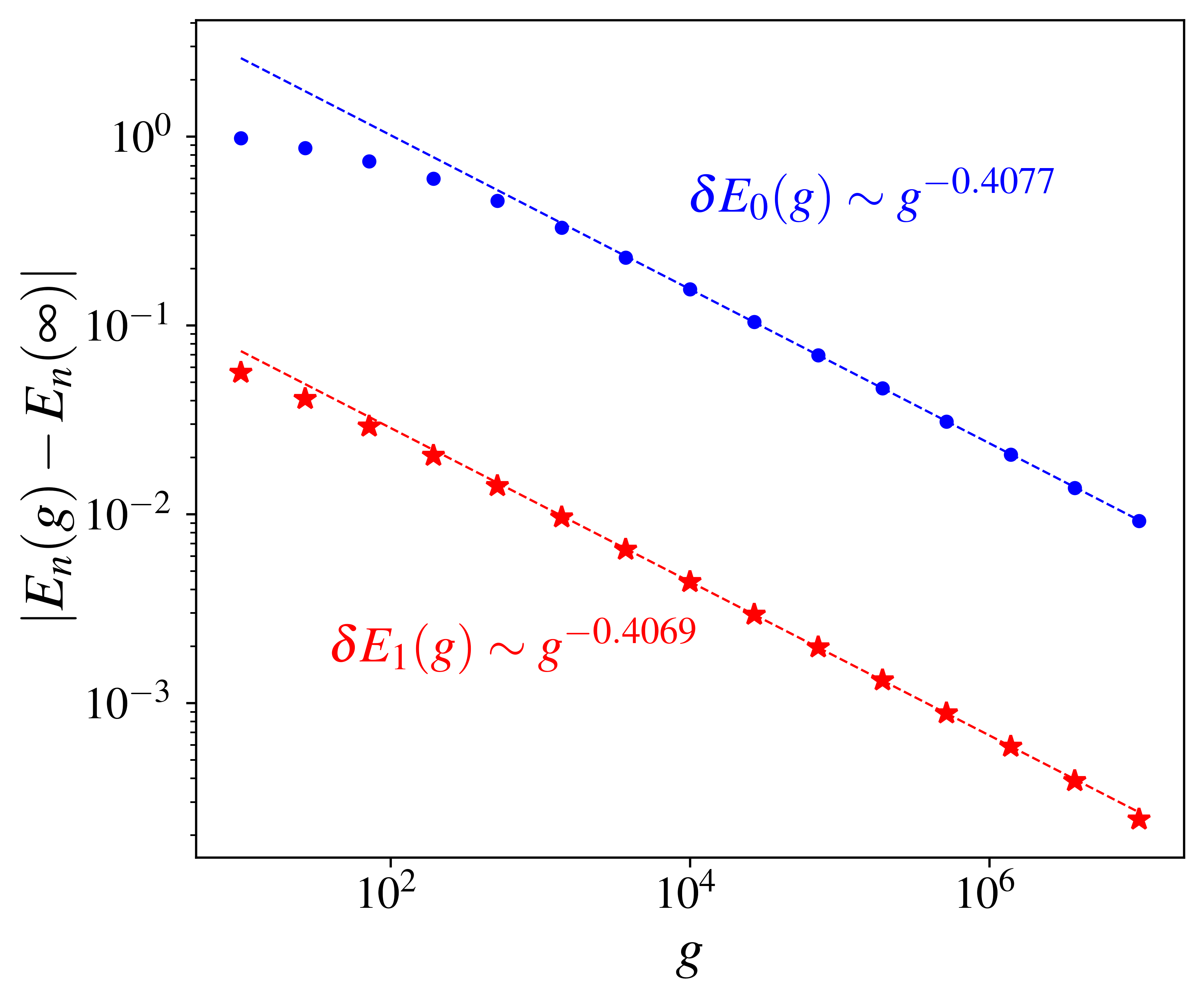}
    \caption{Leading correction at large $g$ for $E_0$ and $E_1$. The exponent is compatible with the prediction in \eqref{eq:spectrum_bosons_large_g}: $\delta E (g) \sim g^{-1/ \sqrt{6}}$. Fitting the coefficient multiplying $g$ gives the Wilson coefficients $w_+ = 5.04 \pm 0.05$, $w_- = 0.135 \pm 0.002$.}
    \label{fig:fit_Eg}
\end{figure}

Clearly the numerical results agree with that from the Hamiltonian approach rather than the naive path integral approach. We will now discuss the correct way to perform the path integral and find the effective action.

\subsection{The path integral approach done right}
\label{subsec: L approach done right}
The path integral approach can be used but one has to carefully determine which variables are to be integrated out. 
The puzzle raised in section \ref{subsec: L approach done wrong} originates from treating $x(\tau)$ as a slow varying function when we integrate out the variable $y(\tau)$. In order to solve this issue it is convenient to split $x(\tau)= x_s(\tau)+x_f(\tau)$ where $x_s$ is slow and $x_f$ is fast. 
The precise split is not very important but for concreteness we can define
\be
x_s(\tau)= \sum_{|n|\leq \Lambda} a_n e^{2\pi i n\tau/\beta} \,,\qquad 
x_f(\tau)= \sum_{|n|>\Lambda} a_n e^{2\pi i n\tau/\beta}\,,
\label{eq:fast and slow}
\ee
where $a_{-n}=a_n^*$.
The cutoff $\Lambda$ should be chosen so that $x_s$ is much slower than the degree of freedom $y$. This leads to the condition 
\be
q\equiv  \frac{g\omega(x_s) \beta}{\Lambda} \gg 1.
\ee
In practice, we will first take   $g\to \infty$ with fixed $q$ and then take $q\to \infty$ at the end.

We  write the action as follows
\begin{align}
    S_E &=  S_{\rm fast}^{\rm kin} +S_{\rm fast}^{\rm int} +S_{\rm slow}
\end{align}
where 
\begin{align}
S_{\rm fast}^{\rm kin} &= \frac{1}{2} \int_{-\beta/2}^{\beta/2} d\tau \left[ y(\tau) \left( -\partial_\tau^2 + g^2 \omega^2(x_s)  \right) y(\tau) +    \dot{x}_f^2\right]  
\\
S_{\rm fast}^{\rm int} &=\frac{1}{2} \int_{-\beta/2}^{\beta/2} d\tau \left[    2 g^2 \omega(x_s) \omega'(x_s) x_f y^2    +  g^2   [\omega'(x_s) ]^2 x_f^2 y^2 -g\omega'(x_s)x_f +\dots \right]
\label{eq: S_fast^int}
\\
S_{\rm slow} &=\frac{1}{2} \int_{-\beta/2}^{\beta/2} d\tau \left( \dot{x}_s^2 +2 V(x_s)-g\omega(x_s) \right).
\end{align}
The first line contains the kinetic terms of the fast variables $y$ and $x_f$. The second line contains the interactions between fast variables that can be treated perturbatively. The last line contains the slow variable $x_s$ that we will not integrate out.
We define the effective action for the slow variable by (notice the difference compared to \eqref{eq: Path int wrong starting point})
\begin{equation}
\begin{split}
 S_{\rm eff}[x_s]\equiv -\log  \int \mathcal{D}y \mathcal{D}x_f\, e^{-S_E} &=  S_{\rm slow} 
  -\log Z_{\rm fast} -\log \langle e^{-S_{\rm fast}^{\rm int}} \rangle_{\rm fast} 
\end{split}
\end{equation}
where $\log Z_{\rm fast}=-\frac{1}{2} \operatorname{log} \left( y_1(\beta/2)+ \dot{y}_2(\beta/2) -2 \right) =-\frac{1}{2} 
\int_{-\beta/2}^{\beta/2} d\tau  g \omega(x_s(\tau)) + \mathcal{O}(1/g)
$ can be computed using Gelfand-Yaglom as explained in Appendix \ref{App: Calculation details}.\footnote{We neglect the  contribution to $\log Z_{\rm fast}$ from $x_f$ because it is independent of $x_s$.}
The last average is taken using the propagators of the fast variables associated with $S_{\rm fast}^{\rm kin}$ that read
\begin{equation}
\begin{split}
\langle x_f(\tau_1) x_f(\tau_2) \rangle &=
\sum_{|n|>\Lambda}  e^{2\pi i n( \tau_1-\tau_2)/\beta }
\frac{\beta}{\left(2\pi  n\right)^2}.
\end{split}
\end{equation}
The propagator of $y$ is more complicated but for nearby times we can use the formulas for the harmonic oscillator
\begin{align}
\label{eq:y propagator}
\langle y(\tau_1) y(\tau_2) \rangle &\approx \frac{1}{2 g \omega(x_s(\tau))}
e^{-g \omega(x_s(\tau)) |\tau_1-\tau_2|},
\end{align}
where $\tau$ can be the average $(\tau_1+\tau_2)/2$ or $\tau_1$ or $\tau_2$ because the slow field $x_s$ does not vary in the time scale $1/(g \omega)$.

Armed with the propagators of the fast variables, we can proceed to compute $ \langle e^{-S_{\rm fast}^{\rm int}}\rangle $ through its expansion in cumulants
\begin{align}
 -\log \langle e^{-S_{\rm fast}^{\rm int}} \rangle_{\rm fast}  =
 \langle S_{\rm fast}^{\rm int}\rangle
 -\frac{1}{2} \left(\langle (S_{\rm fast}^{\rm int} )^2\rangle -
 \langle S_{\rm fast}^{\rm int} \rangle^2
 \right) +\dots
 =\frac{1}{16}\int_{-\beta/2}^{\beta/2} d\tau
   \left[      \frac{\omega' }{\omega }    \right]^2  
   +\dots
\label{eq: cumulant expansion}
\end{align}
where the last equality comes from the computation of $\langle S_{\rm fast}^{\rm int}\rangle$ and $\langle S_{\rm fast}^{\rm int}\rangle^2$ shown in appendix \ref{App: Calculation details}.
Putting everything together, we find
\begin{align}
      S_{\rm eff} &= \int_{-\beta/2}^{\beta/2} d\tau \left( \frac{1}{2} \dot{x}_s^2 + V(x_s)+\frac{1}{16}
      \left[      \frac{\omega'(x_s) }{\omega(x_s) }    \right]^2 + O(1/g)
      \right),
\end{align}
which precisely agrees with the hamiltonian computation!

The attentive reader will notice that $S_{\rm fast}^{\rm int}$ contains more terms than the ones we wrote in \eqref{eq: S_fast^int}. Therefore, it is important to check that no other terms can contribute at leading order $g^0$. In order to address this issue, we notice that  both fast variables scale as $x_f \sim y \sim 1/\sqrt{g} \sim 1/\sqrt{\Lambda}$. This follows from their propagators.
We can use this to quickly estimate which terms in $S_{\rm fast}^{\rm int}$ can contribute at leading order.
Following this reasoning, we notice two terms that can contribute at leading order
\begin{align}
    \frac{1}{2} \int_{-\beta/2}^{\beta/2} d\tau \left[     g^2 \omega(x_s) \omega''(x_s) x_f^2 y^2     -\frac{1}{2} g\omega''(x_s)x_f^2 \right] \subset S_{\rm fast}^{\rm int} 
\end{align}
However, taking the average
\begin{align}
    \frac{1}{4} \int_{-\beta/2}^{\beta/2} d\tau  g\omega''(x_s)
    \langle x_f^2 \rangle
    \left[     2 g \omega(x_s) \langle y^2  \rangle   - 1\right] =0\,,
\end{align}
we find zero.

One may also worry about contributions from higher order cumulants of $S_{\rm fast}^{\rm int}$. 
In this case, it is useful to notice that the two-point function of $y$ decays exponentially with a time scale $1/(g\omega)$. Therefore, for each extra time integral (by taking powers of $S_{\rm fast}^{\rm int}$) one pays a factor of $1/g$. This suppresses all higher cumulants.

In principle, this method can be used to compute the effective action to subleading orders in $1/g$. However, even for this simple toy model this seems difficult, at least, with the hard cutoff used in \eqref{eq:fast and slow} to separate 
fast and slow modes of $x$. 
It would be interesting to streamline this method so that it can be systematically implemented order by orders in the $1/g$ expansion.

As we shall see in the next section, the careful strong coupling expansion of the BMN model using the Hamiltonian approach agrees, at leading order, with the naive path integral approach.
Probably this is due to supersymmetry that leads to cancelations between the contributions of the fast modes of the bosonic and fermionic slow variables. It is not clear to us if the naive path integral approach remains valid at subleading orders.


\section{BMN matrix quantum mechanics at strong coupling}
\label{Sec: free gravitons from strong coupling}


\subsection{Review}
\label{subsec: BMN review}
Let us first review some basics about the BMN model and set up the notation along the way. 
The Hamiltonian \eqref{eq: BMN Hamiltonian with l_P} is given in terms of variables $X$ that correspond  to the transverse coordinates $x$ of the spacetime \eqref{PW}. The BMN model is supersymmetric and it has sixteen dynamical\footnote{The BMN model has another sixteen kinematical supercharges 
\al{
q_\a=\frac{1}{\sqrt{R}}\op{Tr}\hat\Theta_\a\,,
}
which only act in the $U(1)$ sector of the BMN model. See \cite{Dasgupta:2002hx,Kim:2002if} for more details.} real supercharges  
\al{\spl{
Q_\a= & \sqrt{R} \operatorname{Tr}\bigg[
P^I \gamma^I \hat\Theta
-\frac{i}{2\ell_P^3}\left[X^I, X^J\right] \gamma^{IJ} \hat\Theta
-\frac{\mu}{3 R} X^i \gamma^{123} \gamma^i \hat\Theta+\frac{\mu}{6 R} X^p \gamma^{123} \gamma^p \hat\Theta\bigg]_\a\,,
}}
which satisfy \cite{Baake:1984ie,Kim:2002if,Kim:2002zg,Dasgupta:2002hx}
\al{
\left\{Q_\alpha, Q_\beta\right\}=2 \delta_{\alpha \beta} H-\frac{\mu}{3}\left(\gamma^{123} \gamma^{i j}\right)_{\alpha \beta} M^{i j}+\frac{\mu}{6}\left(\gamma^{123} \gamma^{pq}\right)_{\alpha \beta} M^{p q}
+\frac{2R}{\ell_P^3}{\rm Tr}(X^I G)\g^I_{\a\b}\,.
}
Above we have defined $\gamma^{IJ}=\frac12[\gamma^I,\gamma^J]$, the rotation generators
\al{
M^{IJ} = {\rm Tr}
\left(
X^I P^J - X^J P^I +\frac{i}{4}\hat\Theta^\top \g^{IJ} \hat\Theta
\right)\,,
\label{eq: BMN rotation generators}
}
and the $SU(N)$ generators
\al{
G=G_A T_A\,, \quad
G_A = f_{ABC}\left(X^I_B P^I_C - \frac{i}{2}\hat\Theta_{\a B} \hat\Theta_{\a C}\right)\,,
\quad
A=1,\ldots,N^2-1\,,
\label{eq: BMN SU(N) generators}
}
where $f_{ABC}$ are the $SU(N)$ structure constants.

It is also useful to write the supersymmetry generators in the rescaled variables of \eqref{eq: BMN Hamiltonian},
\al{
\spl{
Q_\a/\sqrt{\m} = \operatorname{Tr}
\bigg[
P^I \gamma^I \hat\Theta
-\frac{i}{2}g\left[X^I, X^J\right] \gamma^{IJ} \hat\Theta
- \frac13 X^i \gamma^{123} \gamma^i \hat\Theta
+ \frac16 X^p \gamma^{123} \gamma^p \hat\Theta  
\bigg]_\a\,,
}
\label{eq: real supercharges of BMN}
}
satisfying
\al{
\left\{Q_\alpha, Q_\beta\right\}/\m=2 \delta_{\alpha \beta} H/\m-\frac{1}{3}\left(\gamma^{123} \gamma^{i j}\right)_{\alpha \beta} M^{i j}+\frac{1}{6}\left(\gamma^{123} \gamma^{pq}\right)_{\alpha \beta} M^{pq}
+2 g {\rm Tr}(X^I G)\g^I_{\a\b}\,.
\label{eq: BMN susy algebra}
}

The bosonic potential in the BMN model can be written as 
\al{
V_B/\mu = 
\frac12
{\rm Tr} \left[
\left(\frac13 X^i + ig\e_{ijk} X^j X^k\right)^2
-g^2[X^i,X^p]^2
-\frac{g^2}{2}[X^p,X^q]^2
+ \frac{1}{6^2} (X^p)^2 
\right]\,.
}
Because of the complete squares in the bosonic potential, the BMN model has 
classical vacua \cite{Berenstein:2002jq,Dasgupta:2002hx,Maldacena:2002rb}
\al{\ga{
X^i = \frac{1}{3g} J^i\,, X^p = 0\,,
\quad
(i=1,2,3;\,p=4,5,\ldots9)\,,
\label{eq: BMN vacua}
}}
where $J^i$ form a (not necessarily irreducible) representation of $SU(2)$
\al{
[J^i, J^j] = i \e_{ijk} J^k\,.
}
For matrices of size $N\times N$, each classical vacuum is labelled by an integer partition of $N$ because there are in total $p(N)$ ways to realise the $SU(2)$ algebra (including the trivial representation). 

The classical vacua remain degenerate quantum mechanically \cite{Dasgupta:2002hx,Kim:2002if,Dasgupta:2002ru,Lin:2005nh}. At weak coupling $g\to 0$, 
one can study excitations on top of each individual vacuum. Above the trivial vacuum $X^I=0$, the BMN model simply describes a collection of $9N^2$ bosonic and $8N^2$ fermionic harmonic oscillators.
The spectrum is given by
\begin{align}
   E=\mu \sum_{a,b=1}^N \left[\frac{1}{3} \sum_{i=1}^3 n_{ab}^i
   +  \frac{1}{6}\sum_{p=4}^9 n_{ab}^p
   +\frac{1}{4}\sum_{\alpha=1}^8 n_{ab}^\alpha \right]
\end{align}
where $n_{ab}^i,  n_{ab}^p, n_{ab}^\alpha \ge 0$
are the occupation numbers of each oscillator. For the excitation spectrum on top of vacua corresponding to non-trivial $SU(2)$ representations, see \cite{Dasgupta:2002hx}.

At finite temperature and large $N$, the BMN model has a two-dimensional phase diagram parametrised by the dimensionless coupling $g$ and dimensionless temperature $T/\m$, see Figure \ref{fig: BMN phase diagram}. At weak coupling the BMN model has a Hagerdorn phase transition at $T_c(\l)/\m=1/(12\log3)+O(\l)$ where $\l=g^2 N$ is the `t Hooft coupling \cite{Furuuchi:2003sy}. At strong `t Hooft coupling, there is a Hawking-Page like phase transition in the dual gravitational description with $\lim_{\l\to\infty}T_c(\l)/\mu\approx 0.106$, above which a black hole dominates the thermal ensemble  \cite{Costa:2014wya}. At even stronger coupling, the Gregory-Laflamme instability \cite{Gregory:1993vy} of 11D black strings takes place at $T_c(g)/\m \sim g^{2/3}N^{-2/9}$ \cite{Itzhaki:1998dd}.

\subsection{Effective Hamiltonian at strong coupling}

In the limit $g\to \infty$, the potential term $[X^I,X^J]^2$ suppresses non-commuting matrices. In other words, the diagonal elements of $X^I$ are slow modes with frequency of order $\mu$ while the off-diagonal elements are fast modes with frequency of order $g \mu$. By integrating out the fast modes we can derive an effective Hamiltonian for the slow modes. In the following we will show how this is done by first giving a systematic exposition of the Born-Oppenheimer approach illustrated in Section \ref{Sec: toy models} and then applying it to the BMN model.
The final result is given by \eqref{eq: eff BMN Hamiltonian}.

\subsubsection{The general strategy}
\label{subsubsec: general Born-Oppenheimer strategy}

In the strong coupling limit the Hamiltonian, the supercharge, the wavefunction and the energy can be expanded near $g=\infty$\footnote{The expansion step is $g^{1/2}$ for the BMN model but not necessarily for other models. Throughout this section there are a few more occasions where the discussion is tailored for the BMN model case, but most of the discussions should be adaptable to more generic models.}
\begin{align}
\begin{gathered}
H = g H^{(1)} + g^{1/2} H^{(1/2)}  + \cdots
\\
Q_\a = g^{1/2} Q_\a^{(1/2)} + Q_\a^{(0)}  + \cdots
\\
|\Y\rangle = |\Y^{(0)}\rangle + g^{-1/2} |\Y^{(-1/2)}\rangle + \cdots
\\
E = E^{(0)} + g^{-1/2} E^{(-1/2)} + \cdots
\end{gathered}
\label{eq: schematic large g expansion}
\end{align}
The expansion of the energy above starts from $O(1)$ because we are interested in this low energy regime. 

The main idea is then to solve the Schrodinger equation order by order in $g$. At the leading order, the Hamiltonian $H^{(1)}$ only acts on the fast modes (collectively denoted as $y$) and takes in the slow modes (collectively denoted as $x$) as fixed parameters. As we will see, $H^{(1)}$ is entirely solvable and has zero ground state energy, so the leading order Schrodinger equation is
\al{
O(g):\quad H^{(1)}|\Y^{(0)}\>=0\,,
\label{eq: generic leading order Schrodinger eqn}
}
which instructs us to put the fast modes on their ground states, denoted as $|\W_x(y)\>$. The equation is thus solved by
\al{
|\Y^{(0)}\> = |\y(x)\>|\W_x(y)\>\,,
}
where $|\y(x)\>$ is the slow-mode or reduced wavefunction being completely unconstrained by \eqref{eq: generic leading order Schrodinger eqn}. As a result, the effective Hamiltonian vanishes at $O(g)$, i.e., 
\al{
H^{(1)}_\text{eff}=0\,.
}

At subleading order, it is convenient to define a projector  $P_0\equiv|\W\>\<\W|$ (we abbreviate $|\W_x(y)\>$ by $|\W\>$) \cite{Halpern:1997fv,Lin:2014wka}.
The effective Hamiltonian (and similarly for the effective supercharges) is defined through the projected Schrodinger equations
\al{\ga{
H^{(1/2)}_\text{eff}|\y\> \coloneqq 
\underbrace{\<\W| H^{(1)}}_{=0} |\Y^{(-1/2)}\> + \<\W| H^{(1/2)} |\Y^{(0)}\> = 0\,,\\[5pt]
H^{(0)}_\text{eff}|\y\> \coloneqq 
\underbrace{\<\W| H^{(1)}}_{=0} |\Y^{(-1)}\> + \<\W| H^{(1/2)} |\Y^{(-1/2)}\>  + \<\W| H^{(0)} |\Y^{(0)}\> = E^{(0)}|\y\>\,,
\\
\vdots
}
\label{eq: def of H_eff in general strategy}
}
where the first terms vanish because $|\W\>$ is the ground state with zero energy. These equations need to be solved iteratively starting from the leading order. To find the effective Hamiltonian at order $g^{-n}$ ($n$ is not necessarily an integer) we have a two-step procedure:
\begin{enumerate}
    \item Solve for $|\Y^{(-n-\frac12)}\>$ using the \emph{non}-projected Schrodinger equation at order $g^{-n+\frac12}$, e.g., for $n=0$
    \al{
    |\Y^{(-1/2)}\> = - \frac{1}{H^{(1)}} H^{(1/2)} |\Y^{(0)}\> + |\chi^{(-1/2)}(x)\>|\W\>\,.
    \label{eq: generic inversion of H^1}
    }
    The second term above is in the kernel of $H^{(1)}$ and is fixed by the normalisation condition\footnote{Between the leading and the first subleading order, this is equivalent to imposing wavefunction orthogonality $\<\Y^{(-1/2)}|\Y^{(0)}\>+\<\Y^{(0)}|\Y^{(-1/2)}\>=0$, but in general $\<\Y^{(0)}|\Y^{(-n)}\>+\<\Y^{(-n)}|\Y^{(0)}\>\neq0$ for $n>0$.}
    \al{
    \<\Y|\Y\>_{x,y} = \<\y|\y\>_x\,,
    \label{eq: normalisation condition}
    }
    where the inner product on the LHS involves integral over both $x$ and $y$ while the RHS only involves $x$.
    \item Compute the effective Hamiltonian using the input from Step 1 as well as the projection by $|\W\>$, e.g., for $n=0$
    \al{
    H^{(0)}_\text{eff}|\y\> = \<\W| H^{(1/2)} |\Y^{(-1/2)}\>  + \<\W| H^{(0)} |\Y^{(0)}\>\,.
    }
\end{enumerate}
The second step involves straightforward computations although it gets tedious and technical at higher orders, whereas the first step requires more discussion. In principle, any $|\Y^{(-n)}\>$ can be solved (up to functions in the kernel of $H^{(1)}$) by inverting the action of $H^{(1)}$ on all the previous states and the only unknown is the slow mode wavefunction $|\y\>$. In general, however, solving differential equations like \eqref{eq: generic inversion of H^1} is non-trivial and analytic solutions are not guaranteed a priori. Fortunately, all the leading order Hamiltonian $H^{(1)}$ considered in this paper are (supersymmetric) harmonic oscillators, thus we can use a \emph{polynomial ansatz} for the $y$ dependence of $|\Y^{(-n)}\>$ 
which transforms the differential equations like \eqref{eq: generic inversion of H^1} into easily solvable algebraic equations. This is demonstrated with the toy model in Section \ref{subsec: H approach toy model}.

For the BMN model with supersymmetry, we can adopt a hybrid procedure using both $H$ and $Q$ to further simplify the computation, similar to the ideas used in \cite{Smilga:1986rb} (also see \cite{Anous:2015xah,Anous:2017mwr}). Namely we will apply Step 1 with the Hamiltonian to solve for the states\footnote{Inverting the action of $Q^{(1/2)}$ is more complicated because they change the fermionic excitations.} and then in Step 2 we compute the effective supercharges $Q_\text{eff}$ instead of $H_\text{eff}$ because the former requires solving for fewer $|\Y^{(n)}\>$. Using the supersymmetry algebra, we can then compute the effective Hamiltonian through the anti-commutator of the effective supercharges. In general the Schrodinger equations to solve for BPS states in Step 1 are modified because their energies take specific values. For example, the 1/2 BPS states of the BMN model satisfy
\al{
Q_\a|\Y_{\frac12\text{BPS}}\>=0\quad (\a=1,\ldots,16)\ \Rightarrow \ 
\frac{1}{16}\sum_{\a=1}^{16} Q_\a Q_\a|\Y_{\frac12\text{BPS}}\> =  H|\Y_{\frac12\text{BPS}}\>=0\,.
}
However, up to $O(g^0)$ this hybrid procedure does not distinguish between generic states and BPS states.


\subsubsection{Application to the BMN model}

Let us now apply the strategy explained above to the BMN model. We will set $\m=1$ for simplicity and restore it explicitly for the final result.

First recall that the BMN model has an $SU(N)$ symmetry. We are mainly interested in the $SU(N)$ invariant states, so the extra step compared to the toy model case is to make a change of variables to separate out the $SU(N)$ invariant degrees of freedom. Following \cite{Lin:2014wka} we rewrite the bosonic matrices as  
\al{\spl{
X^I 
&=
U^{-1}
\left(\sum_{a=1}^N r_a^I E_a +  \sum_{a\neq b}^N q_{ab}^I T_{ab} \right) U
\,,
}
\label{eq: change of coordinates by U for X BMN}
}
where $E_a$ and $T_{ab}$ are $N\times N$ matrices with all matrix elements being zero except $(E_a)_{aa}=(T_{ab})_{ab}=1$. 
The unitary matrix $U$ parametrises the $SU(N)$ rotation angles and is chosen such that the off-diagonal elements $q_{a b}^I=( q_{b a}^I )^*$ are complex numbers satisfying the relation\footnote{This constraint guarantees that the fast mode excitations are orthogonal to those of the slow modes. It leads to $9N$ diagonal variables to be identified with slow variables, $8N(N-1)$ off-diagonal variables to be identified with fast variables and the rest $N(N-1)$ variables in $U$. The remaining $N-1$ free variables in $U$ parametrise a residual $U(1)^{N-1}$ symmetry, which is generated by the Cartan subalgebra of $SU(N)$.
The number $8N(N-1)$ equals the total number of off-diagonal elements minus the number of constraints in \eqref{eq: orthogonality condition}.}
\be \sum_{I =1}^9 q_{a b}^I r^I_{ab} = 0 \qquad a,b =1,\ldots,N,\, a \neq b\,,
\label{eq: orthogonality condition}
\ee
where $r_{ab}^I\equiv r_a^I-r_b^I$. In the new coordinates further overall $SU(N)$ transformations only act on $U$. Since the fast variables have a typical scale of order $g^{-1/2}$, it is convenient to define
\al{
q_{ab}\equiv  g^{-1/2} y_{ab}\,,
\label{eq: q and y}
}
such that $y_{ab}=O(1)$. After the change of variables the integral measure becomes 
\al{\spl{
\int \prod_{I=1}^9 \prod_{a,b=1,N} dX_{ab}^I
&=
\int \prod_{a=1}^N d^9\vec{r}_a\, \int [dU] \int \prod_{a\neq b} \frac{d^9\vec{y}_{ab}}{g^{4}} \d(\hat{r}_{ab}\cdot \vec{y}_{ab}) \left(\Delta(r) + \mathcal{O}\left( \frac{1}{\sqrt{g}} \right) \right) \\ &
\equiv
\int \prod_{a=1}^N d^9\vec{r}_a\,
\int [dU] \int [dy]
\,,
}}
where $\hat{r}_{a b}^I \equiv r_{a b}^I/|r_{a b}| $ with $|r_{a b}| \equiv \sqrt{\sum_I r_{a b}^I r_{a b}^I}$, $\D(r)\equiv \prod_{a<b} |r_{ab}|^2$ is the Vandermonde determinant\footnote{There are subleading corrections to the $r$ integral measure, but for the purpose of this section it is enough to consider only the leading order.}, $\int [dU]$ is the integral over the $SU(N)/U(1)^{N-1}$ group (which will always factor out for $SU(N)$ invariant quantities) and the $\d$ function is included to impose the constraint \eqref{eq: orthogonality condition}. 
Under the same transformation the fermionic matrices become
\al{\spl{
\hat{\Theta}_\alpha 
&=
U^{-1}
\left((\theta_a)_\a E_a + (\Theta_{ab})_\a T_{ab} \right) U\,,
}
\label{eq: change of coordinates by U for Theta BMN}
}
with the anti-commutation relations
\al{
\{(\theta_a)_\a,(\theta_b)_\b\} = \d_{ab}\d_{\a\b}\,,
\qquad
\left\{(\Theta_{ab})_\a , (\Theta_{cd})_\b \right\} = \d_{ad}\d_{bc}\d_{\a\b}\,.
}
Note that the new fermions $\theta$ and $\Theta$ are $SU(N)$ invariant but $U$-dependent. See Section \ref{subsec: non-singlet case} for details.

Using the above change of variables and expanding the BMN Hamiltonian \eqref{eq: BMN Hamiltonian} and supercharges \eqref{eq: real supercharges of BMN} in large $g$, we obtain the leading-order terms (see Appendix \ref{App: BMN Heff technicalalities} for details)
\al{
g H^{(1)} &= \sum_{a\neq b}
\left(
-\frac12g\Pi^{IJ}_{ab}\frac{\del}{\del y_{ab}^I}\frac{\del}{\del y_{ba}^J}  
+ \frac12g|r_{ab}|^2 y^I_{ab} y^I_{ba}
+ \frac12 gr_{ab}^I \Theta_{ab}^\top \g^I \Theta_{ba}
\right)\,,
\label{eq: H(1) BMN}
\\[4pt]
\sqrt{g}Q_\a^{(\frac12)}&=
\sum_{a\neq b}
\left(
\sqrt{g}(\g^I \Theta_{ab})_\a \Pi_{ab}^{IJ} \left(-i \frac{\del}{\del y_{ab}^J}\right) 
-
\sqrt{g} i r_{ab}^I y_{ab}^J (\g^{IJ}\Theta_{ba})_\a
\right)\,,
\label{eq: Q 1/2 BMN}
}
where $\Pi_{ab}^{IJ} \equiv \d^{IJ} - \hat{r}_{ab}^I \hat{r}_{ab}^J$ is a projector imposing the orthogonality constraint \eqref{eq: orthogonality condition}. Notice that here $H^{(1)}$ is again a set of supersymmetric harmonic oscillators acting only on the fast variables. The ground state of $H^{(1)}$ is thus (with $|\y(r,\theta)\>$ being unconstrained)
\al{
|\Y^{(0)}\rangle = |\y(r,\theta)\rangle \, |\W\rangle\,,
\qquad
\<\W|\W\> = 1\,,
}
which satisfies
\al{
H^{(1)}|\W\rangle = Q_\a^{(\frac12)}|\W\rangle = 0\,.
}
The explicit form of $|\W\>$ is given in \eqref{eq: full form of ground state of BMN H^1}.

Working out $Q_\a^{(0)}$ (see \eqref{eq: Q 0 BMN}) and using the definition of effective supercharge (similar to \eqref{eq: def of H_eff in general strategy})
\al{
Q_{\text{eff},\a}^{(0)}|\y(r,\theta)\>\coloneqq \<\W|Q_\a^{(0)}|\Y^{(0)}\>\,,
\label{eq: def of effective supercharge}
}
we find
\al{
Q^{(0)}_{\text{eff},\a} 
=
\sum_a \left[\left(-i\frac{\del}{\del r_a^I}\right) (\g^I \theta_a)_\a 
-\frac{1}{3}r_a^i (\g^{123} \g^i \theta_a)_\a 
+\frac{1}{6}r_a^p (\g^{123} \g^p \theta_a)_\a 
\right] \,.
}
By comparing with \eqref{eq: real supercharges of BMN} we see that $Q^{(0)}_{\text{eff},\a}$ is the diagonal supercharge with $g=0$, hence the corresponding effective Hamiltonian can be obtained straightforwardly
\al{
\left\{Q^{(0)}_{\text{eff},\alpha}, Q^{(0)}_{\text{eff},\beta}\right\}
=
2 \delta_{\alpha \beta} H^{(0)}_\text{eff}-\frac{1}{3}\left(\gamma^{123} \gamma^{i j}\right)_{\alpha \beta} M^{(0),ij}_\text{eff}
+\frac{1}{6}\left(\gamma^{123} \gamma^{pq}\right)_{\alpha \beta} M^{(0),pq}_\text{eff}\,.
}
with
\al{\spl{
H_\text{eff}^{(0)}
&= 
\sum_{a=1}^N \left(-\frac12 \frac{\del^2}{\del r_a^I \del r_a^I} + \frac12 \frac{1}{3^2} (r_a^i)^2
+ \frac12 \frac{1}{6^2}(r_a^p)^2
+\frac{i}{8} \theta_a^\top \g^{123} \theta_a\right)
\,,}\label{eq: BMN H_eff}
\\
M^{(0),IJ}_\text{eff} &=
\sum_{a=1}^N \left[
r_a^I\left(-i \frac{\del}{\del r_a^J}\right)
-
r_a^J\left(-i \frac{\del}{\del r_a^I}\right)
+\frac{i}{4} \theta_a^\top  \gamma^{IJ} \theta_a
\right]\,,
}
which are nothing but the $U(1)^N$ 
part of \eqref{eq: BMN Hamiltonian} and \eqref{eq: BMN rotation generators}. Using results given in Appendix \ref{App: BMN Heff technicalalities} we also computed $H_{\rm eff}^{(0)}$ directly through the Hamiltonian instead of the supercharges, and we found the same effective Hamiltonian as above.

Finally, defining
\al{\ga{
b_a^i \coloneqq \sqrt{\frac{1}{6}} \left(r_a^i + 3 \del_{r_a^i}\right)\,,
\quad
b_a^p \coloneqq \sqrt{\frac{1}{12}} \left(r_a^p + 6 \del_{r_a^p}\right)\,,\\
\Pi^{\pm}\coloneqq \frac12 \left(\mathbb{1}\pm i \g^{123}\right)\,,
\quad
(c_{a})_{\a} \coloneqq (\Pi^+ \theta)_\a\,,
\label{eq: switch to oscillator basis for BMN H_eff}
}}
and restoring $\mu$ explicitly, we get the main result\footnote{The (non-trivial) anti-commutation relations for the fermions in \eqref{eq: switch to oscillator basis for BMN H_eff} are
\al{\ga{
\{(c_a)^\dagger_\a, (c_b)_\b\} = \d_{ab} \Pi^+_{\a\b}\,, \quad \a,\b=1,\ldots 16\,,
\\
\{(c_a)^\dagger_{\a'}, (c_b)_{\b'}\} = \frac12 \d_{ab} \d_{\a'\b'}\,, \quad \a',\b'=1,\ldots 8\,,
}}
where in the second line we restricted to the eight independent spinor components. A further rescaling of $(c_a)_{\a'}$ by $1/\sqrt{2}$ leads back to standard anti-commutation relations. We slightly abuse notations for the eight-component fermions in \eqref{eq: BMN H_eff main result} and they should be understood in this sense.
}
\begin{align}
H^{(0)}_\text{eff}
=\sum_{a=1}^N\left(
\frac{\m}{3}\sum_{i=1}^3 (b_a^i)^\dagger b_a^i  
+\frac{\m}{6}\sum_{p=4}^9 (b_a^p)^\dagger b_a^p 
+\frac{\m}{4} \sum_{\alpha=1}^8 (c_{a})^\dagger_\a (c_{a})_{\alpha}
\right) \,.
\label{eq: BMN H_eff main result}
\end{align}
Therefore, to the leading order in the strong coupling limit the BMN model describes $N$ copies of decoupled supersymmetric harmonic oscillators.

\subsubsection{Regime of validity}\label{subsubsec:validity}
Let us comment on the validity of the effective Hamiltonian \eqref{eq: BMN H_eff main result}. Similar to the discussion in Section \ref{subsec: H approach toy model}, the Born-Oppenheimer approximation above breaks down when $|r_{ab}|\sim g^{-1/3}$. In this regime all the elements, diagonal and off-diagonal, in the bosonic matrix $X$ (without $1/\sqrt{g}$ rescaling) are of $O(g^{-\frac13})$. Thus, in order to have a clear separation between fast and slow modes, we need
\al{
\text{Validity regime of BMN }H_\text{eff}:\quad 
\left|X_{aa}-X_{bb}\right| \gg g^{-1/3}\,.
\label{eq: validity regime of BMN H_eff}
}
In terms of the original, dimensionful $X$ matrix in \eqref{eq: BMN Hamiltonian with l_P}, this means $|X_{aa} - X_{bb}|\gg \ell_P$.\footnote{This is the same as the BFSS perturbation regime \cite{Itzhaki:1998dd,Polchinski:1999br} (for finite $N$)
$\frac{g_{\rm YM}^2 }{(r/\a')^3} \ll 1 
\to
r\gg\ell_P$. We thank Xi Yin for discussion on this point.
} 

Now, when $|r_{ab}|\sim g^{-1/3} $, the BMN model has a different effective description. To see this, we rescale $X=g^{-\frac13}\tilde{X}, P=g^{\frac13}\tilde{P}$ and obtain
\al{\spl{
H/\m = & g^{\frac{2}{3}} 
\underbrace{
\operatorname{Tr}\left(\frac{1}{2}\left(\tilde P^I\right)^2
-\frac{1}{4}\left[\tilde X^I, \tilde X^J\right]^2
-\frac{1}{2} \hat\Theta^\top \gamma^I \left[\tilde X^I,\hat\Theta\right]\right)
}_{H_{\mathrm{BFSS}}} \\
& +\operatorname{Tr}\left(\frac{i}{3} \epsilon_{i j k} \tilde X^i \tilde X^j \tilde X^k
+\frac{i}{8} \hat\Theta^\top \gamma^{123} \hat\Theta\right) \\
& +g^{-\frac{2}{3}}\operatorname{Tr}\left(\frac{1}{2} \frac{1}{3^2}\left(\tilde X^i\right)^2
+
\frac{1}{2} \frac{1}{6^2}\left(\tilde X^p\right)^2\right) .
}
\label{eq: BMN with BFSS at leading order}}
This shows that the leading order term in this regime is simply the BFSS Hamiltonian. This fact plays an important role when we discuss bound states in Section \ref{subsubsec:boundstates}.

\subsection{Comparison to supergravity predictions}

\label{sec:comparison SUGRA}

The paper \cite{Kimura:2003um} computes the spectrum of (linearized) 11D supergravity on the plane wave background \eqref{PW}. The energy of a single (super)graviton with a fixed momentum $-p_{-}=k/R>0$ along the compactified null direction is given by
\begin{align}
   \varepsilon_{k}  = \frac{\mu}{3}\sum_{i=1}^3n^i +\frac{\mu}{6}\sum_{p=4}^9 n^p + \frac{\mu}{2} \varepsilon_0
   \label{SUGRAspectrum}
\end{align}
where $n^i,n^p$ are non-negative integers and $\varepsilon_0$ is the vacuum energy which depends on the particular (super)graviton mode:
\begin{center}
\begin{tabular}{ |c|c|c|c|c|c|c|c|c|c| } 
 \hline
 $\varepsilon_0$ & 0 & $\frac{1}{2}$ & 1 & $\frac{3}{2}$ & 2 & $\frac{5}{2}$& 3 & $\frac{7}{2}$ &4 \\ 
  \hline
 d.o.f. & 1 & 8 & 28 & 56 & 70 & 56 & 28 & 8 &1 \\  
 \hline
\end{tabular}
\end{center}
Integer values of $\varepsilon_0$ correspond to bosons and half-integer values correspond to fermions. Notice that, although we added a subscript $k$ to distinguish gravitons with different lightcone momenta, the energy spectrum does \emph{not} depend on $-p_-=k/R$. 

The sector with $N$ units of total momentum along $x^-$ consists of all possible combinations of multi gravitons whose momenta add up to $N$: $k_1 + \dots + k_q =N$. Thus the full spectrum of the linearized 11D supergravity is given by
\begin{align}
    E_N =   \varepsilon_{k_1} + \dots +\varepsilon_{k_q}\,.
    \label{energySUGRA}
\end{align}
Clearly, the number of gravitons obeys $q\le N$. The degeneracy of the multi-graviton states is given by the usual Fock space rules for bosons and fermions.

The BMN Hamiltonian at strong coupling ($g\to \infty$) and finite $N$ should match this energy spectrum. Let us see how this works for $N=1$. In this case, the effective Hamiltonian \eqref{eq: BMN H_eff main result} reduces to 
\begin{align}
    H_1= \frac{\mu}{3}\sum_{i=1}^3 (b^i)^\dagger b^i  
    +\frac{\mu}{6}\sum_{p=4}^9 (b^p)^\dagger b^p 
    + \frac{\mu}{4} \sum_{\alpha=1}^8 c_\alpha^\dagger c_\alpha\,.
    \label{eq:BMN_one_graviton}
\end{align}
This gives precisely the spectrum  \eqref{SUGRAspectrum} with the correct degeneracy. This Hamiltonian describes a particle with 256 internal states moving in an harmonic trap in $\mathbb{R}^9$. In addition,  the gravitons in the PP-wave background and the particles in the harmonic trap have gaussian wavefunctions with the same width (if we use the $X$ variables in \eqref{eq: BMN Hamiltonian with l_P}).

For $N\ge 2$, the effective Hamiltonian \eqref{eq: BMN H_eff main result} is simply $N$ decoupled copies of $H_1$.
This gives part of the SUGRA spectrum, namely the sector with $E_N = \varepsilon_1(1) + \dots +\varepsilon_1(N) $. The degeneracy in this sector also matches on both sides; on the gravity side,  the  $S_N$ permutation symmetry between the $N$ particles is imposed at the quantum level while the same symmetry arises as part of the $SU(N)$ gauge symmetry ($S_N\subset SU(N)$) on the MQM side once the gauge singlet condition is imposed. Now, to obtain the rest of the spectrum, we need to consider bound states on the MQM side. We will discuss this in more detail shortly.

Once the gauge singlet condition is relaxed, we will also have $SU(N)$ non-singlet states on the MQM side. In \cite{Maldacena:2018vsr}, such states were argued to be highly energetic in the large $N$ limit and decouple from the low energy spectrum. By contrast, this does not seem to be the case at finite $N$ as will be discussed in Section \ref{subsubsec:nonsinglet}.



\subsubsection{Bound states}\label{subsubsec:boundstates}

The simplest way to match the supergravity spectrum is to postulate the existence of bound states of several particles in the harmonic trap, which have the same energy spectrum as a single particle. Then, the full Hilbert space can be decomposed into sectors with $q(\leq N)$ bound states, each of which consists of $k_j$ particles such that  $k_1 + \dots + k_q =N$. 
Since the interactions are very short-ranged at strong coupling, the particles that are not bound together move freely and  we reproduce the (linearized) supergravity result \eqref{energySUGRA}. 
A bound state of $k$ particles corresponds to a graviton with momentum $p_- = -k/R$.

Notice that each particle and bound state of particles has 256 states (these 256 states are the fermionic counterpart of the bosonic modes describing the c.o.m motion), one of those being zero energy. Therefore, this scenario correctly predicts a vacuum degeneracy given by integer partitions of $N$. 
 
A bound state of $k$ non-relativistic particles has a mass $k$ times larger than a single particle. This makes the width of the Gaussian wavefunction of the c.o.m. depend on $k$ in the same way as for graviton states with $p_-=-k/R$.

What else can be said about these bound states? Consider the case $N=2$ for simplicity. In this case, it is known that there are two degenerate ground states. In our effective description, the first one corresponds to having two non-interacting particles in the ground state of the harmonic trap. The second one should correspond to a tightly bound state with the c.o.m in the ground state of the harmonic trap. To understand the nature of this bound state, it is useful to recall that, when the particles are close to each other, the dynamics at $g\to\infty$ is governed by the BFSS Hamiltonian as discussed in Section \ref{subsubsec:validity}. It is commonly believed that the BFSS model has a single normalizable ground state, which can be thought of as a threshold bound state of $N$ particles (see \cite{Lin:2014wka,Lin:2023owt} for two  recent references). We thus conclude that the bound state of the BMN model is nothing but the normalizable ground state of the BFSS model  slightly deformed by the presence of the harmonic trap. This predicts the size of the bound state to be of order $g^{-\frac{1}{3}}$ in the dimensionless variables of \eqref{eq: BMN Hamiltonian}
or, equivalently,  of order $\ell_P$ in the original variables of \eqref{eq: BMN Hamiltonian with l_P}.
It also predicts an energy gap for internal excitations of order $\mu g^{2/3}$. This argument can be easily extended to higher $N$: whenever $k(\leq N)$ particles come close ($|X_{aa}-X_{bb}|\sim g^{-1/3}$), their dynamics is approximated by the BFSS model of size $k$ and they together form a bound state of size $g^{-1/3}$. Considering all the possibilities in which the particles form bound states, one can reproduce the full spectrum of the 11D supergravity.

As we decrease the coupling, the bound state grows so that for $g\sim 1$ it has the same size as the harmonic trap. At this point there is no clear distinction between the two degenerate ground states. If we further decrease the coupling and reaches $g\ll 1$, the two degenerate vacua have very different sizes. From \eqref{eq: BMN vacua}, we see that the $2\times 2$ irreducible representation of $SU(2)$   leads to $\langle X^2 \rangle \sim 1/g^2$ and the trivial representation  leads to $\langle X^2 \rangle \sim 1$.

It would be very interesting to confirm (or disprove) these predictions for the BMN MQM. Unfortunately, even the case $N=2$ seems difficult to reach with current numerical methods (like Hamiltonian truncation).

\subsubsection{Non-singlet states}\label{subsubsec:nonsinglet}

There is no low energy state in supergravity that corresponds to $SU(N)$ non-invariant states of the BMN MQM. 
The authors of \cite{Maldacena:2018vsr} proposed that the non-singlets states have high energy at strong coupling. 
Their proposal seems consistent with recent Monte-Carlo studies \cite{Berkowitz:2018qhn,Pateloudis:2022oos}.
Their arguments assumed large $N$ but if we directly translate their conjecture to our conventions, we find
$
    E_\text{non-singlet} \sim \mu g^{2/3}
$ for $g\to \infty$. 
We see no evidence for such scaling at finite $N$. In fact, if we derive the effective Hamiltonian at strong coupling allowing for generic $SU(N)$ charge, we find $
    E_\text{non-singlet} \sim \mu
$.

At strong coupling, we find the following effective Hamiltonian (see Appendix \ref{subsec: non-singlet case} for the details):
\al{
\spl{
H_{\rm eff} =& - \frac{1}{2} \frac{\del^2}{\del r_a^I \del r_a^I} + \frac{1}{2} \left[ \frac{1}{3^2} \sum_{i=1}^3 (r_a^i)^2 + \frac{1}{6^2}  \sum_{p=4}^9 (r_a^p)^2 \right] + \frac{i}{8} \theta_a^T \gamma_{1 2 3} \theta_a 
\\ 
& + \frac{1}{2} \sum_{a \neq b} \frac{1}{|r_{a b}|^2} \left( U \frac{\del}{\del U} \right)'_{a b} \left( U \frac{\del}{\del U} \right)'_{b a} \,,
}
\label{eq: BMN H_eff for non-singlets}
}
where the new term compared to the $SU(N)$ singlet case is in the second line. The matrix $U$ is defined in \eqref{eq: change of coordinates by U for X BMN} and the prime in the differential operator $(U \del/\del U)'$ indicates that it only acts on the explicit $U$ dependence of the reduced wavefunction $|\y(r,\theta,U)\>$ while treating $\theta$ as constant. For the case of $N=2$, the new term is simply $\propto l(l+1)/|r_{12}|^2$ where $l$ is the quantum number of the $SU(2)$ representation.


One may speculate that integrating out the fast d.o.f. leads to the effective Hamiltonian \eqref{eq: BMN H_eff for non-singlets} plus a short range interaction between the particles that depends on the $SU(N)$ representation.
Can this short range interaction lead to 
$
    E_\text{non-singlet} \sim \mu g^{2/3}
$?
We don't think so because we can consider states (not eigenstates) where the wave-functions of the $N$ particles do not overlap. Such states are not affected by the short range interactions and their energy (expectation value of $H_\text{eff}$) can be made of order $\mu$.


It would be interesting to compute (numerically) the energy gap for non-singlet states in the $N=2$ BMN MQM. This may be accessible to the Monte-Carlo methods of \cite{Berkowitz:2018qhn,Pateloudis:2022oos}.


\section{Hamiltonian truncation for the minimal BMN model}
\label{sec:minimalBMN}
After obtaining the main result \eqref{eq: BMN H_eff main result}, ideally now we would like to test it numerically. Unfortunately, the numerical method applied in section \ref{subsec: numerical check of toy model} for the toy model is not practical to upscale to more particles moving in higher dimensions. The Hamiltonian truncation (or Rayleigh-Ritz) method appears to be the more feasible approach.\footnote{For general discussion on Hamiltonian truncation see e.g. \cite{Hogervorst:2014rta,Rychkov:2014eea,Rychkov:2015vap}. For application of Hamiltonian truncation to matrix quantum mechanics problems see e.g. \cite{Balthazar:2016utu,Motycka:2014vra}.} However, the BMN model has a large number of matrices. Simply constructing the $SU(2)$ invariant oscillator basis states below a relatively low energy cutoff, say $\L/\mu=100$, is already challenging. Another option is the matrix quantum mechanics bootstrap \cite{Han:2020bkb}, which for large $N$  has successfully put bounds on the energy spectrum and some expectation values in both simple matrix quantum mechanics models \cite{Han:2020bkb} and the BFSS model \cite{Lin:2023owt} (see also \cite{Bhattacharya:2021btd,Berenstein:2021dyf,Berenstein:2021loy,Nancarrow:2022wdr} for applications to other quantum mechanical systems). 

It would be very interesting to develop the bootstrap techniques to extract the (low-lying) spectrum of the BMN model. In this section, however, we instead use the Hamiltonian truncation method to study a simpler version of the BMN model, the so-called minimal BMN model\footnote{By dimensionally reducing $\NN=1$ SYM in 6$d$, 4$d$, 3$d$ and 2$d$ one can obtain supersymmetric matrix quantum mechanics models similar to the BFSS model with $\NN=8,4,2,1+1$ supersymmetry \cite{Kim:2006wg}. The minimal BMN model is the supersymmetry-preserving mass deformation of the $\NN=2$ matrix quantum mechanics model which has two spatial dimensions before dimensional reduction. For numerical study of mini-BMN and mini-BFSS model ($\NN=4$) see \cite{Anous:2017mwr,Han:2019wue}.} \cite{Kim:2006wg,Rinaldi:2021jbg} for $N=2$. The Hamiltonian of the minimal BMN model with $SU(N)$ symmetry reads\footnote{To connect with the BMN Hamiltonian written in terms of (explicitly real) Majorana fermions \eqref{eq: BMN Hamiltonian} one can use
\al{
\L=(\hat\Theta_{\a=1} + i \hat\Theta_{\a=2})/\sqrt{2}\,,
\quad
\g^1 = -\s^3\,,\quad \g^2 = \s^1\,,\quad \g^{12}=-i\s^2\,,
}
to write the Hamiltonian as
\al{\spl{
H/\m=
&\operatorname{Tr}
\left[
\frac{1}{2} (P^i)^2
-\frac{g^2}{4}\left[X^i, X^j\right]^2
-\frac{g}{2} \hat\Theta^\top \gamma^i\left[X^i, \hat\Theta\right]
+\frac{1}{2} (X^i)^2
-\frac{3 i}{4} \hat\Theta^\top \gamma^{12} \hat\Theta
\right]
-\frac14(N^2-1)\,,
}\label{eq: mBMN hamiltonian in Majorana fermions}
}
with $i,j=1,2$.
}
\al{\spl{
H/\m &= {\rm Tr}\left[
\frac{1}{2}(P^i)^2
-\frac{g^2}{2}[X^1,X^2]^2
-\frac{g}{\sqrt{2}}\L\left[\frac{X^1+iX^2}{\sqrt{2}},\L\right]
-\frac{g}{\sqrt{2}}\L^{\dagger}\left[\frac{X^1-iX^2}{\sqrt{2}},\L^{\dagger}\right]
\right]\\
&\quad
+{\rm Tr}\bigg[\frac{1}{2}(X^i)^2
+\frac{3}{2}\L^{\dagger}\L
\bigg] -(N^2-1)\,,
}
\label{eq: mBMN hamiltonian in X^i}}
where $X^i, P^i$ are $N\times N$ hermitian and \emph{traceless} matrices with $i=1,2$, $\L$ and $\L^\dagger$ are \emph{traceless} complex fermionic matrices. Note that the bosonic zero-point energy is subtracted explicitly. In the rest of this section we set $\m=1$.

Like the the BMN model, the minimal BMN model has one dimensionless coupling constant $g$, and its harmonic potential removes the flat directions leading to a discrete spectrum. The minimal BMN model also has similar though simpler symmetry structure: $SO(2)$ $R$-symmetry, $SU(N)$ symmetry and $\NN=2$ supersymmetry. Applying the method in Section \ref{Sec: free gravitons from strong coupling}, one can find that at large $g$ the minimal BMN model also becomes a collection of supersymmetric harmonic oscillators to the leading order. It is for these reasons that we consider the Hamiltonian truncation study of the minimal BMN model as a useful test of the main result in Section \ref{Sec: free gravitons from strong coupling}. One notable distinction from the BMN model is that there is no Myers term (proportional to $\e_{ijk}X^i X^j X^k$) \cite{Myers:1999ps} in the minimal BMN model, thus it does not have degenerate vacua.

In the rest of this section we will reformulate the minimal BMN Hamiltonian in oscillator basis, explain the Hamiltonian truncation setup, present the numerical results and compare them with explicit analytic predictions for $N=2$.

\subsection{Formulation in oscillator basis}
For the purpose of Hamiltonian truncation it is convenient to use an oscillator basis. For the bosons let us define
\al{
\begin{gathered}
Z \coloneqq \frac{1}{\sqrt{2}} \left(X^1 - i X^2\right)\,,
\quad
\bar{Z} \coloneqq \frac{1}{\sqrt{2}} \left(X^1 + i X^2\right)\,,
\\
P_Z \coloneqq 
\frac{1}{\sqrt{2}} \left(P^1 + i P^2\right)\,,
\quad 
P_{\bar{Z}} \coloneqq 
\frac{1}{\sqrt{2}} \left(P^1 - i P^2\right)\,.
\end{gathered}
\label{eq: U, V variables}
}
The corresponding creation and annihilation operators are
\begin{equation}
    \begin{aligned}
    &A^{\dagger} = \frac{1}{\sqrt{2}}\left(Z-iP_{\bar{Z}}\right)\,,\qquad B^{\dagger} = \frac{1}{\sqrt{2}}\left(\bar{Z}-iP_Z\right)\,,
    \end{aligned}
\label{eq: definition of A and B}
\end{equation}
with the following commutation relations:
\begin{equation}
    \begin{aligned}
    &[A_{ij},A^{\dagger}_{kl}]=\delta_{il}\delta_{jk}-\frac{1}{N}\delta_{ij}\delta_{kl}\,,\qquad [B_{ij},B^{\dagger}_{kl}]=\delta_{il}\delta_{jk}-\frac{1}{N}\delta_{ij}\delta_{kl}\,.
    \end{aligned}
\end{equation}
The fermions satisfy the anti-commutation relations:
\al{
\{\L^{\dagger}_{ij},\L_{kl}\}
=\delta_{il}\delta_{jk}-\frac{1}{N}\delta_{ij}\delta_{kl}\,.
}

Now the Hamiltonian becomes
\al{
\spl{
H&={\rm Tr}\bigg(
A^\dagger A + B^\dagger B +\frac{3}{2}\L^{\dagger}\L 
-\frac{g}{\sqrt{2}}\L[\bar{Z},\L]
-\frac{g}{\sqrt{2}}\L^{\dagger}[Z,\L^{\dagger}]
+ \frac{g^2}{2}[Z,\bar{Z}]^2
\bigg)
\,,
}
\label{eq: mBMN hamiltonian in oscillator basis}
}
and the $SO(2)$ generator can be written as
\al{
M
= \op{Tr}\left(A^{\dagger}A-B^{\dagger}B-\frac{1}{2}\L^{\dagger}\L\right) 
\,,
\label{eq: mBMN SO(2) charge}
}
which satisfies 
\begin{equation}
    \begin{aligned}
    [M,H] = 0\,,\quad
    [M, A^\dagger_{ij}] = A^\dagger_{ij}\,,\quad [M,B^\dagger_{ij}] = -B^\dagger_{ij}\,,\quad
    [M,\L^\dagger_{ij}] = -\frac{1}{2}\L^\dagger_{ij}\,.
    \end{aligned}
\end{equation}
Therefore, $A^\dagger$ and $B^\dagger$ carry charge $+1$ and $-1$, respectively, and the fermions $\L^{\dagger}$ carry charge $-1/2$.

The only complex supersymmetry generator is
\al{
Q
&=
{\rm Tr}\left(2i B \L^{\dagger}+i g[Z,\bar{Z}]\L\right)\,,
}
and the anti-commutation relation reads
\al{\begin{gathered}
\{Q,Q^{\dagger}\} = 2\left(H - M\right)\,.
\end{gathered}
\label{eq: mBMN superalg}}
The supercharge can be constructed from the real supercharges through $Q=Q_{\a=1}+i Q_{\a=2}$ where $Q_\a=Q_\a^\dagger$, and similarly for its hermitian conjugate. In addition, the supercharges transforms under $SO(2)$ rotations as
\begin{align}
\ga{
\big[H,Q\big] = \frac{1}{2}Q
+ ig{\rm Tr}\left(\L G\right)
\,,
\qquad 
\big[H,Q^\dagger\big] = -\frac{1}{2}Q^{\dagger}
+ ig{\rm Tr}\left(\L^\dagger G\right)
\,,
\\
[M,Q] = \frac{1}{2}Q\,,
\qquad [M,Q^{\dagger}] = -\frac{1}{2}Q^{\dagger}\,,
}
\end{align}
where $G^A$ are the $SU(N)$ generators 
\al{
G = G_A T_A\,,
\quad
G_A = f_{ABC}\left(X_B^i P_C^i - i \L_B \L_C^\dagger\right)\,,
\quad
A=1,\ldots,N^2-1\,.
}
In the following we will set $G^A=0$ because we are only interested in $SU(N)$ invariant states. The BPS states are states annihilated by both supercharges, they satisfy 
\begin{align}
    Q|\Y_{\rm BPS}\rangle=Q^{\dagger}|\Y_{\rm BPS}\rangle=\left(H-M\right)|\Y_{\rm BPS}\rangle = 0\,.
\label{eq: BPS condition of minimal BMN}
\end{align}

\subsection{Hamiltonian truncation}

\subsubsection{Setup}
Let us briefly state the idea of Hamiltonian truncation. For more details see, for example, \cite{Hogervorst:2014rta,Hogervorst:2021spa}. Given a Hamiltonian $H(g)$ depending on a parameter $g$, in general we can split the full Hamiltonian into two parts $H(g) = H_0 + H_{\rm int}(g)$, where $H_0$ is exactly diagonalisable. In our case, a simple choice is $H_0 \equiv H(g=0)$. The full Hilbert space of the interacting system can be then expanded in the eigenbasis of $H_0$, denoted as $\{|i\>\}$ (in general the states are  labelled by the energy eigenvalue under $H_0$ and other quantum numbers). One can then define the Hamiltonian matrix $\Tilde{H}(g)$ as
\begin{align}
    H(g) |i\rangle = \sum_j \Tilde{H}^j_{\ i}(g)|j\rangle\,.
\end{align}
We will assume the set $\{|i\>\}$ is ordered with respect to its energy under $H_0$. To perform Hamiltonian truncation, we need to find all states with energy below a given cutoff $\Lambda$: $\{|i\>\}_{i=1}^{n(\Lambda)}$, $n(\Lambda)$ finite. Then, the matrix $\Tilde{H}(g)^j_{\ i}$ is square and its eigenvalues are real for any $g$.\footnote{
Notice that $\tilde H^j_{\ i}$ is not hermitian in general. One can use the Gram matrix $G_{ij}$ to transform it to a hermitian matrix \cite{Hogervorst:2014rta}
\al{
\tilde H_{ij}\equiv \<i|H|j\> = G_{ik} \tilde H^k_{\ j}\,,
\qquad G_{ij}\equiv \<i|j\>\,.
}
} In the limit $\Lambda\to\infty$, the eigenvalues of $\Tilde{H}(g)^j_{\ i}$ converge towards those of $H(g)$. 

In the case of the minimal BMN model (as well as the BMN model with any $N$), at $g=0$ the Hamiltonian is a collection of harmonic oscillators 
\al{\spl{
H_0 &={\rm Tr}\bigg(
 A^\dagger A +  B^\dagger B +\frac{3}{2}\L^{\dagger}\L 
\bigg)
\,.
\label{eq: mBMN H_0}
}}
Note that this is different from the collection of harmonic oscillators at $g\to\infty$ \eqref{eq: BMN H_eff main result}.

We will now focus on $SU(2)$ invariant states and build up the basis $\{|i\>\}$, which can be constructed by acting multi-trace operators on the vacuum. 
The three basic matrix operators are $A^\dagger,B^\dagger$ and $\L^\dagger$ and the vacuum satisfy
\al{
A_{ij}|0\>=B_{ij}|0\>=\L_{ij}|0\>=0\,.
}

The above oscillator basis is not optimal for the Hamiltonian truncation of the minimal BMN model for $g>1$. It is numerically more efficient to use a collection of oscillators whose widths are variable as functions of $g$ and tune them close to the typical widths of the actual wavefunctions at the corresponding coupling. In this paper we choose the simplest implementation of variable oscillator frequency. To this extent we redefine 
\al{\spl{
H_{0,\w} &={\rm Tr}\bigg(
\omega   A_{\omega}^\dagger A_{\omega} +  \omega   B_{\omega}^\dagger B_{\omega} +\frac{3}{2}\L^{\dagger}\L 
\bigg)
\,.
}}
where $\omega $ is a variable oscillator frequency that is optimized numerically for each $g$ to minimize the ground state energy. The creation and annihilation operators become
\begin{equation}
    \begin{aligned}
    &A_{\omega}^{\dagger} = \sqrt{\frac{\omega}{2}}\left(Z-\frac{i}{\omega}P_{\bar{Z}}\right)\,,\qquad B_{\omega}^{\dagger} = \sqrt{\frac{\omega}{2}}\left(\bar{Z}-\frac{i}{\omega}P_Z\right)\,,
    \end{aligned}
\label{eq: definition of A_w and B_w}
\end{equation}
and the supersymmetry generator
\al{
Q
&=
{\rm Tr}\left(\frac{i}{\sqrt{\omega}} ((1+\omega)B_{\omega}+(1-\omega)A_{\omega}^{\dagger}) \L^{\dagger}+i g[Z,\bar{Z}]\L\right)\,.
}
All commutation relations are essentially unchanged. $H_{0,\w}$ now depends on $g$ but it is still exactly diagonalisable.
The interaction part becomes
\al{\spl{
H_{{\rm int},\w} &={\rm Tr}\bigg(-\frac{1-\omega^2}{2\omega}Z\bar{Z}
-\frac{g}{\sqrt{2}}\L[\bar{Z},\L]
-\frac{g}{\sqrt{2}}\L^{\dagger}[Z,\L^{\dagger}]
+ \frac{g^2}{2}[Z,\bar{Z}]^2
\bigg)+(\omega -1)(N^2-1)\,.
}}
Of course, the full Hamiltonian as well as the supercharges are independent of $\omega$.
We will use the oscillator basis built out of $(A_\w)_{ij}^{\dagger}, (B_\w)_{ij}^{\dagger},\L_{ij}^{\dagger}$ which diagonalizes $H_{0,\w}$. In principle, one can adopt a series of different frequencies. For example, it would be interesting to use different frequencies for the diagonal and the off-diagonal matrix elements and this should improve the numerics especially at large $g$. For clarity and simplicity, the discussion in the rest of this section assumes using the simplest oscillator basis from \eqref{eq: mBMN H_0}.

For finite $N$ not all single-trace operators with arbitrary length are independent due to trace relations. For example, for a single $N\times N$ matrix $M$, $\op{Tr}\left(M^{N+1}\right)$ can be written as a combination of products of shorter traces. The construction of independent $SU(N)$ invariant single-/multi-trace operators can be implemented using the so-called Hilbert series (see e.g. \cite{Aharony:2003sx,Jenkins:2009dy,Henning:2017fpj} for reviews on this subject).\footnote{We thank Brian Henning for helpful discussions related to the Hilbert series.} The (multi-graded) Hilbert series for the $SU(2)$ minimal BMN model is
\al{
h(a,b,f) = 
\frac{1+ fa + fb + fab + f^3 + f^2 a + f^2 b + f^2 ab}{(1-a b) \left(1-a^2\right) \left(1-b^2\right) }\,.
\label{eq: Hilbert series}
}
Through the identification
\al{
a \leftrightarrow A^\dagger\,, \quad
b \leftrightarrow B^\dagger\,, \quad
f \leftrightarrow \L^\dagger\,
\label{eq: hilbert series identification}
}
each term in the function $h(a,b,f)$ can be identified with a single-/multi-trace operator (see Table \ref{tab:build})\footnote{
The identification is not unique. For example, there are three (non-zero) candidate operators to be associated with the term $f^2ab$
\begin{align*}
\op{Tr}(\L^\dagger \L^\dagger A^\dagger B^\dagger)\,,
\quad
\op{Tr}(\L^\dagger \L^\dagger B^\dagger A^\dagger)\,,
\quad
\op{Tr}(\L^\dagger A^\dagger)
\op{Tr}( \L^\dagger  B^\dagger)\,.
\end{align*}
The fact that the term $f^2ab$ has coefficient 1 in the Hilbert series predicts that there is actually only one independent trace operator (in the $SU(2)$ case). This can indeed be shown using the Caley-Hamilton relations (see e.g. \cite{Dempsey:2022uie}), or computing the rank of the Gram matrix built out of these three states.
} and $h(a,b,f)$ itself is interpreted as the generating function for all $SU(2)$ invariant single-/multi-trace operators. When expanded in small $a, b, f$, the numerical coefficient in front of a generic term $a^m b^n f^l$ dictates the number of independent invariant states with the corresponding matrix content. Notice that each term in the denominator can be freely generated to arbitrary powers, while the terms in the numerator can appear at most once. This is consistent with the Bose and the Fermi statistics. The building blocks are summarized in Table \ref{tab:build}. In other words, any basis state can be written as
\begin{align}
  \sum_{\{n_i\}} c_{\{n_i\}}  \prod_{i=1}^{10}T_i^{n_i}|0\rangle
  \label{eq: general N=2 mBMN states from T}
\end{align}
with the restriction $\sum_{i=1}^7n_i \in \{0,1\}$,\footnote{
In Table \ref{tab:build} we make it manifest that $T_7 = T_1T_2$ and the restriction here should be understood as counting $T_7$ as if it is independent of $T_1$ and $T_2$.
} 
and the $c_{\{n_i\}}$ are numbers. 
\begin{table}[!h]
    \centering
    \renewcommand{\arraystretch}{1.2}
    \begin{tabular}{|c|c|c|c|c|}
     \hline
     Label & Content & Hilbert series term & $E_{g=0}$ & $SO(2)$ charge\\ \hline\hline
       $ T_1$ & $\text{Tr}\left(\L^{\dagger}A^{\dagger}\right)$ & $fa$ & $5/2$ & $1/2$ \\ \hline
        $T_2$ & $\text{Tr}\left(\L^{\dagger}B^{\dagger}\right)$ & $fb$ & $5/2$ & $-3/2$ \\ \hline
       $ T_3$ & $\text{Tr}\left(\L^{\dagger}A^{\dagger}B^{\dagger}\right)$ & $fab$ & $7/2$ &$ -1/2$ \\ \hline
        $T_4$ & $\text{Tr}\left(\L^{\dagger}\L^{\dagger}\L^{\dagger}\right)$ & $f^3$ & $9/2$ & $-3/2$ \\ \hline
       $ T_5$ & $\text{Tr}\left(\L^{\dagger}\L^{\dagger}A^{\dagger}\right)$ & $f^2a$ & $4$ & $0$ \\ \hline
       $ T_6$ & $\text{Tr}\left(\L^{\dagger}\L^{\dagger}B^{\dagger}\right)$ & $f^2b$ & $4$ & $-2$ \\ \hline
      $T_7(=T_1 T_2)$ & $\text{Tr}\left(\L^{\dagger}A^\dagger\right)\op{Tr}\left(\L^{\dagger}B^{\dagger}\right)$ & $f^2ab$ & $5$ &$-1$ \\ \hline
       $ T_8$ & $\text{Tr}\left(A^{\dagger}B^{\dagger}\right)$ & $ab$ &$2$ & $0$ \\ \hline
      $  T_9$ & $\text{Tr}\left(A^{\dagger}A^{\dagger}\right)$ & $a^2$ & $2$ & $2$ \\ \hline
       $ T_{10} $& $\text{Tr}\left(B^{\dagger}B^{\dagger}\right)$ & $b^2$ & $2$ & $-2$ \\ \hline
    \end{tabular}
    \caption{Building blocks for the basis states of $N=2$ minimal BMN. In terms of the oscillator basis \eqref{eq: definition of A_w and B_w}, one simply replaces $A^\dagger\to A^\dagger_\w$, $B^\dagger\to B^\dagger_\w$ and the numbers in the fourth column should be understood as eigenvalues of $H_{0,\w}$ measured in units of $\w$.}
    \label{tab:build}
\end{table}
Notice that each building block in Table \ref{tab:build} is labelled by free oscillator energy $E_{g=0}$ and $SO(2)$ charge $M$. Since $[H,M]=0$ we will study sectors with definite $M$ separately.

So far the procedure described is generic for Hamiltonian truncation. Additionally the $N=2$ minimal BMN model has a few more noteworthy properties by itself including the $\NN=2$ supersymmetry, which help simplify the numerics. We summarise the key points here and relegate the details to Appendix \ref{App: N=2 mBMN properties}.

\begin{itemize}
    \item  The Hilbert space can be divided into four types of $SO(2)$ charge sectors with 
    \begin{align*}
    M=2n,2n+1/2,2n+1,2n+3/2,\qquad n\in\mathbb{Z}\,.
    \end{align*}
    The latter two types of sectors do not see the Yukawa interaction terms so the dynamics is essentially governed by a purely bosonic Hamiltonian. 
    This is because the states in the $M=2n+1$ sectors always have the form $T_7 T_8^{n_8} T_9^{n_9} T_{10}^{n_{10}}|0\>$ while the states in the $M=2n+3/2$ sectors always have the form $T_3 T_8^{n_8} T_9^{n_9} T_{10}^{n_{10}}|0\>$, and one can verify explicitly that the Yukawa terms annihilate these states. The energy spectrum in these sectors grows as $g^{2/3}$ for large $g$ and decouples from the energy regime we are interested in. Therefore, we will not focus on these two types of sectors.
    
    \item For the remaining two types of sectors, there is one and only one BPS state in each $M=2n$ sector if $n\geq0$ and no BPS state in the $M=2n+1/2$ sectors. This can be easily verified at $g=0$ by recalling the BPS condition \eqref{eq: BPS condition of minimal BMN} and using Table \ref{tab:build}. Since the BPS states are protected from the interactions the conclusion generalises to any $g$.
    
    \item The $\NN=2$ supersymmetry relates the states in the $M=2n+1/2$ sectors with 
    states in $M=2n$ sectors through the action of the supercharge
    \al{
    |\Y_{E(>M),M(=2n)}\>\  \overset{Q}{\underset{Q^\dagger}{\rightleftarrows}} \ |\Y_{E+1/2,M+1/2}\>\,,\qquad (\text{for any } g)\,.
    }
    Each pair of such states forms an $\NN=2$ SUSY long multiplet, whereas the BPS states are short multiplets (with only one state in each short multiplet). However, supersymmetry also relates the $M=2n,\,2n+1/2$ sectors to the $M=2n+1,\,2n+3/2$ sectors, so within the former two types of sectors there are states that also decouple in the large $g$ limit. This leads to level crossing phenomena in the energy spectrum, see Figure \ref{fig: mBMN main results} and \ref{fig: susy long multiplet}.
\end{itemize}
To sum up, we can mainly focus on the $M=2n$ sectors of the $SU(2)$ minimal BMN model for the Hamiltonian truncation.

\subsubsection{Numerical results}
\begin{figure}[t]
    \centering
    \includegraphics[width=\textwidth]{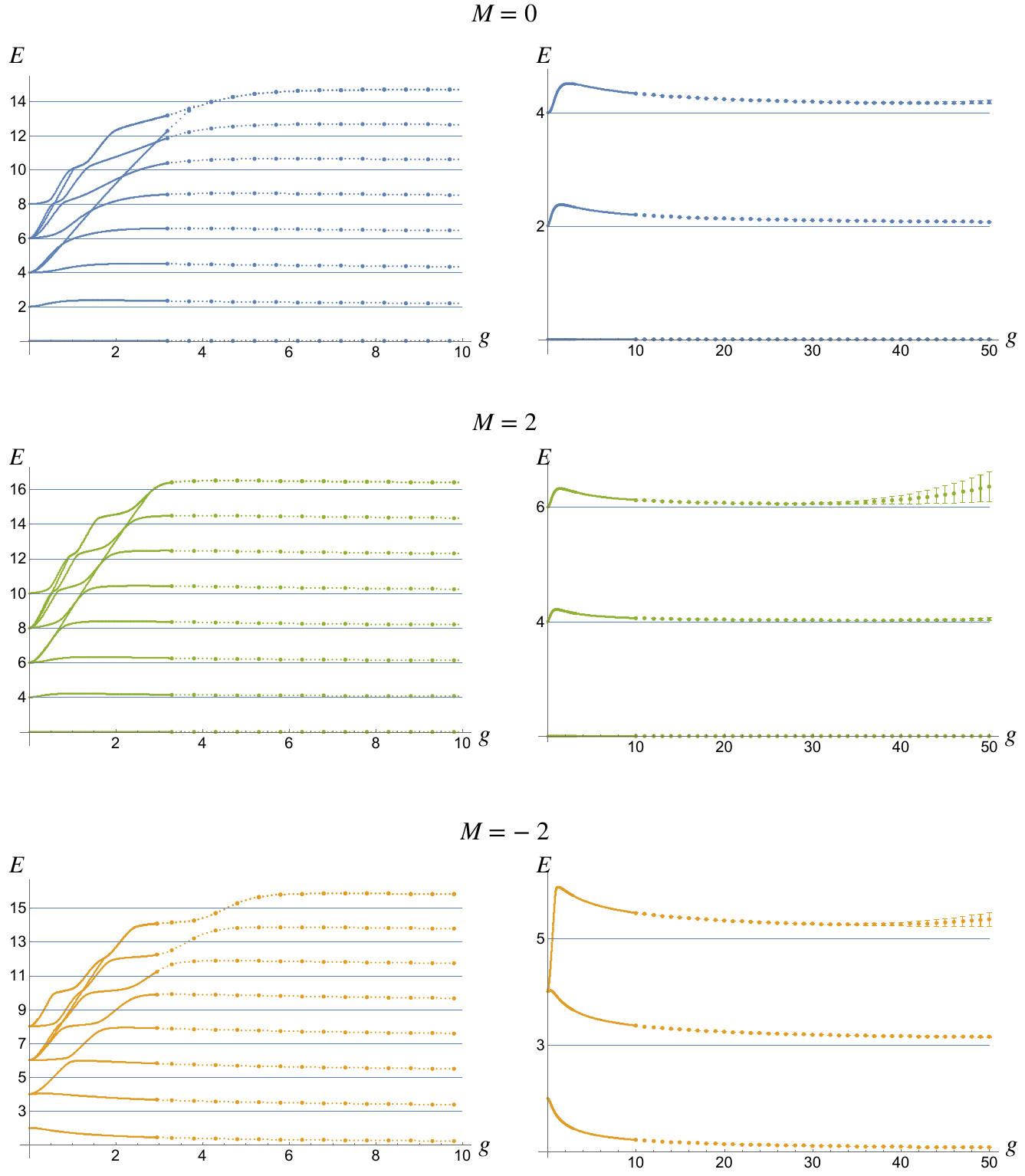}
    \caption{The energy spectrum of the $N=2$ minimal BMN model at various $SO(2)$ charge ($M$) sectors. The left column shows the first eight lowest energy levels with $0.2\leq g \leq 10$ (nine or ten levels are shown near $g=0$ because of level crossing). The sampling step in $g$ is 0.01 for $g\leq3$ and 0.1 afterwards. The right column shows the first three lowest energy levels with $g$ extended up to 50. The energy cutoff of the oscillator basis is $\L=200$ and the dots represent the results from extrapolation to infinite cutoff. The error bars which are only visible for $g\gtrsim30$ indicate the differences between $\L=200$ and extrapolation to $\L=\infty$. The thin straight lines indicate the analytic prediction at $g\to\infty$ \eqref{eq: SU(2) mBMN analytic spectrum}.}
    \label{fig: mBMN main results}
\end{figure}

The main results of the Hamiltonian truncation are given in Figure \ref{fig: mBMN main results} (also see Figure \ref{fig: susy long multiplet}). The energy cutoff is $\L/\m=200$. An extrapolation to infinite $\L/\m$ was made to bypass finite cutoff effects.

In Figure \ref{fig: mBMN main results} the left end of the spectrum can be understood from (perturbation away from) free theory. More importantly, the right end agrees with prediction given in \eqref{eq: SU(2) mBMN analytic spectrum}. As $g$ increases, the states on the left end get ``diluted'' towards the right end and the degeneracy gets lifted. Notice that BPS states are present only in even positive charge sectors. Unfortunately, given the available data for relatively small $g$ we are unable to extract $1/g$ correction to the spectrum  (in Section \ref{subsec: numerical check of toy model} the maximal $g$ for the toy model is $10^7$). Adopting a more optimised variable oscillator frequency as discussed previously should improve the numerics and allow for obtaining results at higher $g$.

\subsection{Analytic results for $N=2$}
Using the same method in Section \ref{Sec: free gravitons from strong coupling} for the BMN model, we find that the minimal BMN model (including the centre-of-mass d.o.f.) also becomes a collection of supersymmetric harmonic oscillators in the strong coupling limit
\al{
H_\text{eff}^{(0)}
&= 
\sum_{a=1}^N \left(-\frac12 \frac{\del^2}{\del r_a^i \del r_a^i} + \frac12 (r_a^i)^2
+\frac{i}{4} \theta_a^\top \g^{12} \theta_a\right)
-\frac{N^2}{4}
\,,
\label{eq: mBMN H_eff}
}
where we have adopted the flat integration measure and subtracted the zero-point energy, see \eqref{eq: mBMN hamiltonian in Majorana fermions}.

In the following we set $N=2$. In order to remove the centre-of-mass degrees of freedom we define
\al{
\begin{cases} 
\Vec{r}_1 \equiv \frac{1}{\sqrt{2}} (\Vec{r}_{\rm CM} + \Vec{r}) \\ 
\Vec{r}_2 \equiv \frac{1}{\sqrt{2}} (\Vec{r}_{\rm CM} - \Vec{r}) \end{cases} \,,
\qquad  
\begin{cases} 
\theta_1 = \frac{1}{\sqrt{2}} (\theta_{\rm CM} + \theta) \\ \theta_2 = \frac{1}{\sqrt{2}} (\theta_{\rm CM} - \theta)  
\end{cases}\,.
}
We further use polar coordinates and switch to complex fermions 
\al{
\Vec{r} = r \bmat \cos \alpha 
\\ 
\sin \alpha \emat \,,
\quad 
\theta = \frac{1}{\sqrt{2}} \bmat \l + \l^\dagger 
\\ 
-i (\l - \l^\dagger) \emat \,,
\quad 
\theta_{\rm CM} = \frac{1}{\sqrt{2}} \bmat \l_{\rm CM} + \l_{\rm CM}^\dagger 
\\ 
-i (\l_{\rm CM} - \l_{\rm CM}^\dagger) \emat  \, . 
}
With these definitions the effective Hamiltonian can be split into relative motion and centre-of-mass motion
\al{ 
H_{\mathrm{eff}}^{(0)} &= 
H_{\mathrm{eff},\,{\rm rel}}^{(0)} + H_{\mathrm{eff},\,{\rm CM}}^{(0)}
\\
H_{\mathrm{eff},\,{\rm rel}}^{(0)}&= 
-\frac{1}{2r}\del_r\left(r \del_r\right)
-\frac{1}{2r^2}\del_\a^2
+\frac{1}{2}r^2
+\frac{3}{2}\l^\dagger \l -\frac{3}{2}
\,,
\label{eq: mBMN H_eff in radial coordinates}
\\
H_{\mathrm{eff},\,{\rm CM}}^{(0)}  &= - \frac{1}{2} \left(\frac{\del}{\del r_{\rm CM}^i}\right)^2 + \frac{1}{2} (r_{\rm CM}^i)^2  + \frac{3}{2} \l_{\rm CM}^\dagger \l_{\rm CM} - 1  
\,. 
}

To compare with the numerics let us focus on the relative motion part and drop the ``rel'' subscript for simplicity. The $SO(2)$ generator \eqref{eq: mBMN SO(2) charge} now becomes
\al{
M= i \del_\a - \frac12 \l^\dagger \l - \frac12\,,
}
where the subtraction of $1/2$ is to be consistent with the subtraction of the fermionic zero-point energy from the Hamiltonian and it is crucial to give the correct final result. The reduced wavefunction reads
\al{
|\y_\pm(\vec r)\> = \y(r)e^{-im\a} |\pm\>\,,
}
where $|+\>$ denotes the single fermion created by acting $\l^\dagger$ on the fermionic vacuum $|-\>$. For $M=2n+1/2$ sectors we use $|\y_+(\vec r)\>$ because only fermions have half-integer $SO(2)$ charges and it has $SO(2)$ charge $M_+=m-1$; for $M=2n$ sectors we use $|\y_-(\vec r)\>$ and it has $SO(2)$ charge $M_-=m-1/2$.

Finally the effective Hamiltonian in $M=2n$ and $M=2n+1/2$ sectors read, respectively,
\al{
\ga{
H^{(0)}_{\text{eff},\,M_-}
=
-\frac{1}{2r}\del_r\left(r \del_r\right)
+\frac{(M_-+1/2)^2}{2r^2}
+\frac{1}{2}r^2
-\frac{3}{2}
\quad \left(M_-\in 2\mathbb{Z}\right)\,,
\\
H^{(0)}_{\text{eff},\,M_+}
=
-\frac{1}{2r}\del_r\left(r \del_r\right)
+\frac{(M_++1)^2}{2r^2}
+\frac{1}{2}r^2
\quad \left(M_+\in 2\mathbb{Z}+1/2\right)\,.
}
\label{eq: mBMN H_eff split into M=2n,2n+1/2}
}
To the leading order, the energy eigenvalues are
\al{\ga{
E_{M_{-}}=
\left|M_- +\frac12\right|-\frac12+2k_- 
+ (\text{subleading})
\quad \left(M_-\in 2\mathbb{Z}\right)\,,
\\[5pt]
E_{M_+}=
\left|M_+ +1\right|+1+2k_+
+ (\text{subleading})
\quad \left(M_+\in 2\mathbb{Z}+1/2\right)\,,
}
\label{eq: SU(2) mBMN analytic spectrum}
}
where $k_\pm=0,1,2...$ denote the energy levels.


\section{Discussion}
\label{sec:discussion}

In this paper, we have revisited the BMN model and studied its strong coupling limit at finite $N$.
This regime leads to an effective Hamiltonian where $N$ particles move freely in an harmonic trap, up to very short range interactions that lead to tightly bound states. If this picture (described in detail in section \ref{sec:comparison SUGRA}) is correct, then the spectrum exactly matches the spectrum of perturbative supergravity on the pp-wave background. The possibility of having such a weakly-coupled gravitational description at finite $N$ is surprising in view of other examples of the AdS/CFT correspondence, where one needs to take the large $N$ limit to make gravity weakly-coupled. Our proposal is in line with the strong form of the Matrix Theory conjecture by Susskind \cite{Susskind:1994vu}. However, as compared to the AdS/CFT correspondence, this conjecture is on less firm footing and it is therefore important to further examine its validity. In this context, it is worth noting that the recent paper \cite{Herderschee:2023pza} found that the BFSS MQM with finite $N$ correctly reproduces the three graviton coupling.

For the future, it is important to explore the BMN duality in different regimes (see figure \ref{fig: BMN phase diagram}) and understand the relation between the regime studied in this paper and the regime relevant for the standard AdS/CFT correspondence (i.e. the 't Hooft limit):

\begin{figure}[t]
    \centering 
    \includegraphics[width=14cm]{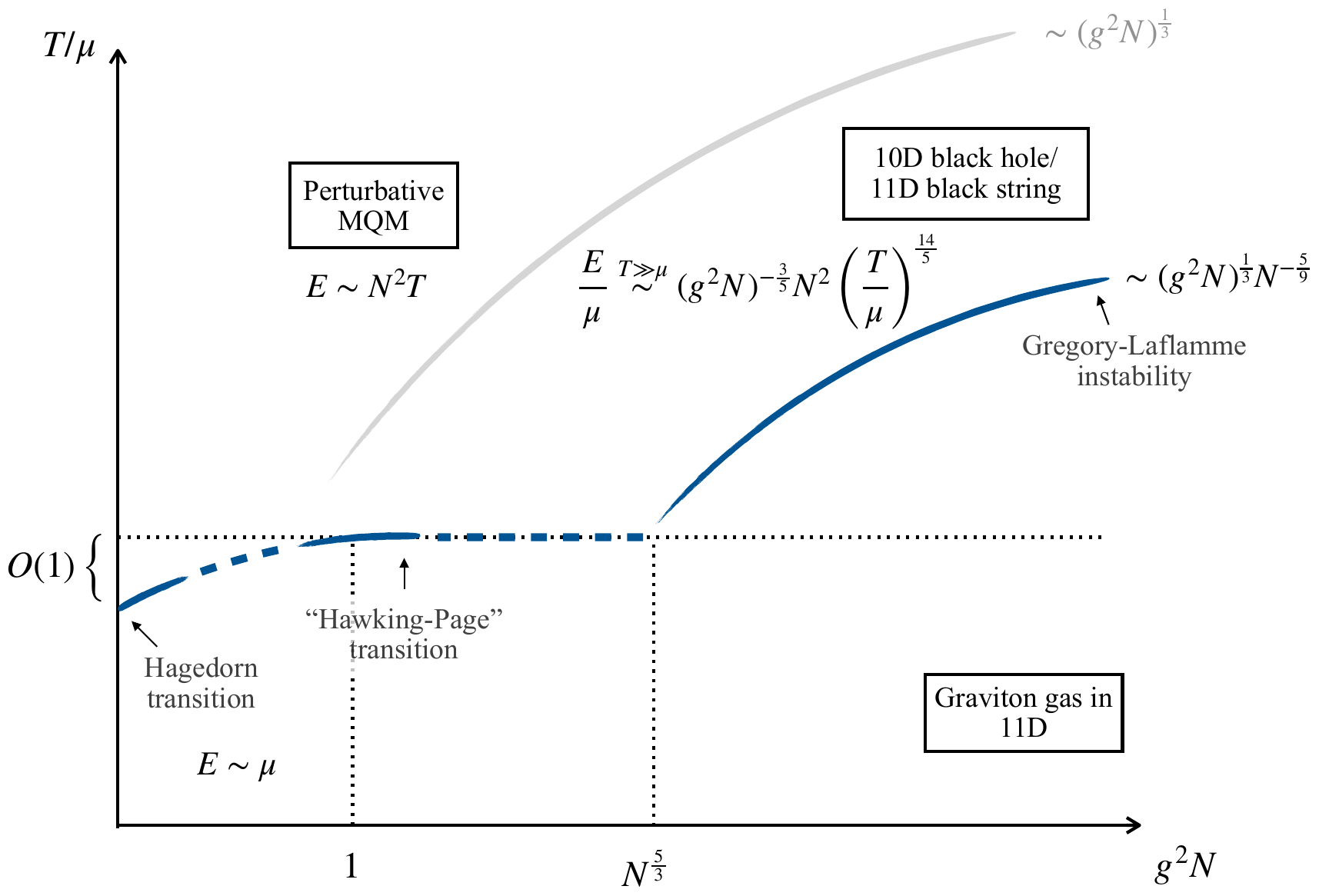}
    \caption{A sketch of the BMN phase diagram at large $N$. See Section \ref{subsec: BMN review} for exposition.
    }
    \label{fig: BMN phase diagram}
\end{figure}

\begin{itemize}

\item At finite $N$ and $g\to \infty$, it is urgent to confirm the bound state picture described in  section \ref{sec:comparison SUGRA}. This should be accessible to numerical methods, at least for $N=2$. In addition, it would be very instructive to clarify the fate of the non-singlet states in this regime (see section \ref{sec:comparison SUGRA}).

\item  The asymptotic expansion in $1/g$ at finite $N$ corresponds to the perturbative expansion in supergravity.
In fact, does the perturbative expansion of supergravity in DLCQ make sense? 
This is a crucial point so we discussed it below in detail.
If the supergravity expansion makes sense, does it  match order by order with the strong coupling expansion of MQM?
If this is the case, it could be used to determine the higher derivative corrections to supergravity in M-theory. Unfortunately, at the moment, these computations seem out of reach. Even the next correction to the effective Hamiltonian \eqref{eq: eff BMN Hamiltonian} is very challenging to compute using the methods of this paper. In addition, it is important to understand the boundary condition effects when two particles come close to each other. These can be the leading effects as exemplified in the toy model of section \ref{Sec: toy models}.

\item At finite $N$ and non-perturbatively in $g$, we could imagine a high energy scattering experiment between particles in the trap. On the gravitational side, if the energy is high enough we would form a black hole. How can this be described by a quantum system with finitely many d.o.f.?

\item At large $N$ and strong coupling, the backreaction of the gravitons on the PP-wave background cannot be neglected.\footnote{More precisely, writing $\bar{r} \equiv r \sqrt{\mu /R}$, the radial distance in units of the width of the gaussian ground state wavefunction, the backreaction produces a type IIA 10D geometry whose curvature (in units of the string length) and string coupling (dilaton) are small in the window \begin{equation} \frac{N^{1/7}}{g^{1/3}} \ll \bar{r} \ll \frac{N^{1/3}}{g^{1/3}}. \end{equation} This gives a region where type IIA supergravity should be a good description. When $\bar{r}$ is larger the curvature becomes large and a good description should be in terms of type IIA string theory. When $\bar{r}$ is smaller the dilaton becomes large, and a good description is given by 11D supergravity by uplifting the IIA geometry. For even smaller $\bar{r}$ the physics depends on the vacua and there might not be any geometrical description in general. These scalings can be  derived from the asymptotic expansion of the Lin--Maldacena  geometries. 
} This gives rise to the geometries constructed by Maldacena and Lin \cite{Lin:2005nh}. There is one such geometry for each vacuum of the BMN model. On the MQM side, we expect the interaction between the particles in the harmonic trap to become important in this regime. It would be fascinating to understand how these interactions create a collective state of the particles that can be described by the Lin--Maldacena geometries.\footnote{It is encouraging that the electrostatic equations governing the Lin--Maldacena geometries were already derived from a localization analysis of the BMN MQM \cite{Asano:2014vba, Asano:2014eca}.}
Notice that these geometries have a discrete spectrum of excitations that can be reliably computed within supergravity. Can we derive this spectrum from MQM?

\item For high enough temperature, the gravitational description is a black hole geometry \cite{Costa:2014wya}.
This black hole is similar to the asymptotically flat D0-brane black hole. The main difference is that the harmonic trap reflects back Hawking radiation and allows the black hole to be in thermodynamic equilibrium. 
Naively, increasing the temperature excites the off-diagonal modes and this generates an effective attraction	between the diagonal elements of the matrices. Can we develop an intuitive picture of the black hole state for particles in the harmonic trap?

\end{itemize}


\subsection{Validity of BMN duality at finite $N$}
In this paper, we presented evidence that the BMN MQM at finite $N$ and $g^2\to\infty$ reproduces the spectrum of DLCQ of 11D supergravity. Let us now discuss the validity and potential subtleties of this claim. 

\paragraph{Objections and counterarguments:}One immediate objection one may have is that 11D supergravity description should break down in the presence of a circle with a zero proper length and there should exist extra light degrees of freedom (such as wrapped branes) that are not included in DLCQ of 11D supergravity. However, if this were the case, one would already see them at the level of the spectrum at $g^2\to \infty$. In this sense, our results by themselves provide counterarguments to such objections. Another objection could be that DLCQ of 11D supergravity may be ill-defined; for (nonsupersymmetric) QFTs, Hellerman and Polchinski \cite{Hellerman:1997yu} demonstrated that individual Feynman diagrams diverge as one takes the direction of the compactification to be lightlike. However, they also showed that such divergences are often absent in supersymmetric QFTs. We can therefore hope that the same is true for 11D supergravity. 

That said, both of these counterarguments may seem rather superficial. Thus, below we examine and summarize the status of this duality in more detail, with the hope of clarifying some aspects. 

\paragraph{Examining the status of the duality:} Let us, for the moment, examine the validity of the following (seemingly) stronger claim:
\begin{itemize}
\item[1.] The BMN model at finite $N$ and {\it finite} $g^2$ is dual to DLCQ of M-theory on the PP-wave with $N$ units of momenta.
\end{itemize}
This is basically the PP-wave analog of the stronger version of the BFSS conjecture proposed by Susskind \cite{Susskind:1997cw}. In the case of the BFSS model, supporting arguments for this claim were given in \cite{Seiberg:1997ad} and \cite{Sen:1997we} based on a careful double scaling limit of spatial compactification and infinite boost. The same arguments can be applied to PP-wave backgrounds. However, one issue with these arguments is that there is no obvious independent definition of DLCQ of M-theory, making it difficult to test the claim as a {\it duality} between two different descriptions\footnote{In other words, in the absence of other definitions, the BFSS model {\it defines} DLCQ of M-theory rather than being dual to it.}. 

For the BFSS model, one can make concrete progress by compactifying one spatial direction and viewing it as the M-theory circle. This converts the BFSS duality into a duality\footnote{For details, see the arguments given in \cite{Seiberg:1997ad,Sen:1997we}. } between the so-called matrix string theory \cite{Motl:1997th,Banks:1996my,Dijkgraaf:1997vv} and DLCQ of type IIA superstring at finite $g_s$\footnote{The BFSS conjecture is recovered in the limit of $g_s\to\infty$.}. One can then take the small $g_s$ limit in which both sides have well-defined descriptions: the matrix string theory is believed to become a supersymmetric version of symmetric product orbifold of $\mathbb{R}^{8}$ (deformed by irrelevant operators) while the other side simply becomes DLCQ of perturbative IIA superstring. The explicit computation shows that DLCQ of perturbative IIA superstring is well-defined (i.e.~amplitudes do not diverge) \cite{Bilal:1998vq,Bilal:1998ys} and reproduces the expectations from the matrix string theory \cite{Arutyunov:1997gi}. In addition, DLCQ of type IIA superstring theory is known to be T-dual to the so-called nonrelativistic string theory \cite{Klebanov:2000pp,Danielsson:1996uw,Gomis:2000bd,Bergshoeff:2018yvt,Oling:2022fft}, which can be formulated directly in the conformal gauge without taking any limits. This guarantees that  standard rules of string perturbation theory apply\footnote{This is in contrast to the lightcone gauge, which is normally used to study DLCQ of type IIA superstring. There one needs to introduce extra operator insertions \cite{Mandelstam:1974hk} at the interaction vertices of strings in order to retain  Lorentz invariance, at every order in perturbation theory.} and there is no pathology at the level of perturbation theory. Furthermore, there is evidence that DLCQ is well-defined beyond perturbation theory \cite{Bilal:1998ys}. All in all, there are at least reasons to hope that DLCQ of IIA superstring in flat space is well-defined at finite $g_s$. One can then take the $g_s\to \infty$ limit and use it as a definition of DLCQ of M-theory in flat space.

For the BMN model, the same argument cannot be applied directly owing to the lack of flat transverse directions, although it is not far-fetched to think that DLCQ is still well-defined given that we are only modifying the structure of the transverse space. One possible path forward is to define it as a limit of M-theory on AdS backgrounds that have holographic duals (see e.g.~\cite{Kovacs:2013una, Dorey:2023jfw}). However details still need to be worked out.

Let us now come back to the conjecture discussed in this paper:
\begin{itemize}
\item[2.] The BMN model at finite $N$ and $g^2\to\infty$ is dual to DLCQ of 11D supergravity on the PP-wave with $N$ units of momenta.
\end{itemize}
This may seem like a special case of the conjecture above. However, strictly speaking, one cannot deduce the conjecture 2 from the conjecture 1 (even if it is correct). This is because the usual relation between M-theory at $\ell_P\to 0$ and 11D supergravity is established only when it is compactified on a manifold of size $L\gg \ell_P$. Here we do not have other supporting arguments apart from the match of the spectrum that we observed in this paper. Nevertheless, one can study a similar question in the type IIA limit. Namely, one can ask the following:
\begin{itemize}
\item[] Does the infinite tension limit of DLCQ of type IIA superstring coincide with DLCQ of type IIA supergravity?
\end{itemize}
As far as we know, this question has not been fully addressed in the literature although it is in principle tractable, in particular if one uses the T-dual description, namely nonrelativistic string\footnote{At the level of the spectrum, the infinite tension limit of DLCQ IIA superstring corresponds to a double-scaling limit of $\alpha^{\prime}$ and the radius of compactification of nonrelativistic string, in which winding modes have finite energy while momentum modes become infinitely heavy. In the DLCQ picture, this corresponds to keeping all the Kaluza-Klein modes and discarding winding strings, and is consistent with DLCQ of IIA supergravity.}.
\paragraph{Questions for the future:} As summarized above, at the moment, no definite conclusion can be made as to whether  the BMN duality at finite $N$ (and the stronger version of the BFSS conjecture by Susskind)  holds or not. However, there are a few concrete directions, apart from computing the $1/g^2$ correction of the BMN model mentioned earlier, in which one can hope to make progress and shed light on the status of the duality:
\begin{itemize}
\item Scrutinize the relation between matrix string theory and IIA superstring in the perturbative regime using the T-dual description, i.e.~nonrelativistic string.
\item Come up with a constructive definition of DLCQ of M-theory and IIA superstring on the PP-wave background through a limit of AdS backgrounds with holographic duals.
\item Study the infinite tension limit of DLCQ of IIA superstring and show that it coincides with DLCQ of IIA supergravity.
\end{itemize}
Any progress in these directions is welcome as it could illuminate these important dualities.

\subsection{Beyond M-theory?}

Can we use DLCQ in PP-wave backgrounds to construct  UV completions for other gravitational theories? Notice that  there is a PP-wave solution of Einstein-Maxwell theory in four spacetime dimensions, \emph{i.e.}  the low-energy theory of the real world.\footnote{This can be obtained starting from an extremal Reissner-Nordström black hole, taking the near-horizon limit, which is $AdS_2\times S^2$ and then taking a Penrose limit near a null geodesic going around an equator of the $S^2$.}
One can certainly study the spectrum of photons and gravitons on top of this background. This will again be a set of harmonic oscillators.
However, it is unclear if we can study interactions perturbatively in DLCQ since this theory is not supersymmetric (see discussion above).

Even if we focus on 11D supergravity on the PP-wave background \eqref{PW}, can we define other quantum mechanical systems that are also UV completions? 
For example, do we really need the off-diagonal d.o.f. of MQM? It seems that we just a need a quantum system where the diagonal d.o.f. have the appropriate interactions when $g\to \infty$. What are the minimal ingredients needed for the emergence of a gravitational description?





\subsection{PP-waves for UV complete QFTs}
It would also be interesting to explore whether one can use DLCQ of PP-waves to study non-perturbative properties of higher-dimensional QFTs. 

DLCQ of QFTs in {\it flat space} has been studied extensively for theories in 1+1 dimensions. In particular, this was successfully applied to two-dimensional QCDs and the nonperturbative spectrum has been computed through the numerical diagonalization of the lightcone Hamiltonian (see e.g.~\cite{Bhanot:1993xp,Dempsey:2021xpf}). However one cannot simply extend this approach to higher dimensions because of the existence of non-compact flat transverse directions. One idea which potentially overcomes this problem is to replace DLCQ of flat space with DLCQ of the PP-wave background.

In this context, the general goal is to study Renormalization Group (RG) flows that start in a UV CFT deformed by relevant operators. The first step in this direction would be to analyze in detail the properties of DLCQ of CFTs in the PP-wave background.
As discussed in \cite{Brecher:2000pa,Ishizeki:2008dw, Dorey:2023jfw}, the Weyl transformation maps uncompactified PP-wave backgrounds to flat Minkowski space and  DLCQ of PP-waves (in even spacetime dimensions) to a limit of $R_t \times S^{d-1}/\mathbb{Z}_L$; also known as lens spaces\footnote{$R_t \times$ (lens space) is one of a few examples of homogeneous (locally) conformally flat manifolds \cite{Alekseevsky:2009zz} (other examples include $S^1 \times R^{d-1}$, $RP^{d}$ and Euclidean AdS). As the conformal group acts transitively on such manifolds, the spacetime dependence of one-point functions of local operators is fully fixed by the symmetry (i.e.~no conformal cross ratios). Thus studying CFTs on such manifolds is potentially interesting on its own. We thank Nadav Drukker for pointing this out.}. Understanding the details of the map and using it to study general CFTs on PP-waves is an important future direction.

Note that the problems pointed out by Hellerman and Polchinski on the DLCQ limit of individual perturbative Feynman diagrams in non-supersymmetric QFTs do not necessarily preclude the existence of DLCQ at the non-perturbative level, as already mentioned in their paper \cite{Hellerman:1997yu}. To some extent, this problem is reminiscent of ``infrared divergence'' in interacting massless scalar fields in de Sitter. Also there, individual diagrams are logarithmically divergent at late time due to strongly-coupled long wave-length modes, invalidating the application of na\"{i}ve perturbation theory. Fortunately, a resolution is known in that context; there exists an improved perturbation theory based on the stochastic approach \cite{Starobinsky:1986fx,Starobinsky:1994bd,Gorbenko:2019rza}, which treats short modes perturbatively while solving the dynamics of long modes at the non-perturbative level\footnote{In spirit, this is somewhat similar to what have been attempted in the literature on DLCQ of QCD, in which one first solves a classical non-linear equation of motion of zero modes, plug the solution back into the Lagrangian and perform perturbation theory, although the regime of validity of this approximation is not clear.}.  In the light of recent developments on QFTs in de Sitter, it might be interesting to revisit the issue of the DLCQ limit of field theory. See also \cite{Fitzpatrick:2018ttk,Fitzpatrick:2018xlz} for recent discussions on non-perturbative treatment of zero modes.

\section*{Acknowledgements}

We thank T.~Anous, D.~Berenstein, G.~Cuomo, N.~Drukker, S.~Hartnoll, B.~Henning, J.~Hoppe, J.~Maldacena, J.~Matos, J.~Vilas Boas, A.~Wallberg, X.~Yin and participants of the CERH-TH Workshop on Matrix Quantum Mechanics for M-theory Revisited.

This work was supported by the Simons Foundation grant 488649 (Simons Collaboration on the Nonperturbative Bootstrap) and by the Swiss National Science Foundation through the project
200020\_197160 and through the National Centre of Competence in Research SwissMAP.

\appendix


\section{Calculational details for the path integral approach}
\label{App: Calculation details}
In this appendix we explain how to compute the effective Hamiltonian for the toy model \eqref{eq: toy model with general potential} using the path integral formalism. 
\subsection{The Gelfand Yaglom method}

The idea is to use the path integral formulation of quantum mechanics and naïvely perform the Gaussian integration over the fast variables $y$. In practice it is convenient to put the system at temperature $1/\beta$ and compute its partition function given by the Euclidean path integral over periodic paths of period $\beta$. 
\begin{equation}
    Z(\beta) = \int \mathcal{D}x \mathcal{D}y e^{-S_E[x,y]} \equiv \int \mathcal{D}x  e^{-S_{E,\text{eff}}[x]},
\end{equation}
where the second equality essentially defines what we mean by the effective action. The main advantage of the Euclidean formalism is that the Gaussian integral over $y$ is given by a functional determinant that can be computed with the Gelfand-Yaglom theorem
\begin{equation}
    \operatorname{det} \left( -\partial_\tau^2+W(\tau) \right) = y_1(\beta/2) +\Dot{y}_2(\beta/2) -2,
\end{equation}
where 
\begin{equation}
 \Ddot{y}_{1,2}(\tau) = W(\tau) y_{1,2}(\tau), \qquad \begin{pmatrix} y_1 & y_2 \\ \Dot{y}_1 & \Dot{y}_2 \end{pmatrix}(-\beta/2) = \begin{pmatrix} 1 & 0 \\ 0 & 1 \end{pmatrix}, 
 \end{equation}
i.e. $y_1$ and $y_2$ solve the Euclidean classical equations of motion with appropriate boundary conditions. We will use this approach for the model \eqref{eq: toy model with general potential} for which the Euclidean action reads 
\begin{equation}
S_E[x,y] = \frac{1}{2} \int_{-\beta/2}^{\beta/2} d\tau  y(\tau) \left( -\partial_\tau^2 + g^2 \omega^2(x)  \right) y(\tau) + \frac{1}{2} \int_{-\beta/2}^{\beta/2} d\tau \left( \dot{x}^2 +2 V(x)-g\omega(x) \right),
\end{equation}
which identifies the potential $W(\tau) = g^2 \omega^2(x(\tau))$. Therefore using Gelfand-Yaglom theorem we get 
\be S_{E,\text{eff}} = \frac{1}{2} \int_{-\beta/2}^{\beta/2} d\tau \left( \dot{x}^2 +2V(x)-g\omega(x) \right) + \frac{1}{2} \operatorname{log} \left( y_1(\beta/2)+ \dot{y}_2(\beta/2) -2 \right). \ee 

To evaluate the second term we need the solution of $\Ddot{y}(\tau) = g^2 \omega^2(x(\tau)) y(\tau)$ which cannot be written for generic paths $x(\tau)$. However we can use a WKB-like approximation (thinking of $g$ as $1/\hbar$) to solve it perturbatively. The idea is to change variables 
\begin{equation}
\label{eq:lagrangian_eom}
 y=e^S \implies \frac{\Ddot{y}}{y}=\Ddot{S}+ \Dot{S}^2 = g^2 \omega^2(x), 
\end{equation}
and solve the equations of motion perturbatively in $g$. To solve \eqref{eq:lagrangian_eom} the next step is to ansatz a perturbative expansion for $S(\tau)$. The simplest possibility is 
\be S(\tau) = g S^{(1)}(\tau) + S^{(0)}(\tau) + \frac{1}{g} S^{(-1)}(\tau) + \mathcal{O}(1/g^2). \ee Collecting powers of $g$ in the equation of motion we get 
\be (\dot{S}^{(1)})^2 = \omega^2(x), \qquad 2 \dot{S}^{(0)} \dot{S}^{(1)} +\Ddot{S}^{(1)}=0, \qquad \Ddot{S}^{(0)} + (\Dot{S}^{(0)})^2 + 2 \Dot{S}^{(-1)} \Dot{S}^{(1)}  = 0. \ee 
Each order is a first order ODE if we use the solution given by the previous order, therefore we can write the result in a closed form. It reads 
\begin{equation}
\begin{gathered}
S^{(1)} = \pm \int_{-\beta/2}^\tau d \tau' \omega(x(\tau')), \quad S^{(0)} = -\frac{1}{2} \int_{-\beta/2}^\tau d \tau' \frac{\Ddot{S}^{(1)}(\tau')}{\Dot{S}^{(1)}(\tau')},\\ S^{(-1)} = -\frac{1}{2} \int_{-\beta/2}^\tau d \tau' \frac{\Ddot{S}^{(0)}(\tau')+(\Dot{S}^{(0)})^2(\tau')}{\Dot{S}^{(1)}(\tau')}. 
\end{gathered}
\end{equation}
Defining $S^{(1)}$ to be the solution above with the "+" sign, the two independent general solutions to the equations of motion can therefore be written
\be y_{(+)}(\tau) \approx  e^{gS^{(1)} + S^{(0)} + \frac{1}{g} S^{(-1)}}, \qquad y_{(-)}(\tau) \approx e^{-gS^{(1)} + S^{(0)} - \frac{1}{g} S^{(-1)}}, \ee 
where the second solution follows from the first with the replacement $S^{(1)} \to -S^{(1)}$.
We can therefore write 
\be y_1 = a_1 y_{(+)} + a_2 y_{(-)}, \qquad y_2 = b_1 y_{(+)} + b_2 y_{(-)}, \ee 
where $a_1,a_2,b_1,b_2$ are completely fixed by the initial conditions.
Namely,
\begin{equation}
    \begin{split}
    a_1&=\frac{1}{2}+ \frac{\omega'(x(-\beta/2)) \dot{x}(-\beta/2)}{4g \omega(x(-\beta/2))^2} + \mathcal{O}(1/g^2), \qquad a_2\ =\frac{1}{2}- \frac{\omega'(x(-\beta/2)) \dot{x}(-\beta/2)}{4g \omega(x(-\beta/2))^2} + \mathcal{O}(1/g^2), \\
    b_1&=\frac{1}{2 g \omega(x(-\beta/2))}+O(1/g^2) =-b_2\,.
    \end{split}
\end{equation}
We can then compute 
\begin{equation}
    \begin{split}
    \operatorname{log} \left( y_1(\beta/2)+ \dot{y}_2(\beta/2) -2 \right) &=  \int_{-\beta/2}^{\beta/2} d\tau \left[ g \omega(x) - \frac{1}{2} \frac{ \omega'(x)}{\omega(x)} \dot{x} + \frac{1}{8g} \left( \frac{ \omega'(x)^2 \dot{x}^2}{ \omega(x)^3} \right) \right] \\ &+  \mathcal{O}(1/g^2).
    \end{split}
\end{equation} 
The second term is a total derivative giving 
\be 
\operatorname{log} \frac{\omega(x(\beta/2))}{\omega(x(-\beta/2))} = 0\,. 
\ee 
that vanishes because the Euclidean path integral is over paths with periodic boundary conditions. Expanding in $1/g$ the final result is 
\be S_{E,\text{eff}} =  \int_{-\beta/2}^{\beta/2} d\tau \left[ \frac{1}{2}\dot{x}^2 +V(x)  + \frac{1}{16g} \left( \frac{ \omega'(x)^2 \dot{x}^2}{ \omega(x)^3} \right)+ O(1/g^2) \right]\,. \ee 
Wick rotating to real time and performing the Legendre transform the effective Hamiltonian is 
\be 
H_{\text{eff}} = \frac{1}{2} p_x^2+ V(x) - \frac{1}{16g} \left( \frac{ \omega'(x)^2 p_x^2}{ \omega(x)^3} \right) \,.
\ee 
Note that the ordering of the last term can not be computed with this approach and was chosen arbitrarily. Plugging in \eqref{eq:omega and V} we get 
\be 
H_{\text{eff}} = \frac{1}{2} \left( p_x^2+ x^2-1 \right) - \frac{1}{16g} \frac{  p_x^2}{ x^3}  \,,
\ee 
which is different from \eqref{eq: final H_eff of toy model} at both $O(1)$ and $O(1/g)$. In addition, it is not clear how the operators should be ordered.

\subsection{Fast and slow modes}
Here we give details for the (correct!) calculation of the effective action using the path integral approach. First the propagators of the fast variables are computed as
\begin{equation}
\begin{split}
\langle x_f(\tau_1) x_f(\tau_2) \rangle &=
\sum_{|n|>\Lambda}\sum_{|m|>\Lambda} e^{2\pi i/\beta (n\tau_1+m\tau_2)}
\langle a_n a_m \rangle \\
&=
\sum_{|n|>\Lambda}\sum_{|m|>\Lambda} e^{2\pi i/\beta (n\tau_1+m\tau_2)}
\delta_{n+m} \frac{\beta}{\left(2\pi  n\right)^2} \\
&=
\sum_{|n|>\Lambda}  e^{2\pi i n( \tau_1-\tau_2)/\beta }
\frac{\beta}{\left(2\pi  n\right)^2}.
\end{split}
\end{equation}
In particular,
\begin{align}
\langle \left[x_f(\tau)\right]^2  \rangle  
&=
\sum_{|n|>\Lambda}  
\frac{\beta}{\left(2\pi  n\right)^2}
= 2 \int_\Lambda^\infty  dn \frac{\beta}{\left(2\pi  n\right)^2} +O(1/\Lambda^2)
 =   \frac{\beta}{2\pi^2  \Lambda}   +O(1/\Lambda^2).
\end{align}
The propagators of the $y$ variables were given in \eqref{eq:y propagator}.
We then need to compute the terms $\langle S_{\rm fast}^{\rm int}\rangle$ and $\langle S_{\rm fast}^{\rm int}\rangle^2$ appearing in the expansion in cumulants. We have
\begin{equation}
\begin{split}
 \langle S_{\rm fast}^{\rm int} \rangle &=\frac{1}{2} \int_{-\beta/2}^{\beta/2} d\tau \left[      g^2   [\omega'(x_s) ]^2 \langle x_f^2 y^2\rangle    +\dots \right]\\
 &= \int_{-\beta/2}^{\beta/2} d\tau \left[      \frac{q}{8\pi^2 }    \left(\frac{\omega'(x_s)}{\omega(x_s)} \right)^2    +\dots \right]
\label{eq: <S_fast^int>}
\end{split}
\end{equation}
Notice that this is a large correction because $q\gg 1$. 
Then, defining $\tilde{y} =y/\sqrt{g\omega(x_s)}$, we can write
\begin{align} 
   S_{\rm fast}^{\rm int} &=\frac{1}{2} \int_{-\beta/2}^{\beta/2} d\tau 
   g   \omega'
   \left[   ( 2  \tilde{y}^2-1)  x_f    +    \frac{\omega' }{\omega }  \tilde{y}^2    x_f^2 +\dots \right]  .
\end{align}
 Hence we compute
\begin{equation}
\begin{split}
 \langle (S_{\rm fast}^{\rm int} )^2\rangle 
 &=
\frac{1}{4} \int_{-\beta/2}^{\beta/2} d\tau_1 d\tau_2
   (g   \omega')_1(g   \omega')_2 
\\
&\left\<
\left[   ( 2  \tilde{y}^2-1)  x_f    +    \frac{\omega' }{\omega }  \tilde{y}^2    x_f^2 +\dots \right]_1 
\left[   ( 2  \tilde{y}^2-1)  x_f  
   +    \frac{\omega' }{\omega }  \tilde{y}^2    x_f^2 +\dots \right]_2 \right\>
\\
&=
\frac{1}{4} \int_{-\beta/2}^{\beta/2} d\tau_1 d\tau_2
   (g   \omega')_1(g   \omega')_2 
\\
&\left[ \left\<
   \left[ ( 2  \tilde{y}^2-1)  x_f     \right]_1 
   \left[   ( 2  \tilde{y}^2-1)  x_f     \right]_2 \right\>
+
\left\< 
\left[     \frac{\omega' }{\omega }  \tilde{y}^2    x_f^2  \right]_1 
   \left[      \frac{\omega' }{\omega }  \tilde{y}^2    x_f^2   \right]_2 \right\> 
+\dots
\right],
\end{split}
\end{equation}
where $[f]_i \equiv f(\tau_i)$.  Defining the new variable $u=g\omega(x_s)(\tau_1-\tau_2)$ we get
\begin{equation}
\begin{split}
     \langle (S_{\rm fast}^{\rm int} )^2\rangle &= \langle S_{\rm fast}^{\rm int} \rangle^2+
\frac{1}{4} \int_{-\beta/2}^{\beta/2} d\tau
   \left[      \frac{\omega' }{\omega }    \right]^2 g\omega  \\
   & \int du e^{-2|u|} \left[ 2 \langle x_f(0) x_f(u/(g\omega) ) \rangle +
  \frac{1}{2}  \langle x_f^2(0) x_f^2(u/(g\omega) ) \rangle
   \left[     \frac{\omega' }{\omega }   \right]^2   +\dots
   \right].
\end{split}
\end{equation}
The $x_f$ propagators are evaluated as 
\begin{equation}
\begin{split}
\langle x_f(0) x_f(u/(g\omega) ) \rangle &=   \sum_{|n|>\Lambda}  e^{2\pi i n  u/(q \Lambda) } \frac{q \Lambda}{g \omega \left(2\pi  n\right)^2}, \\
 \langle x_f^2(0) x_f^2(u/(g\omega) ) \rangle &= 2 \sum_{|n|>\Lambda} \sum_{|m|>\Lambda}   e^{2\pi i (n+m)  u/(q \Lambda) }
\frac{q \Lambda}{g\omega \left(2\pi  n\right)^2}\frac{q \Lambda}{g \omega \left(2\pi  m\right)^2}.
\end{split}
\end{equation}
Integrating over $u$ we then get
 \begin{equation}
\begin{split}
      \langle (S_{\rm fast}^{\rm int})^2 \rangle &= \langle S_{\rm fast}^{\rm int} \rangle^2 + \frac{1}{2} \int_{-\beta/2}^{\beta/2} d\tau
   \left[      \frac{\omega' }{\omega }    \right]^2 \left( \sum_{|n|>\Lambda} 
   \frac{q^2\Lambda^2}{q^2\Lambda^2+\pi^2n^2} \frac{q \Lambda}{\left(2\pi  n\right)^2} \right) \\ 
&+\frac{1}{4}\int_{-\beta/2}^{\beta/2} d\tau
   \left[      \frac{\omega' }{\omega }    \right]^2 \left( \frac{q^2\Lambda^2}{q^2\Lambda^2+\pi^2(n+m)^2}
\frac{q \Lambda}{g\omega \left(2\pi  n\right)^2}\frac{q \Lambda}{\left(2\pi  m\right)^2}
   \left[     \frac{\omega' }{\omega }   \right]^2   +\dots
   \right).
\end{split}
 \end{equation}
The second term is of order $1/\Lambda\sim 1/g $ (recall that $q$ is fixed).
The first term can be approximated by an integral up to errors that vanish as $g \to \infty $,
\begin{align} 
\spl{
 \langle (S_{\rm fast}^{\rm int} )^2\rangle 
   &= \langle S_{\rm fast}^{\rm int} \rangle^2+
  \int_{-\beta/2}^{\beta/2} d\tau
   \left[      \frac{\omega' }{\omega }    \right]^2         \int_1^\infty dx
   \frac{q^2}{q^2+\pi^2x^2}
\frac{q }{\left(2\pi  x\right)^2}    +\dots\\
 &= \langle S_{\rm fast}^{\rm int} \rangle^2+
  \int_{-\beta/2}^{\beta/2} d\tau
   \left[      \frac{\omega' }{\omega }    \right]^2         
   \left(\frac{q}{4\pi^2}
   -\frac{1}{8}+O(1/q)
   \right)+\dots \, ,
}
\end{align}
where in the cumulant expansion \eqref{eq: cumulant expansion} the first term from the $\t$ integral above cancels exactly with the leading term in $\langle S_{\rm fast}^{\rm int} \rangle$ given in \eqref{eq: <S_fast^int>}.


\section{Numerics for the toy model}
\label{App: numerics}
In this appendix we describe our numerical implementation to diagonalize the Hamiltonian \eqref{eq: toy model with general potential}.
\subsection{Numerical Setup for the Chebyshev method}

We use coordinates for which the system lives in $[-1,1] \times [-1,1]$. Let us first describe the one dimensional case with only one interval, discretized on a Chebyshev grid with $M$ points 
\begin{equation}
    \theta_i = - \operatorname{cos} \frac{i \pi}{M-1}, \qquad i=0,1,...,M-1.
\end{equation}
Using $\mathbb{Z}_2$ symmetry it is sufficient to map to the portion of physical space with $x>0$, and therefore a possible map is 
\be 
x(\theta) = \operatorname{tan} \frac{\pi}{4}(\theta+1).
\ee
Any function on that space can be written in a basis of interpolating Lagrange polynomials 
\be \psi(\theta) \approx \sum_{j=0}^{M-1} v_j p_j(\theta), \qquad p_j(\theta) \equiv \prod_{k \neq j} \frac{\theta-\theta_k}{\theta_j-\theta_k}, \qquad p_j(\theta_i)= \delta_{ij}.  \ee 
The main advantage is that derivatives are given by matrix multiplication making use of information on the full space (summation over Greek indices is implied)
\be \psi'(\theta_k) \approx \sum_{j=0}^{M-1} v_j p_j'(\theta_k) = D_{k\mu} v_\mu, \qquad D_{kj} \equiv p'_j(\theta_k). \ee 
The Hamiltonian then reads
\be (\partial^2_x + U(x_k)) \psi(x_k)  =  \sum_{j,l} \frac{ d \theta}{dx}(\theta_k) D_{k j} \frac{d\theta}{dx}(\theta_j) D_{jl} \psi(\theta_l) + U(x_k)  \psi(\theta_k), \ee
where $x_k \equiv x(\theta_k)$, and therefore Schrödinger's problem is reduced to the diagonalization of the $M \times M$ matrix $H$ whose elements are 
\be H_{kl} = \sum_j \frac{ d \theta}{dx}(\theta_k) D_{kj} \frac{d\theta}{dx}(\theta_j) D_{jl} + \delta_{kl} U(x_l). \ee 
To specify the problem we also have to impose boundary conditions. Even wavefunctions are specified by Neumann boundary conditions at $\theta_0$ (mapped to $x=0$) while odd wavefunctions are specified by Dirichlet boundary conditions. Both cases have Dirichlet boundary conditions at $\theta_{M-1}$ (mapped to $x=\infty)$.

Dirichlet boundary conditions are automatically implemented by setting the first and last rows/columns of $H$ to zero. On the other hand for Neumann boundary conditions we use
\be &D_{0\mu} v_\mu = 0 \implies v_0 = -\frac{1}{D_{00}} \sum_{i \neq 0} D_{0i} v_i \implies  D_{i \mu}v_\mu = \sum_{j \neq 0}( D_{ij} - \frac{1}{D_{00}} D_{i0}D_{0j}) v_j, \ee 
and therefore we have to shift the derivative matrix elements as prescribed by the last equality 
\be D^{(N)}_{ij} = D_{ij} - \frac{1}{D_{00}} D_{i0}D_{0j}. \ee

The two dimensional case is implemented in a similar way by taking Kronecker products of all matrices. One difference is that we expect the wavefunction to be squeezed in the $y$ direction, which we can turn in our advantage by compactifying the coordinates as 
\be y(\theta) = \frac{1}{ \sqrt{g}} \operatorname{tan} \frac{\pi}{4}(\theta+1). \ee
The matrix that we have to diagonalize will be a $M^2 \times M^2$ matrix.


\section{Technical details for finding the BMN $Q_\text{eff}$ and $H_\text{eff}$}
\label{App: BMN Heff technicalalities}
This appendix summarises various technical details for finding the effective supercharges and the effective Hamiltonian of the BMN model. In this appendix we set $\m=1$.

\subsection{Change of coordinates}
\label{subsec: change of coordinates BMN}
Recall that the BMN model has an $SU(N)$ symmetry. As a first step, it is convenient to make a change of variables to separate out the $SU(N)$ invariant degrees of freedom in the matrices. Following \cite{Lin:2014wka} we parametrise the bosonic matrices as\footnote{The matrices $X^I$ live in the algebra of $U(N)$, the symmetry of the Hamiltonian is $SU(N)$, and the coordinates $U$ are elements of the group $U(N)/U(1)^N = SU(N)/U(1)^{N-1}$}  
\al{\spl{
X^I 
&=
U^{-1}
\left(\sum_{a=1}^N r_a^I E_a +  \sum_{a\neq b}^N q_{ab}^I T_{ab} \right) U
\,,
}
\label{eq in app: change of coordinates by U for X BMN}
}
where $E_a$ and $T_{ab}$ are $N\times N$ matrices with all matrix elements being zero except $(E_a)_{aa}=(T_{ab})_{ab}=1$. The unitary matrix $U$ is chosen such that the off-diagonal elements $q_{a b}^i$ are complex numbers satisfying the relation
\be \sum_{I =1}^9 q_{a b}^I r^I_{ab} = 0 \qquad a,b =1,\ldots,N,\, a \neq b\,,
\label{eq in app: orthogonality condition}
\ee
where $r_{ab}^I\equiv r_a^I-r_b^I$. The fast variables have a typical scale of order $g^{-1/2}$, thus it is convenient to define
\al{
q_{ab}\equiv  g^{-1/2} y_{ab}\,,
\label{eq: q and y}
}
such that $y_{ab}$ are of order 1. After the change of variables the integral measure becomes
\al{\spl{
\int \prod_{I=1}^9 \prod_{a,b=1,N} dX_{ab}^I
&=
\int \prod_{a=1}^N d^9\vec{r}_a\, \int [dU] \int \prod_{a\neq b} \frac{d^9\vec{y}_{ab}}{g^{4}} \d(\hat{r}_{ab}\cdot \vec{y}_{ab}) \left(\Delta(r) + \mathcal{O}\left(\frac{1}{\sqrt{g}}\right)\right) \\ &
\equiv
\int \prod_{a=1}^N d^9\vec{r}_a\,
\int [dU] \int [dy]
\,,
}}
where $\D(r)\equiv \prod_{a<b} |r_{ab}|^2$ with $|r_{a b}| \equiv \sqrt{\sum_I r_{a b}^I r_{a b}^I}$ is the Vandermonde determinant\footnote{There are subleading corrections to the $r$ integral  measure, but for the purpose of this paper it is enough to consider only the leading order.}, $\int [dU]$ is the integral over the $U(N)/U(1)^N$ group (which will always factor out for $SU(N)$ invariant quantities) and the $\d$ function is included to impose the constraint \eqref{eq in app: orthogonality condition}.

Under the same transformation the fermionic matrices become
\al{\spl{
\hat{\Theta}_\alpha 
&=
U^{-1}
\left((\theta_a)_\a E_a + (\Theta_{ab})_\a T_{ab} \right) U\,,
}
\label{eq in app: change of coordinates by U for Theta BMN}
}
with the anti-commutation relations
\al{
\{(\theta_a)_\a,(\theta_b)_\b\} = \d_{ab}\d_{\a\b}\,,
\qquad
\left\{(\Theta_{ab})_\a , (\Theta_{cd})_\b \right\} = \d_{ad}\d_{bc}\d_{\a\b}\,.
}
Note that the new fermions $\theta$ and $\Theta$ are $SU(N)$ invariant but $U$-dependent. See Section \ref{subsec: non-singlet case} for details.

The momentum matrix $P^I$ in the new coordinates \eqref{eq: change of coordinates by U for X BMN} reads, up to $O(1)$ in large $g$ expansion (here $\DD$ and $J$ do not act on $U$, see Section \ref{subsubsec: change of coordinate at any order} for the general expansion),
\al{
P^I = U^{-1}\left(
-i \DD_a^I E_a - i J_{ab}^I  T_{ab}
\right) U 
+O\left(g^{-1/2}\right)
\,,
\label{eq: change of coordinates by U for P BMN}
}
where $\DD_a^I$ is the interior $r$-derivative
\al{
\mathcal{D}_a^I \equiv \frac{\partial}{\partial r_a^I} + \sum_{b \neq a} \biggl(\frac{y_{b a}^I \hat r_{b a}^J}{|r_{a b}|} \frac{\partial}{\partial y_{b a}^J} - \frac{y_{a b}^I \hat r_{a b}^J}{|r_{a b}|} \frac{\partial}{\partial y_{a b}^J} \biggr)\,,
\label{eq: interior r derivative}
}
satisfying $\DD_a^I(r_{cd}^J y_{cd}^J)=0$. The benefit of $\DD_a^I$ is that it can act on $r$ and $y$ variables as if they are independent regardless of the constraint \eqref{eq in app: orthogonality condition}. The off-diagonal terms are
\al{
\spl{
J_{b a}^I & \equiv 
\sqrt{g} \Pi_{a b}^{IJ} \frac{\partial}{\partial y_{a b}^J}
-\frac{\hat{r}_{a b}^I}{\left|r_{a b}\right|} \sum_{c \neq a, b}\left[y_{c a}^K \Pi_{c b}^{KJ} \frac{\partial}{\partial y_{c b}^J}-y_{b c}^K \Pi_{a c}^{KJ} \frac{\partial}{\partial y_{a c}^J}\right] 
\\
&\quad+
\frac{\hat{r}_{a b}^I}{\left|r_{a b}\right|} \left[U \frac{\partial}{\partial U}\right]_{b a}
+O\left(g^{-1/2}\right)\,.
}\label{eq: K_{ba}}}
Above we have defined $\hat{r}_{a b}^I \equiv r_{a b}^I/|r_{a b}|$ and a projection operator
\al{
\Pi_{ab}^{IJ} \equiv \d^{IJ} - \hat{r}_{ab}^I \hat{r}_{ab}^J\,,
}
which is used to construct the interior $y$-derivative $\Pi_{ab}^{IJ}\del/\del y_{ab}^J$ and it satisfies
\al{
\Pi_{ab}^{IJ} y_{ab}^J = y_{ab}^I\,,
\qquad
\Pi_{ab}^{IJ} r_{ab}^J = 0\,.
}

In the new coordinates \eqref{eq: change of coordinates by U for Theta BMN} the $SU(N)$ generators become
\al{
\tilde{G}_{ab} &\equiv (U G U^{-1})_{ab} = \left(U\frac{\del}{\del U}\right)_{ab} - \tilde{G}_{ab}^F\,,\\
\tilde{G}^F_{ab} &= 
\sum_{c \neq a, b} \frac12 [(\Theta_{a c})_\a, (\Theta_{c b})_\a]
+
(\theta_{ab})_\a (\Theta_{a b})_\a
\,,
\label{eq:fermion_gauge_generator}
}
with $\theta_{ab}\equiv\theta_a-\theta_b$. When acting $J_{ba}^I$ on the $SU(N)$ invariant wavefunctions we can make the replacement \cite{Lin:2014wka} 
\al{
\left(U\frac{\del}{\del U}\right)_{ab} \to \tilde{G}^F_{ab}\,.
\label{eq: replacement rule for SU(N) singlet}
}
In this way the $U$ matrices disappear in all computations that follow. 

\subsection{The momentum operator in the new coordinates at any order}
\label{subsubsec: change of coordinate at any order}
In this section, we partially follow and refine \cite{Lin:2014wka} to compute the momentum operator $P_{b a}^I$ in the $SU(N)$-rotated coordinates, which is given by
\al{
\spl{
& i \left(U P^I U^{-1}\right)_{ba}  = U_{b c} U^{-1}_{d a} \frac{\del}{\del X_{d c}^I} \\ & = U_{b c} U^{-1}_{d a} \left( \frac{\del r_e^J}{\del X_{d c}^I} \frac{\del}{\del r_e^J} + \frac{\del q_{e f}^J}{\del X_{d c}^I} \frac{\del}{\del q_{e f}^J} + \frac{\del U_{e f}}{\del X_{d c}^I} \frac{\del}{\del U_{e f}} \right) \Biggl|_{\sum_{K} q_{a' b'}^K r^K_{a'b'} = 0} \, . \label{eq:formula_change_of_coordinates}
}
}
In the following, we will leave sums over indices implicit (except when ambiguous), and expressions of the type $q_{ab}^I$, $\frac{1}{|r_{a b}|^2}$,... are implicitly considered absent when $a = b$. Starting from \eqref{eq: change of coordinates by U for X BMN}, we compute the differential
\al{
\spl{
U dX^I U^{-1} & = dr_a^I E_a + dq_{a b}^I T_{a b} + q_{a b}^I (dU U^{-1})_{b a} (E_a - E_b) \\ & + \Bigl( r_{a b}^I (dU U^{-1})_{a b} +  q_{a c}^I (dU U^{-1})_{c b} - (dU U^{-1})_{a c} q_{c b}^I \Bigr) T_{a b} \, . \label{eq:differential_dX}
}
}
To obtain this result we used $U d U^{-1} = -dU U^{-1} $ and the following commutation relations
\al{
[E_a,T_{c d}] = (\delta_{a c}- \delta_{a d})T_{c d} \,,
\qquad [T_{a b},T_{c d}] = T_{a d} \delta_{b c} - T_{c b} \delta_{a d} + (E_a - E_b) \delta_{b c} \delta_{a d}\,.
} 
It is then possible to isolate $dr_{a}^I$ and $dq_{a b}^I$ in terms of $dX^I$ and $dU U^{-1}$ by taking respectively diagonal and off-diagonal components of the above expression. Doing so, we find
\be dr_a^I = (U dX^I U^{-1})_{a a} - \sum_{b \neq a} \bigl( q_{a b}^I (dU U^{-1})_{b a} - q_{b a}^I (dU U^{-1})_{a b} \bigr) \, , \label{eq:differential_dr} \ee
\be dq_{a b}^I = (U dX^I U^{-1})_{a b} - r_{a b}^I (dU U^{-1})_{a b} + \sum_{c \neq a,b} \Bigl(  q_{a c}^I (dU U^{-1})_{c b} - (dU U^{-1})_{a c} q_{c b}^I \Bigr) \, , \label{eq:differential_dq} \ee
\be dU_{a b} = \sum_{c \neq a} (dU U^{-1})_{a c} U_{c b} \, . \label{eq:differential_dU} \ee
To finish the computation of \eqref{eq:formula_change_of_coordinates}, we need to find an expression of $dU U^{-1}$ in terms of $dX^I$. To obtain an equation for $dU U^{-1}$, we consider two projections of \eqref{eq:differential_dX}. Firstly, we project the $c,d$ component on $r_{c d}^I$ ($c,d$ are fixed, $I$ is summed), and use $r_{c d}^I dq_{c d}^I = -q_{c d}^I (dr_c^I - dr_d^I)$ which follows from $r_{c d}^I q_{c d}^I = 0$, obtaining
\al{
\spl{
r_{c d}^I [U dX^I U^{-1}]_{c d} & =  - q_{c d}^I (dr_c^I - dr_d^I)  \\ & + |r_{c d}|^2 (dU U^{-1})_{cd} + r_{c d}^I \left( q_{c b}^I (dU U^{-1})_{b d} - (dU U^{-1})_{c b} q_{b d}^I \right) \, . \label{eq:COC_dUU_1}
 }
}
Secondly, we project the difference of diagonal components $cc - dd$ on $q_{c d}^I$, obtaining
\al{
\spl{
q_{c d}^I \left( [U dX^I U^{-1}]_{c c} - [U dX^I U^{-1}]_{d d} \right) = q_{c d}^I (dr_c^I - dr_d^I) & + q_{c d}^I \Bigl( q_{c a}^I (dU U^{-1})_{a c} - (dU U^{-1})_{c a} q_{a c}^I \\ & - q_{d a}^I (dU U^{-1})_{a d} + (dU U^{-1})_{d a} q_{a d}^I \Bigr)
\label{eq:COC_dUU_2}
}
}
Summing these two expressions, we eliminate the $dr,dq$ dependence and obtain a recursive relation which we write as
\be (dU U^{-1})_{a b} = F_{ab} + K_{a b;c d} (dU U^{-1})_{c d} \label{eq: recursion relation for dUU^-1}\ee
where (remembering $q_{a b}^I = \frac{y^I_{a b}}{\sqrt{g}}$)
\be F_{a b} \equiv \frac{r_{a b}^I}{|r_{a b}|^2} (U dX^I U^{-1})_{a b} + \frac{1}{\sqrt{g}} \frac{y_{a b}^I}{|r_{a b}|^2} \Bigl( (U dX^I U^{-1})_{a a} - (U dX^I U^{-1})_{b b} \Bigr) \, , \ee
\be K_{a b; c d} \equiv - \frac{1}{\sqrt{g}} \frac{r_{a b}^I}{|r_{a b}|^2} (y_{a c}^I \delta_{d b} - \delta_{a c} y_{d b}^I) - \frac{1}{g} \frac{y_{a b}^I}{|r_{a b}|^2} (\delta_{a d} - \delta_{a c} - \delta_{b d} + \delta_{b c}) y_{d c}^I \, . \ee
The above equation takes the form $(dU U^{-1}) = F + K (dU U^{-1})$ where $dU U^{-1}$ and $F$ are considered as vectors with a double-index $(ab)$ and $K$ is a matrix with two double-indices $(ab)$,$(cd)$. The solution is simply the geometric series
\be (dU U^{-1})_{a b} = t_{a b; c d} F_{c d} \, , \label{eq:dUU_solution} \ee
where 
\be t_{ab ; cd} \equiv \sum_{n = 0}^\infty (K^n)_{ab;cd} = \delta_{a c} \delta_{b d} + K_{a b; cd} + K_{ab; g_1 h_1} K_{g_1 h_1; cd}  + ... \, . \label{eq: change of coordinates, definition of the matrix products} \ee

Plugging \eqref{eq:dUU_solution} in \eqref{eq:differential_dr}, \eqref{eq:differential_dq} and \eqref{eq:differential_dU}, we simplify\footnote{To obtain this manifestly interior expression, one needs to use $t_{ab; cd} = \delta_{ac} \delta_{b d} + K_{ab;ef} t_{ef;cd}$ which follows from the recursion relation \eqref{eq: recursion relation for dUU^-1}.} \eqref{eq:formula_change_of_coordinates} and obtain
\al{
\spl{
& i (U P^I U^{-1})_{b a} =  \ \sqrt{g} \Pi_{a b}^{I J} \frac{\del}{\del y_{a b}^{J}} - \frac{r_{a b}^I}{|r_{a b}|^2} (y_{ce}^J t_{ed;ab} - t_{ce;ab}y_{e d}^J) \Pi_{c d}^{J K} \frac{\del}{\del y_{cd}^K} \\ & \qquad \qquad \qquad + \frac{r_{a b}^I}{|r_{a b}|^2} t_{dc; ab} \left[ U \frac{\partial}{\partial U} \right]_{c d} - \frac{1}{\sqrt{g}} \frac{r_{a b}^I}{|r_{a b}|^2} y_{c d}^J t_{d c; ab} (\mathcal{D}_c^J - \mathcal{D}_d^J)  \\ & \qquad + \delta_{a b} \Biggl\{ \mathcal{D}_a^I - \frac{1}{\sqrt{g}} \left( \frac{y_{a h}^I}{|r_{a h}|^2} (y_{c e}^J t_{ed; ah} - t_{ce;ah} y_{ed}^J) - (a \leftrightarrow h) \right) \Pi_{cd}^{J K} \frac{\del}{\del y_{cd}^K}  \\ & \qquad + \frac{1}{\sqrt{g}} \left( \frac{y_{a h}^I}{|r_{a h}|^2} t_{dc;ah} - (a \leftrightarrow h) \right) \left[ U \frac{\partial}{\partial U} \right]_{c d}  - \frac{1}{g} \left( \frac{y_{a h}^I}{|r_{a h}|^2} t_{d c; a h} - (a \leftrightarrow h) \right) y_{c d}^J (\mathcal{D}_c^J - \mathcal{D}_d^J)\Biggr\} \, . \label{eq:solution_change_of_coord} 
}
}
where we defined $\left(\frac{\del}{\del U}\right)_{a b} \equiv \frac{\del}{\del U_{b a}}$ and the interior $r$-derivative $\mathcal{D}_a^I$ is defined in \eqref{eq: interior r derivative}. This operator $P_{b a}^I$ is manifestly interior, $P_{b a}^I f \bigl( r_{cd}^J y_{cd}^J \bigr) = 0$ and can in principle be computed to any order through the polynomial expansion of $t_{ab;cd}$. One can check that the result \eqref{eq: change of coordinates by U for P BMN} is now easily obtained from $t_{ab; cd} = \delta_{a c} \delta_{b d} - \frac{1}{\sqrt{g}} \frac{r_{a b}^I}{|r_{a b}|^2} (y_{a c}^I \delta_{d b} - \delta_{a c} y_{d b}^I) + \mathcal{O} (1/g)$.

\subsubsection{Computation of the Laplacian}
We now compute the kinetic part of the Hamiltonian  $H_{\mathrm{kin}} = \frac{1}{2} \mathrm{Tr}((P^I)^2) = -\frac{1}{2} \left(\frac{\del}{\del X^I}\right)_{c d} \left(\frac{\del}{\del X^I}\right)_{d c}$ up to order one. Let us write 
\be \left(i P^I\right)_{d c} = U^{-1}_{d b} U_{a c} ( D_{b a}^I + L_{b a}^I) \, ,  \ee 
where $D_{a b}^I$ are the $\frac{\del}{\del y}$ and $\frac{\del}{\del r}$ derivative terms of \eqref{eq:solution_change_of_coord}, whereas $L_{a b}^I$ are the $SU(N)$-variable derivative terms $\frac{\del}{\del U}$. With this split, the kinetic Hamiltonian reads
\be H_{\mathrm{kin}} = - \frac{1}{2} \Biggl(U^{-1}_{c e} U_{f d} (D_{e f}^I + L_{e f}^I)\Biggr) \Biggl(U^{-1}_{d b} U_{a c} (D_{b a}^I + L_{b a}^I)\Biggr)  \, .  \ee
Since we restrict ourselves to $SU(N)$ invariant wave functions, any $L$ that acts on the wave function must have its $\left(U\frac{\del}{\del U}\right)_{a b}$ replaced by $\tilde{G}^F_{a b}$, which was defined in \eqref{eq:fermion_gauge_generator}. 
More precisely, when $L_{b a}^I$ acts on the wave function, it is replaced by $\Tilde{L}_{b a}^I$ where
\al{
\spl{
\Tilde{L}_{b a}^I & \equiv \frac{\hat{r}_{a b}^I}{|r_{a b}|} \tilde{G}^F_{b a} - \frac{1}{\sqrt{g}} \frac{r_{a b}^I}{|r_{a b}|^2} \sum_{c \neq a,b} \left[ \frac{r_{c b}^J y_{c a}^J}{|r_{c b}|^2 }  \tilde{G}^F_{b c} - \frac{r_{a c}^J y_{b c}^J}{|r_{a c}|^2} \tilde{G}^F_{c a} \right] \\ &  + \frac{1}{\sqrt{g}} \delta_{a b} \sum_{c \neq a} \left[\frac{y_{a c}^I}{|r_{a c}|^2} \tilde{G}^F_{c a} - \frac{y_{c a}^I}{|r_{a c}|^2} \tilde{G}^F_{a c} \right] + O\left(\frac{1}{g}\right) \, . 
}
}
Noting also that $L_{e f}^I$ can act on $U^{-1}_{d b} U_{a c}$, the kinetic Hamiltonian is
\al{
\spl{
 H_{\mathrm{kin}} = - \frac{1}{2} \Biggl\{& D_{a b}^I D_{b a}^I + U^{-1}_{c e} U_{f d} L_{e f}^I \bigl( U^{-1}_{d b} U_{a c} \bigr) D_{b a}^I \\ & + D_{a b}^I \Tilde{L}_{b a}^I + \Tilde{L}_{a b}^I D_{b a}^I + \Tilde{L}_{a b}^I \Tilde{L}_{b a}^I \Biggr\}  \, , \label{eq:Hkin_starting_point} 
}
}
where in the notation $L_{ef}^I(...)$, the $L_{ef}^I$ derivatives only act on $(...)$.
In order to compute each of these terms, we use the convenient fact that $\Pi_{a b}^{I J} r_{a b}^J = 0$, and that off-diagonal and diagonal terms do not mix in the trace. This simplifies the products drastically, allowing to obtain
\al{
\spl{
H_{\rm kin} = & \ - \frac{g}{2} \sum_{a \neq b} \Pi_{a b}^{I J} \frac{\partial}{\partial y_{a b}^I} \frac{\partial }{\partial y_{b a}^J} - \frac{1}{2} \frac{1}{\Delta(r)} \mathcal{D}_a^I \biggl( \Delta(r) \mathcal{D}_a^I  \biggr)  + \sum_{a \neq b} \frac{2}{|r_{a b}|^2} y_{a b}^I \Pi_{a b}^{I J} \frac{\del}{\del y_{a b}^J} \\ & + \frac{1}{2} \sum_{a \neq b} \sum_{c\neq a,b} \sum_{d \neq a,b} \frac{1}{ |r_{a b}|^2} \left( y_{c a}^I \Pi_{c b}^{I J} \frac{\del}{\del y_{c b}^J} - y_{b c}^I \Pi_{a c}^{I J} \frac{\del}{\del y_{a c}^J} \right) \left( y_{d b}^K \Pi_{d a}^{K L} \frac{\del}{\del y_{d a}^L} - y_{a d}^K \Pi_{b d}^{K L} \frac{\del}{\del y_{b d}^L} \right) \\ & + \sum_{a \neq b \neq c} \frac{\hat{r}_{c b}^J \hat{r}_{a b}^I}{|r_{a b}| |r_{b c}|} \Pi_{a c}^{I K}  y_{c a}^J \frac{\del}{\del y_{c a}^K} + \frac{1}{2} \sum_{a \neq b} \frac{1}{|r_{a b}|^2} \tilde{G}^F_{a b} \tilde{G}^F_{b a}      \\ & - \sum_{a \neq b \neq c} \frac{1}{|r_{a b}|^2} \left( y_{c a}^I \Pi_{b c}^{I J} \frac{\del}{\del y_{c b}^J} - y_{b c}^I \Pi_{a c}^{I J} \frac{\del}{\del y_{a c}^J} \right) \tilde{G}^F_{a b} + \mathcal{O}\left(\frac{1}{\sqrt{g}}\right) \, .
}
\label{eq: BMN H_kin}
}
Above we have replaced $U\frac{\del}{\del U}$ by $\tilde{G}^F$.

\subsection{Expansion of the operators}
Using the results from previous subsections, we can now rewrite the Hamiltonian, the supercharge and the rotation generators in the new coordinates and expand them in large $g$.

\paragraph{Expansion of the Hamiltonian}
Recall the BMN Hamiltonian is
\al{
\spl{
H &= {\rm Tr} \left[\frac{1}{2} (P^I)^2 -\frac{g^2}{4}[X^I,X^J]^2 - \frac{g}{2}\hat{\Theta}^\top\gamma^I [X^I,\hat{\Theta}] \right] 
 \\
&+\frac{1}{2} \,{\rm Tr} \left[  \frac{1}{3^2} \sum_{i=1}^3 (X^i)^2 
+ \frac{1}{6^2}  \sum_{p=4}^9 (X^p)^2 
+i\frac{1}{4}\hat{\Theta}^\top\gamma^{123} \hat{\Theta}
+i\frac{2g}{3}\epsilon_{i j k} X^i X^j X^k
\right] \, .
}
}
Under the change of coordinates \eqref{eq: change of coordinates by U for X BMN}, \eqref{eq: change of coordinates by U for Theta BMN} and \eqref{eq: change of coordinates by U for P BMN}, we get 
\al{
H=H_{\rm kin} + V + H_F\,,
}
where $H_{\rm kin}$ is given in \eqref{eq: BMN H_kin} above and
\al{
\spl{
V = & \ \frac{g}{2} \sum_{a \neq b} |r_{a b}|^2 y_{a b}^I y_{b a}^I - 2 \sqrt{g} \sum_{a \neq b} \sum_{c \neq a,b} r_{a b}^I y_{a b}^J y_{b c}^I y_{c a}^J \\ &  - \frac{1}{2} \sum_{a,b} \sum_{c \neq a,b} \sum_{d \neq a,b} ( y_{a c}^I y_{c b}^J y_{b d}^I y_{d a}^J - y_{a c}^I y_{c b}^J y_{b d}^J y_{d a}^I) \\ & + \frac{1}{2} \left[ \frac{1}{3^2} \sum_{i=1}^3 (r_a^i)^2 + \frac{1}{6^2}  \sum_{p=4}^9 (r_a^p)^2 \right] + \frac{i}{2} \epsilon_{i j k} \sum_{a \neq b} y_{a b}^i y_{b a}^j r_{a b}^k + \mathcal{O}\left(\frac{1}{g}\right)\,.
}
}
\al{
\spl{
H_F = & - g \sum_{a < b} r_{a b}^I \Theta_{a b}^\dagger \gamma^I \Theta_{a b} - \sqrt{g} \sum_{a < b} \left( y_{b a}^I (\theta_b^T - \theta_a^T) \gamma^I \Theta_{a b} + y_{a b}^I \Theta_{a b}^\dagger \gamma^I (\theta_b - \theta_a) \right) \\ & - \sqrt{g} \sum_{a < b < c} (y_{a b}^I \Theta_{a c}^\dagger \gamma^I \Theta_{b c} + y_{a c}^I \Theta_{a b}^\dagger \gamma^I \Theta_{b c}^* + y_{b c}^I \Theta_{a b}^T \gamma^I \Theta_{a c}^* + h.c.) \\ & + \frac{i}{4} \sum_{a < b} \Theta_{a b}^\dagger \gamma_{1 2 3} \Theta_{a b} + \frac{i}{8} \theta_a^T \gamma_{1 2 3} \theta_a \,.
}
}
In particular, at the leading order we have
\al{
g H^{(1)} &= \sum_{a\neq b}
\left(
-\frac{1}{2}g\Pi^{IJ}_{ab}\frac{\del}{\del y_{ab}^I}\frac{\del}{\del y_{ba}^J}  
+ \frac12 g |r_{ab}|^2 y^I_{ab} y^I_{ba}
+ \frac12 g r_{ab}^I \Theta_{ab}^\top \g^I \Theta_{ba}
\right)\,,
}
which are supersymmetric harmonic oscillators in terms of the off-diagonal matrix elements.

\paragraph{Expansion of the supercharge}

Recall the supercharges are
\al{
\spl{
Q_\a = \operatorname{Tr}
\bigg[
P^I \gamma^I \hat\Theta
-\frac{i}{2}g\left[X^I, X^J\right] \gamma^{IJ} \hat\Theta
- \frac13 X^i \gamma^{123} \gamma^i \hat\Theta
+ \frac16 X^p \gamma^{123} \gamma^p \hat\Theta  
\bigg]_\a\,.
}
}
Under the change of coordinates \eqref{eq: change of coordinates by U for X BMN}, \eqref{eq: change of coordinates by U for Theta BMN} and \eqref{eq: change of coordinates by U for P BMN}, we get
\al{
Q_\a = \sqrt{g} Q_\a^{(\frac12)} + Q_\a^{(0)} + O(g^{-1/2})\,,
}
with
\al{
Q_\a^{(\frac12)}=
\sum_{a\neq b}
\left[
(\g^I \Theta_{ab})_\a \Pi_{ab}^{IJ} \left(-i \frac{\del}{\del y_{ab}^J}\right) 
-
i r_{ab}^I y_{ab}^J (\g^{IJ}\Theta_{ba})_\a
\right]\,,
\label{eq: Q 1/2 BMN}
}
and
\al{\spl{
Q_\a^{(0)} 
= 
&\sum_a (-i\g^I \theta_a)_\a
\left(
 \frac{\partial}{\partial r_a^I}
 + \sum_{b \neq a} \biggl( \frac{y_{b a}^I \hat{r}_{b a}^J}{|r_{a b}|} \frac{\partial}{\partial y_{b a}^J} - \frac{y_{a b}^I \hat{r}_{a b}^J}{|r_{a b}|} \frac{\partial}{\partial y_{a b}^J} \biggr)\right)
\\
&+
\sum_a \left[
-\frac{1}{3}r_a^i (\g^{123} \g^i \theta_a)_\a 
+\frac{1}{6}r_a^p (\g^{123} \g^p \theta_a)_\a 
\right]
\\
&+  \sum_{a\neq b}\left[ 
- \frac{i}{2} y_{ab}^I y_{ba}^J (\g^{IJ} \theta_{ab})_\a 
+ 
(\g^I \Theta_{ab})_\a \frac{- i\hat{r}_{a b}^I}{\left|r_{a b}\right|} \tilde{G}^F_{b a}
\right]
\\
&
+ \sum_{a\neq b\neq c} 
\left\{
(-i) y_{ac}^I y_{cb}^J (\g^{IJ} \Theta_{ba})_\a
- (\g^I \Theta_{ab})_\a  \frac{\hat{r}_{a b}^I}{\left|r_{a b}\right|} 
(y_{c a}^J \d_{bd} - y_{bd}^J \d_{ac}) \Pi_{c d}^{JK} \left(-i\frac{\partial}{\partial y_{c d}^K}\right)
\right\}\,.
}\label{eq: Q 0 BMN}}
Above we have replaced $U\frac{\del}{\del U}$ by $\tilde{G}^F$.

\paragraph{Expansion of the rotation generators}
Recall the rotation generators are
\al{
M^{IJ} = {\rm Tr}
\left(
X^I P^J - X^J P^I +\frac{i}{4}\hat\Theta^\top \g^{IJ} \hat\Theta
\right)\,.
}
Under the change of coordinates \eqref{eq: change of coordinates by U for X BMN}, \eqref{eq: change of coordinates by U for Theta BMN} and \eqref{eq: change of coordinates by U for P BMN}, we get
\al{\spl{
M^{IJ} 
&=
\sum_{a} \left[
r_a^I\left(-i \DD_a^J\right)
-
r_a^J\left(-i \DD_a^I\right)
+\frac{i}{4} \theta_a^\top  \gamma^{IJ} \theta_a
\right]
\\
&\quad
+
\sum_{a\neq b} \left[
y_{ab}^I\Pi_{ab}^{JK}\left(-i\frac{\del}{\del y_{ab}^K}\right)
-
y_{ab}^J\Pi_{ab}^{IK}\left(-i\frac{\del}{\del y_{ab}^K}\right)
+\frac{i}{4}  \Theta_{a b}^{\top}  \gamma^{IJ} \Theta_{ba}
\right]
\\&
\quad+
O(g^{-1/2})\,.
}}

\subsection{Finding the effective supercharges and the effective Hamiltonian}
The methodology of finding the effective supercharges and the effective Hamiltonian is explained in Section \ref{Sec: free gravitons from strong coupling}. In this subsection, we combine all the previous technical results to finish the final computation.

\subsubsection{Order $g^{1/2}$}
At the leading order our goal is to determine the ground state of $H^{(1)}$ given in \eqref{eq: H(1) BMN}.
Note that $H^{(1)}$ is just a set of supersymmetric harmonic oscillators acting only on the fast modes, so it can be diagonalised exactly and has zero ground state energy. Let us split it into bosonic and fermionic part
\al{\ga{
H^{(1)} = H_B^{(1)} + H_F^{(1)}\,,\\
H_B^{(1)} =\sum_{a\neq b}\left( -\frac12\Pi^{IJ}_{ab}\frac{\del}{\del y_{ab}^I}\frac{\del}{\del y_{ba}^J}  
+ \frac12|r_{ab}|^2 y^I_{ab} y^I_{ba} \right)\,,
\quad
H_F^{(1)} =\sum_{a\neq b}\left( \frac12 r_{ab}^I \Theta_{ab}^\top \g^I \Theta_{ba} \right) \,.
}}
The bosonic part $H_B^{(1)}$ is diagonalised by the Gaussian wavefunction $N(r)\varphi_r(y)$\footnote{The normalisation factor is fixed up to the leading order by imposing for the bosonic part of the wavefunction
\al{
\int \prod_{a=1}^N d^D\vec{r}_a \Delta(r) \int \prod_{a\neq b} \frac{d^D\vec{y}_{ab}}{g^4}  \d(\hat{r}_{ab}\cdot \vec{y}_{ab})
\left|\Y(r,y)\right|^2
=
\int \prod_{a=1}^N d^D\vec{r}_a
\left|\y(r)\right|^2\,,
}
where $\D(r)$ is the Vandermonde determinant.
}
\al{
N(r)\equiv \frac{1}{\sqrt{\Delta(r)}}
\prod_{a<b}
\left(\frac{2g|r_{ab}|}{\pi}\right)^{\frac{D-1}{2}}
\,, \quad
\varphi_r(y)\equiv e^{-\frac12\sum_{a\neq b}|r_{ab}|y_{ab}^I y_{ba}^I}\,, \quad
}
such that
\al{
g H_B^{(1)} N(r)\varphi_r(y) = g (D-1) \sum_{a<b} |r_{ab}| N(r)\varphi_r(y)\qquad (D=9)\,.
}
For the fermionic part let us define \cite{deWit:1988xki}
\al{
\tilde{\Theta}_{ab}\equiv D_{ab}\Theta_{ab}
\,,
}
with
\al{\ga{
D_{ab}\equiv\frac{1}{\sqrt{2}}
\left(
\sqrt{1+\hat{r}_{a b}^{D}}\,\mathbb{1} 
-  
\frac{\left(\sum_{i=1}^{D-1}\hat{r}_{a b}^i \gamma^i\right) \gamma^D}{\sqrt{1+\hat{r}_{a b}^{D}}}
\right)\,,
}}
which satisfies $D^\dagger_{ab}= D^\top_{ab} = D_{ab}^{-1}$. This allows us to write the fermionic term as
\al{
H_F^{(1)} = - \sum_{a<b} |r_{ab}|
(\tilde\Theta_{ab})^\dagger \g^D \tilde\Theta_{ab}\,,
}
where we have denoted $\Theta_{ba} = \Theta_{ab}^\dagger$.
We can further split the fermions into ``+'' and ``-'' modes by defining\footnote{With our convention of gamma matrices we have the following anti-commutation relations
\al{\ga{
\left\{(\tilde\Theta_{ab}^\pm)_\a , (\tilde\Theta_{ab}^\pm)^\dagger_\b \right\} = (P_\pm)_{\a\b}
\qquad
(\a,\b=1,\ldots,16)\,,
\\
\left\{(\tilde\Theta_{ab}^\pm)_{\a'} , (\tilde\Theta_{ab}^\pm)^\dagger_{\b'} \right\} = \d_{\a'\b'}
\qquad
(\a',\b'=1,\ldots,8)\,,
}}
where $\a'$ runs over the nonvanishing components of $(\tilde\Theta_{ab}^\pm)_\a$ and similarly for other fermionic spinors.
}
\al{
\tilde{\Theta}^\pm_{ab}
\equiv
P_{\pm} \tilde\Theta_{ab}
\,,
\qquad
P_{\pm} \equiv \frac{\mathbb{1}\pm \g^D}{2}\,,
}
then the fermionic term becomes
\al{
H_F^{(1)} = - \sum_{a<b} |r_{ab}|\left[(\tilde\Theta^+_{ab})^\dagger \tilde\Theta^+_{ab} - (\tilde\Theta^-_{ab})^\dagger \tilde\Theta^-_{ab} \right]\,.
}
The fermionic ground state is defined as being filled with all ``+'' mode excitations
\al{
\big|\xi_0(\hat r,\Theta)\big\rangle 
\equiv 
\prod_{a<b}\prod_{\a'=1}^8 (\tilde\Theta_{ab}^+)^\dagger_{\a'} |0\rangle_F\,,
\qquad
\big\<\xi_0(\hat r,\Theta)\big|\xi_0(\hat r,\Theta)\big\rangle = 1\,,
}
where $|0\>_F$ is the Fock vacuum. The fermionic ground state energy is
\al{
g H_F^{(1)} \big|\xi_0(\hat r,\Theta)\big\rangle = -g(D-1) \sum_{a<b}|r_{ab}| \big|\xi_0(\hat r,\Theta)\big\rangle\,,
}
which cancels exactly with the bosonic one.

Altogether the ground state of $H^{(1)}$ is
\al{
|\Y^{(0)}\rangle = |\y(r,\theta)\rangle \, |\W\rangle\,,
\qquad
|\W\rangle \equiv N(r) \varphi_r(y) \big|\xi_0(\hat r,\Theta)\big\rangle\,, 
\qquad
\<\W|\W\> = 1\,,
\label{eq: full form of ground state of BMN H^1}
}
which satisfies
\al{
H^{(1)}|\W\rangle = Q_\a^{(\frac12)}|\W\rangle = 0\,.
\label{eq: ground state of fast modes}
}
Above the slow mode wave function is denoted as $|\y(r,\theta)\rangle$ and is completely unspecified. Since nothing acts on the slow modes, we conclude that
\al{
H_\text{eff}^{(1)}=Q_{\text{eff},\a}^{(\frac12)}=0\,.
}

\subsubsection{Order $1$}
At order 1 the effective supercharge is defined as
\al{
Q^{(0)}_{\text{eff},\a} |\y(r,\theta)\rangle
\coloneqq
\underbrace{\langle \W | Q_\a^{(\frac12)}}_{=0} |\Y^{(-\frac12)}\rangle 
+
\langle \W | Q_\a^{(0)} |\Y^{(0)}\rangle 
\,.
}
This is analogous to the first line of \eqref{eq: def of H_eff in general strategy}. Note that at this order the effective supercharge is the same for BPS states or general states, because we are interested in the energy regime $E=O(1)$.


To compute the above (partial) matrix element, notice that all the terms in $Q_\a^{(0)}$ that are linear in $\Theta$ do not contribute and we are left with
\al{\spl{
Q_\a^{(0)} 
\supset
&
\sum_a \left[
 (-i\g^I \theta_a)_\a \DD_a^I
-\frac{1}{3}r_a^i (\g^{123} \g^i \theta_a)_\a 
+\frac{1}{6}r_a^p (\g^{123} \g^p \theta_a)_\a 
\right]
\\
&+  \sum_{a\neq b}\left[ 
(\g^I \Theta_{ab})_\a \frac{- i\hat{r}_{a b}^I}{\left|r_{a b}\right|} \theta_{ba}^\top \Theta_{ba}
\right]\,.
}}
Let us first consider the contribution from the interior derivative $\DD_a^I$:
\al{\ga{
\big\langle\xi_0(\hat r,\Theta)\big|\DD_a^I \big|\xi_0(\hat r,\Theta)\big\rangle
=
\sum_{c<d} 
\big\langle\xi_0(\hat r,\Theta)\big|
\left(\tilde{\Theta}_{c d}^{-}\right)^{\dagger} 
D_{c d} 
\frac{\partial D_{c d}^{\dagger}}{\partial r_a^I} 
\tilde{\Theta}_{c d}^{+}
\big|\xi_0(\hat r,\Theta)\big\rangle
= 0\,,
\\
\int [dy] N(r)\varphi^2_r(y) \DD_a^I N(r) + \int [dy] N^2(r)\varphi_r(y) \DD_a^I \varphi_r(y)= -
 \sum_{b\neq a} \frac{\hat{r}_{ab}^I}{|r_{ab}|}  \, .
\label{eq: cancellation d/dr's action}
}}
The first line above can be verified explicitly:
\al{
\spl{
\frac{\partial}{\del r_a^I} |\xi_0 \rangle 
& = 
\frac{\partial}{\partial r_a^I} \left[ \prod_\alpha \prod_{c < d} (D_{c d}^\dagger P_+)_{\gamma \alpha} (\Theta_{c d}^\dagger)_\gamma \right] | 0 \rangle_F 
\\ 
& =
\sum_{c < d}  \op{Tr}(D_{c d} \frac{\partial D_{c d}^\dagger}{\partial r_a^I} P_+) |\xi_0 \rangle + \sum_{c < d} \left((\Tilde{\Theta}_{c d}^-)^\dagger D_{c d} \frac{\del D_{c d}^\dagger}{\partial r_a^I} \Tilde{\Theta}_{c d}^+ \right) | \xi_0 \rangle 
\\ 
& = 
\sum_{c < d} \left((\Tilde{\Theta}_{c d}^-)^\dagger D_{c d} \frac{\del D_{c d}^\dagger}{\partial r_a^I} \Tilde{\Theta}_{c d}^+ \right) 
| \xi_0 \rangle
}
\label{eq: d/dr on xi_0}
}
where we used that the first trace vanishes, which can either be checked explicitly, or using the reality of this trace:
\al{
\op{Tr}\left(D_{c d} \frac{\partial D_{c d}^\dagger}{\partial r_a^I} P_+\right) 
= 
\op{Tr}\left(P_+ \frac{\partial D_{c d} }{\partial r_a^I} D_{c d}^\dagger \right) 
= 
- \op{Tr}\left(P_+  D_{c d} \frac{\partial D_{c d}^\dagger }{\partial r_a^I}\right) = 0 \, .
} 
\eqref{eq: d/dr on xi_0} indicates that $\del/\del_{r_a^I}$ acts as rotation generator in the fermion space since it annihilates a $\tilde\Theta^+$ and creates a $\tilde\Theta^-$. Also see Section 2.3 of \cite{Lin:2014wka}.

Finally, the matrix element of the $\Theta$ bilinear is
\al{
\big\langle\xi_0(\hat r,\Theta)\big|
\sum_{a\neq b} 
(\g^I \Theta_{ab})_\a \frac{-i\hat{r}_{a b}^I}{\left|r_{a b}\right|} 
\theta_{ba}^\top \Theta_{b a}
\big|\xi_0(\hat r,\Theta)\big\rangle 
=-i\sum_{a\neq b}\frac{\hat r_{ab}^I}{|r_{ab}|}(\g^I\theta_a)_\a
\,.
\label{eq: two Theta term for Q_eff^0}}
Altogether, the contributions \eqref{eq: two Theta term for Q_eff^0} and \eqref{eq: cancellation d/dr's action} cancel each other and we have
\al{
Q^{(0)}_{\text{eff},\a} = 
\sum_a \left[ (\g^I \theta_a)_\a \left(-i\frac{\del}{\del r_a^I}\right)
-\frac{1}{3}r_a^i (\g^{123} \g^i \theta_a)_\a 
+\frac{1}{6}r_a^p (\g^{123} \g^p \theta_a)_\a 
\right] \,.
}
By comparing with \eqref{eq: real supercharges of BMN} we see that $Q^{(0)}_{\text{eff},\a}$ is the diagonal supercharges with $g=0$, hence the corresponding effective Hamiltonian can be obtained straightforwardly
\al{
\left\{Q^{(0)}_{\text{eff},\alpha}, Q^{(0)}_{\text{eff},\beta}\right\}
=
2 \delta_{\alpha \beta} H^{(0)}_\text{eff}-\frac{1}{3}\left(\gamma^{123} \gamma^{i j}\right)_{\alpha \beta} M^{(0),ij}_\text{eff}
+\frac{1}{6}\left(\gamma^{123} \gamma^{pq}\right)_{\alpha \beta} M^{(0),pq}_\text{eff}\,.
}
with
\al{\spl{
H_\text{eff}^{(0)}
&= 
\sum_{a=1}^N \left(-\frac12 \frac{\del^2}{\del r_a^I \del r_a^I} + \frac12 \frac{1}{3^2} (r_a^i)^2
+ \frac12 \frac{1}{6^2}(r_a^p)^2
+\frac{i}{8} \theta_a^\top \g^{123} \theta_a\right)
\,,}\label{eq: BMN H_eff}
}
and
\al{
M^{(0),IJ}_\text{eff} &=
\sum_{a=1}^N \left[
r_a^I\left(-i \frac{\del}{\del r_a^J}\right)
-
r_a^J\left(-i \frac{\del}{\del r_a^I}\right)
+\frac{i}{4} \theta_a^\top  \gamma^{IJ} \theta_a
\right]\,,
}
which are nothing but the $U(1)^N$ part of \eqref{eq: BMN Hamiltonian} and \eqref{eq: BMN rotation generators}. Therefore, to the leading order in the strong coupling limit the BMN model describes $N$ copies of decoupled supersymmetric harmonic oscillators.

With the results given in the previous subsections, one can also compute the effective Hamiltonian directly through
\al{
\tilde{H}_{\rm eff}^{(0)}|\y\> \coloneqq 
\<\W|H^{(\frac12)}|\Y^{(-\frac12)}\>+\<\W|H^{(0)}|\Y^{(0)}\> = E^{(0)} |\y\>\,.
\label{eq in app: sketch of direct approach to H_eff}
}
This requires solving for $|\Y^{(-\frac12)}\>$ and we found that it gives the same effective Hamiltonian as above.

\subsection{Non-singlet states}
\label{subsec: non-singlet case}
In this subsection we consider $SU(N)$ non-singlet states and compute the corresponding effective Hamiltonian. 

In the $SU(N)$ singlet case we have the following constraint
\al{
G_{ab}|\Y\> = 0 \ \to\ \tilde{G}_{ab} |\Psi^{(0)} (r,\theta;y,\Theta) \rangle = 0\,,
\label{eq in app: SU(N) singlet condition}
}
where $\tilde{G}$ is the rotated $SU(N)$ generator $\tilde{G}\equiv U G U^\dagger = \left( U \frac{\del}{\del U} \right) - \tilde{G}^{F}$.
Let us explicitly derive this because it is informative for the non-singlet case. First notice that the $\theta,\,\Theta$ fermions defined in \eqref{eq: change of coordinates by U for Theta BMN} are $U$-dependant and $SU(N)$ invariant. More precisely,
\al{
\spl{
\left[\left( U \frac{\del}{\del U} \right)_{ab}, \theta_c \right] & = (\delta_{b c} - \delta_{a c}) \Theta_{a b} \,,
\\ 
\left[\left( U \frac{\del}{\del U} \right)_{ab}, \Theta_{c d} \right] &  = \delta_{b c} \Theta_{a d} - \delta_{a d} \Theta_{c b} - \delta_{b c} \delta_{a d} (\theta_c - \theta_d) \, .
}
}
Using these commutation relations, we can verify that
\be [\tilde{G}_{a b}, \theta_c] = [\tilde{G}_{a b}, \Theta_{c d}] = 0\,. \ee
As a result, without an explicit $U$ dependence of $\Psi^{(0)}$, $\tilde{G}$ can be commuted up to the Fock vacuum and annihilate the state, leading to the condition \eqref{eq in app: SU(N) singlet condition}.

In the non-singlet case, the reduced wave function can have an explicit $U$-dependence
\be | \Psi^{(0)} \rangle = | \Psi^{(0)}(r,\theta(U),U;y,\Theta(U)) \rangle = |\psi(r,\theta(U),U)\rangle |\Omega(y,\Theta(U))\rangle \, . \ee
Since $\tilde{G}_{a b}$ commutes with all the fermions, the only non-trivial action is on the explicit $U$-dependence in the reduced wave-function,
\be \Tilde{G}_{a b} | \Psi^{(0)} \rangle = \left[\left( U \frac{\del}{\del U} \right)'_{a b} |\psi(r,\theta,U) \rangle \right] |\Omega(y,\Theta) \rangle \, , \ee
where we denoted by $(U \frac{\del}{\del U})'_{a b}$ the derivatives with respect to $U$ which treat the $\theta, \Theta$ fermions as constants. 

To determine the effect of non-singlet states on the effective Hamiltonian, it turns out to be more convenient to adopt the computation explicitly using the Hamiltonian as per \eqref{eq in app: sketch of direct approach to H_eff}. It is enough to focus on the terms in the Hamiltonian that depend on the $U$-derivatives which appear in $O(g^0)$ term of $H_{\rm kin}$ \eqref{eq: BMN H_kin}. Without the replacement by the fermionic $SU(N)$ generator \eqref{eq: replacement rule for SU(N) singlet}, we have
\al{
\spl{
H_{\rm kin} \supset H_U \equiv \frac{1}{2} \sum_{a \neq b} \frac{1}{|r_{a b}|^2} & \left(U \frac{\del}{\del U}\right)_{a b} \left(U \frac{\del}{\del U}\right)_{b a} \\ & - \sum_{a \neq b \neq c} \frac{1}{|r_{a b}|^2} \left( y_{c a}^I \Pi_{b c}^{I J} \frac{\del}{\del y_{c b}^J} - y_{b c}^I \Pi_{a c}^{I J} \frac{\del}{\del y_{a c}^J} \right) \left(U \frac{\del}{\del U}\right)_{a b}\,.
}
}
We now use $\left(U \frac{\del}{\del U}\right)_{a b} = \Tilde{G}_{a b} + \tilde{G}^F_{a b}$ and $[\tilde{G}_{a b}, \tilde{G}^F_{c d}] = 0$ to write
\al{
\spl{
H_U = \frac{1}{2} & \sum_{a \neq b} \frac{1}{|r_{a b}|^2} (\Tilde{G}_{a b} \Tilde{G}_{b a} + 2  \tilde{G}^F_{a b} \Tilde{G}_{b a}  + \tilde{G}^F_{a b} \tilde{G}^F_{b a} ) \\ & - \sum_{a \neq b \neq c} \frac{1}{|r_{a b}|^2} \left( y_{c a}^I \Pi_{b c}^{I J} \frac{\del}{\del y_{c b}^J} - y_{b c}^I \Pi_{a c}^{I J} \frac{\del}{\del y_{a c}^J} \right) (\Tilde{G}_{a b} + \tilde{G}^F_{a b})\,.
}
}
It is now easy to check that
\be \langle \Omega | H_U | \Psi^{(0)} \rangle = \frac{1}{2} \sum_{a \neq b} \frac{1}{|r_{a b}|^2} \langle \Omega | \tilde{G}^F_{a b} \tilde{G}^F_{b a} | \Psi^{(0)} \rangle + \frac{1}{2} \sum_{a \neq b} \frac{1}{|r_{a b}|^2} \left( U \frac{\del}{\del U} \right)'_{a b} \left( U \frac{\del}{\del U} \right)'_{b a} |\psi \rangle \,. \ee
The first term also appeared in the singlet case, but we now get a new contribution,
\al{
\spl{
H_{\rm eff} = - \frac{1}{2} & \frac{\del^2}{\del r_a^I \del r_a^I} + \frac{1}{2} \left[ \frac{1}{3^2} \sum_{i=1}^3 (r_a^i)^2 + \frac{1}{6^2}  \sum_{p=4}^9 (r_a^p)^2 \right] + \frac{i}{8} \theta_a^T \gamma_{1 2 3} \theta_a \\ & + \frac{1}{2} \sum_{a \neq b} \frac{1}{|r_{a b}|^2} \left( U \frac{\del}{\del U} \right)'_{a b} \left( U \frac{\del}{\del U} \right)'_{b a} \, .
}
}

\subsubsection{Explicit check of $N=2$ case}
As an explicit check let us consider the case with $N=2$. Remember that $U \in SU(2)/U(1)$. We thus need an explicit parametrization of $SU(2)$, e.g. in terms of Euler angles,
\be U_{SU(2)} = \begin{pmatrix}
e^{i \psi/2} & 0 \\0 & e^{-i \psi/2} 
\end{pmatrix} \begin{pmatrix}
\cos(\theta/2) & \sin(\theta/2) \\ - \sin(\theta/2) & \cos(\theta/2)
\end{pmatrix} \begin{pmatrix}
e^{i \phi/2} & 0 \\0 & e^{-i \phi/2} 
\end{pmatrix} \ee
where $\phi \in [0,4\pi)$, $\psi \in [0,2 \pi)$, and $\theta \in [0, \pi)$.
Remember that $SU(2)/U(1)$ is defined by identifying $U \sim U_0 U$ where $U_0= \mathrm{diag}(e^{i \alpha/2}, e^{-i \alpha/2})$. Using this identification, we can gauge fix $\psi \to \phi$.
This gives
\be U = \begin{pmatrix} \cos(\frac{\theta}{2}) e^{i \phi } & \sin(\frac{\theta}{2})  \\ - \sin(\frac{\theta}{2}) & \cos(\frac{\theta}{2}) e^{- i \phi}
\end{pmatrix} \ee
where $\phi \in [0,2\pi)$, $\theta \in [0,\pi)$.
We now compute derivatives, using
\be \del_{\theta} = \frac{\del U_{a b}}{\del \theta}  U^\dagger_{b c} \left( U \frac{\del}{\del U} \right)_{c a} \ee
and similarily for $\phi$. In doing so, we set $(U \frac{\del}{\del U})_{1 1}, (U \frac{\del}{\del U})_{2 2} \to 0$ because they are related to the dependence on the Cartan variables. (Also note that this is consistent with the fact that those derivatives never appear in our computations). Inverting the relations, we obtain
\al{
\spl{
\left( U \frac{\del}{\del U} \right)_{1 2} & = \frac{i}{\sin \theta} e^{i \phi} (\del_\phi + i \sin \theta \del_\theta) \\ \left( U \frac{\del}{\del U} \right)_{2 1} & = \frac{i}{\sin \theta} e^{- i \phi} (\del_\phi - i \sin \theta \del_\theta)
}
}
The $U$-derivative term in the effective Hamiltonian becomes 
\al{
\spl{
 \frac{1}{2} & \sum_{a \neq b} \frac{1}{|r_{a b}|^2} \left( U \frac{\del}{\del U} \right)'_{a b} \left( U \frac{\del}{\del U} \right)'_{b a} 
 \\ & = - \frac{1}{|r_{1 2}|^2} \left(\del_\theta^2 + \frac{\cos \theta}{\sin \theta} \del_\theta + \frac{1}{\sin^2 \theta} \del_\phi^2\right) = - \frac{1}{ |r_{12}|^2} \Delta_{S^2} = \frac{l (l+1)}{ |r_{12}|^2}
}
}
where $\Delta_{S^2}$ is the Laplacian on the 2-sphere with eigenvalues $-l(l+1)$.

\section{Supplementary details for the $N=2$ minimal BMN model}
\label{App: N=2 mBMN properties}
In this appendix we give further details on the properties of the $N=2$ minimal BMN model that helps simplifying the Hamiltonian truncation study. 



First, the Hamiltonian truncation in practice is performed by using all symmetries available. In our case, we want to build basis states with an energy below a given cutoff and with a fixed $SO(2)$ charge $M$. Notice that there can be at most one single-trace operator containing fermions in a given basis state as can be seen from \eqref{eq: Hilbert series}. After having specified which operator is present, the charge can still be changed by acting with products of operators labelled by $T_9$ and $T_{10}$ (see Table \ref{tab:build}). In order to stay in a fixed $M$ sector, one can act with powers of $T_8$ or $T_9\times T_{10}$ which only increases the energy.

The sectors built out of $T_3$ and $T_7$ have $SO(2)$ charge $M=2n+3/2$ and $M=2n+1$, respectively. They are disconnected from each other and also from any other types of sectors, in that their $SO(2)$ charges can never be reached from other types of sectors by shifts of even integers carried by $T_8$, $T_9$ or $T_{10}$. More importantly, these two sectors do not feel the presence of the Yukawa term in the Hamiltonian \eqref{eq: mBMN hamiltonian in X^i}. This is because only the Yukawa terms can change the fermionic content while the $T_3$ and $T_7$ sectors are isolated so their fermionic content is fixed. One can also verify explicitly that the Yukawa terms annihilate $T_3|0\>$ and $T_7|0\>$. These sectors are less relevant for the purpose of this paper as their physics is that of a purely bosonic Hamiltonian\footnote{The large $N$ limit (with $g^2N$ fixed) of the bosonic model was studied in \cite{Han:2020bkb} using matrix quantum mechanics bootstrap.} with a constant shift due to the fermionic number operator:
\al{
H=\op{Tr}\bigg[
\frac{1}{2}(P^i)^2 + \frac{1}{2}(X^i)^2 - \frac{g^2}{2} [X^1,X^2]^2 
\bigg]
+\op{Tr}\left(\frac32 \L^\dagger \L\right)
\,,
\label{eq: bosonic two-matrix model}
}
It is symmetric under $X^1 \leftrightarrow X^2$, which sends $M \to -M-2$ for $M=2n+1$ and $M\to -M-1$ for $M=2n+3/2$. In addition, adjacent sectors whose $SO(2)$ charges differ by $1/2$ have the same spectrum up to an overall shift due to difference in fermionic numbers: $E_{M=2n+1} = E_{M=2n+3/2}+3/2$. The spectrum of this system grows as $g^{2/3}$ in the large-$g$ limit\footnote{This can be seen by considering the toy model 
\begin{align*}
H=\frac{1}{2}(p_x^2+p_y^2+ g^2 x^2 y^2)
\end{align*} 
of \eqref{eq: bosonic two-matrix model}. From the $x\leftrightarrow y$ symmetry we deduce that the typical length scales are $x\sim y\sim g^{-1/3}$ (see Section \ref{subsec: H approach toy model}). After rescaling $x=g^{-1/3}\tilde x,\,y=g^{-1/3}\tilde y$, the Hamiltonian becomes $H=\frac{g^{2/3}}{2}(\tilde p_x^2+\tilde p_y^2+\tilde x^2 \tilde y^2)$ so the energy is of $O(g^{2/3})$.
} and there are no low energy states at large coupling.

\begin{figure}[t]
    \centering
    \includegraphics[width=\textwidth]{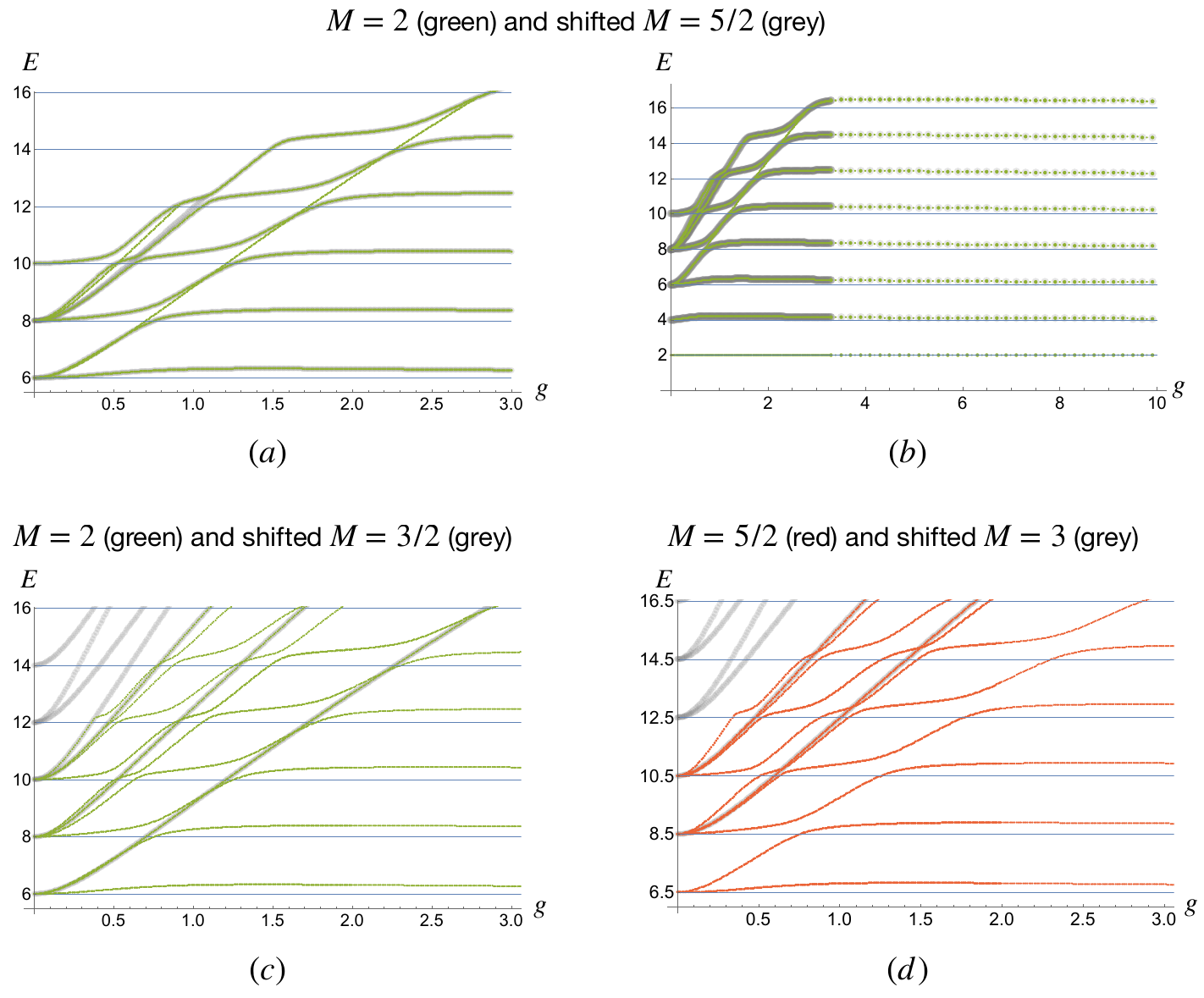}
    \caption{Comparison of spectrum from sectors whose $SO(2)$ charges differ by 1/2. Figure $(a)$ is a zoomed-in version of Figure $(b)$. The spectrum in the $M=3/2(3)$ sectors are shown with an overall $1/2(-1/2)$ shift, in order to facilitate straightforward comparison. This is the finite $g$ version of the structure illustrated in Table \ref{tab: example of susy multiplet pattern}, and is also the case in the middle of Figure \ref{fig: general pattern mBMN}.}
    \label{fig: susy long multiplet}
\end{figure}

The remaining sectors can be classified in two categories: $M=2n$ sectors built from $T_5$, $T_6$ or without fermions and $M=2n+1/2$ sectors built from $T_1$, $T_2$ or $T_4$. 
There are only BPS states in the non-negative, even charge sectors. To see this, consider the BPS equation:
\begin{align}
    \{Q,Q^{\dagger}\} = 2(H-M) \geq 0
\end{align}
At $g=0$ the full list of BPS states can be easily found to be
\begin{align}
    |\Y_\text{BPS}\rangle = T_9^k|0\rangle\,,\qquad k\in \mathbb{N}\,,
\end{align}
and there is one BPS state at each $M=2n$ sector with $n$ non-negative. Since the BPS states remain BPS for all $g$, we deduce that there is exactly one BPS state for each even non-negative charge sector.

The $\NN=2$ supersymmetry relates the states in the $M=2n+1/2$ sectors with those in the non-BPS states in $M=2n$ sectors through the action of the supercharge
\al{
|\Y_{E(>M),M(=2n)}\>\  \overset{Q}{\underset{Q^\dagger}{\rightleftarrows}} \ |\Y_{E+1/2,M+1/2}\>\,,\qquad (\text{for any } g)\,.
}
Each pair of such states forms an $\NN=2$ SUSY long multiplet, whereas the BPS states are short multiplets (with only one state in each short multiplet). To illustrate this SUSY multiplet structure, in Figure \ref{fig: susy long multiplet} $(a)$ and $(b)$ we show the spectrum of $M=2$ and $M=5/2$ sector from Hamiltonian truncation. Notice that we have shifted the entire $M=5/2$ spectrum by $-1/2$ to make the comparison manifest. We see that leaving out the BPS state with $E=2$ (for any $g$), the spectrum from the two sectors align exactly, except for two energy levels in the $M=2$ sector and one in the $M=5/2$ sector near the left end of Figure \ref{fig: susy long multiplet} $(a)$.

In fact, the total number of states with $M=2n+1/2$ at a certain energy $E+1/2$ does not equal the total number of states with $M=2n$ and energy $E$ in general. This is because supersymmetry connects all four types of charge sectors. For example, at $g=0$ the connection among the lowest energy levels of $M=3/2,2,5/2,3$ sectors through supersymmetry is given in Table \ref{tab: example of susy multiplet pattern}, where we have denoted the free theory supercharge by $Q_0$ and used the labels in Table \ref{tab:build}. The structure of how the states are connected continues to hold at finite $g$, as shown in Figure \ref{fig: susy long multiplet}. Since the energies in $M=3/2$ and $M=3$ sectors grow as $g^{2/3}$, this leads to the level crossing observed in Figure \ref{fig: susy long multiplet}.

\begin{table}[t]
\centering
\renewcommand{\arraystretch}{1.5}
\begin{tabular}{c | c c c c c c c c}
     \backslashbox{$E_{g=0}$}{$M$} & $\frac32$ & & $2$ & & $\frac{5}{2}$ & & $3$ 
     \\[5pt]\hline\hline
     $\cdot|2|\cdot|\cdot$ &  &  & $T_9|0\>\ (\text{BPS})$ & & & & & 
     \\[5pt]\hline
     $\cdot|4|\frac92|\cdot$ & & & $T_8 T_9|0\>$ & $\overset{Q_0}{\underset{Q_0^\dagger}{\rightleftarrows}}$ & $T_1 T_9|0\>$ & &
     \\[5pt]\hline
     $\cdot|6|\frac{13}{2}|\cdot$
     & & & $T_8 T_8 T_9|0\>$ & $\overset{Q_0}{\underset{Q_0^\dagger}{\rightleftarrows}}$ & $T_1 T_8 T_9|0\>$
     \\[5pt]
     $\cdot|6|\frac{13}{2}|\cdot$ &  &  & $T_9 T_9 T_{10}|0\>$ & $\overset{Q_0}{\underset{Q_0^\dagger}{\rightleftarrows}}$ & $T_9 T_9 T_2|0\>$
     \\[5pt]
     $\frac{11}{2}|6|\cdot|\cdot$ & $T_3 T_9|0\>$ & $\overset{Q_0}{\underset{Q_0^\dagger}{\rightleftarrows}}$ & $T_5 T_9|0\>$ &  &
     \\[5pt]\hline
     $\cdot|8|\frac{17}{2}|\cdot$
     &  &  & $T_6 T_9 T_9 |0\>$ & $\overset{Q_0}{\underset{Q_0^\dagger}{\rightleftarrows}}$ & $T_4 T_9 T_9 |0\>$
     \\[5pt]
     $\cdot|8|\frac{17}{2}|\cdot$ &  &  & $T_8 T_8 T_8 T_9|0\>$ & $\overset{Q_0}{\underset{Q_0^\dagger}{\rightleftarrows}}$ & $T_1 T_8 T_8 T_9|0\>$
      \\[5pt]
     $\cdot|8|\frac{17}{2}|9$ &  &  & $T_8 T_9 T_9 T_{10}|0\>$ & $\overset{Q_0}{\underset{Q_0^\dagger}{\rightleftarrows}}$ &
     \bigg\{\!\begin{tabular}{c}
     $T_1 T_9 T_9 T_{10}|0\>$  \\ $T_8 T_9 T_9 T_2|0\>$
     \end{tabular}\!\!\bigg\}
     & $\overset{Q_0}{\underset{Q_0^\dagger}{\rightleftarrows}}$ & $T_1 T_2 T_9 T_9 |0\>$
     \\[5pt]
     $\frac{15}{2}|8|\cdot|\cdot$ & $T_3 T_8 T_9|0\>$ & $\overset{Q_0}{\underset{Q_0^\dagger}{\rightleftarrows}}$ & $T_5 T_8 T_9|0\>$ & & 
\end{tabular}
\caption{An example to illustrate the SUSY multiplet pattern in the $N=2$ minimal BMN model with $g=0$. The two states inside the curly brackets form two independent linear combinations, with one in the SUSY long multiplet also containing $T_8 T_9 T_9 T_{10}|0\>$ and the other in the long multiplet also containing $T_1 T_2 T_9 T_9 |0\>$.}
\label{tab: example of susy multiplet pattern}
\end{table}

\begin{figure}[t]
    \centering
    \includegraphics[width=\textwidth]{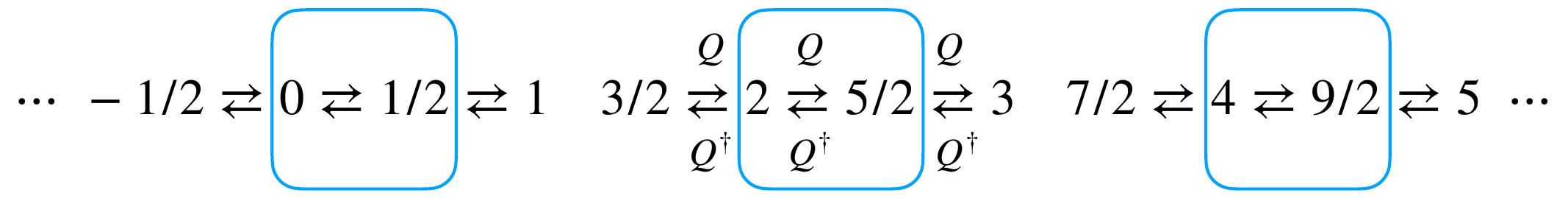}
    \caption{General spectrum matching pattern among $SO(2)$ charge sectors. The numbers indicate $SO(2)$ charges and the energies are suppressed. The arrows should be understood in the sense illustrated in Table \ref{tab: example of susy multiplet pattern}. In particular, $Q$ or $Q^\dagger$ cannot act on any state more than once. The blue boxes indicate SUSY long multiplets involving $M=2n$ and $M=2n+1/2$ sectors and this structure is illustrated in Figure \ref{fig: susy long multiplet} $(a)$ and $(b)$.}
    \label{fig: general pattern mBMN}
\end{figure}

The above discussion on matching among the (shifted) spectrum from different $SO(2)$ charge sectors can be generalised straightforwardly, and the pattern is summarised in Figure \ref{fig: general pattern mBMN}. Notice that the supercharges do not connect adjacent $M=2n+1$ and $M=2n+3/2$ sectors. The reason is the following. As explained above, the states in the $M=2n+3/2$ sectors can only be of the form $T_3 T_8^{n_8} T_9^{n_9} T_{10}^{n_{10}}|0\>$ and all these building blocks are annihilated by $Q_0^\dagger$, being the lowest weight states. Similarly, the states in the $M=2n+1$ sectors are of the form $T_1 T_2 T_8^{n_8} T_9^{n_9} T_{10}^{n_{10}}|0\>$ and are annihilated by $Q_0$, being the highest weight states. Therefore, acting $Q_0$ on a certain $M=2n+1$ sector cannot bring us to the $M=2n+3/2$ sector (and acting with $Q_0^\dagger$ instead decreases the $SO(2)$ charge). This structure should continue to hold for non-zero $g$ and is also partially observed numerically. A rigorous proof should be straightforward but is irrelevant for the purpose of this paper, because we focus on the part of the spectrum that remain $O(1)$ in the large $g$ limit, which only appear in $M=2n$ and $M=2n+1/2$ sectors.

\bibliographystyle{utphys} 
\bibliography{refs}
\end{document}